\documentclass[12pt]{article}

\usepackage{amstext,amsmath,amssymb,amsfonts,bbm,slashed}
\usepackage[latin1]{inputenc}
\usepackage{epsfig}
\usepackage{hyperref}
\usepackage{amsthm}
\usepackage{subfigure}
\usepackage{color}
\usepackage{multirow}
\usepackage{calc}
\usepackage{comment}

\usepackage{stmaryrd}

\usepackage{cancel}
\usepackage{graphicx}

\usepackage{float}

\usepackage{soul}

\usepackage[normalem]{ulem}
\usepackage{amsmath}
\newcommand{\stkout}[1]{\ifmmode\text{\sout{\ensuremath{#1}}}\else\sout{#1}\fi}

\usepackage[T1]{fontenc}
\usepackage{array}
\usepackage{makecell}
\newcolumntype{x}[1]{>{\centering\arraybackslash}p{#1}}
\usepackage{tikz}

\usepackage{pbox}

\newcommand{\kin}{{\rm{kin\,}}}
\newcommand{\ext}{{\rm{ext\,}}}
\newcommand{\inter}{{\rm{int\,}}}

\newcommand{\lam}{\lambda}
\newcommand{\laren}{\lambda_{\rm ren} }
\newcommand{\lapren}{ \lambda_{+; {\rm ren}} }
\newcommand{\lac}{\lambda^{(c)}}
\newcommand{\lacren}{\lambda^{(c)}_{\rm ren}}
\newcommand{\lapc}{ \lambda^{(c)}_+ }
\newcommand{\lapcren}{\lambda^{(c)}_{\rm +;ren}}

\newcommand{\pet}{\lambda_{+}}

\newcommand{\rhop}{\rho_{+}}
\newcommand{\rhot}{\rho_{\times}}

\newcommand{\omp}{\omega_{{\rm d};+}}
\newcommand{\omt}{\omega_{{\rm d};\times}}

\def\C{{\mathbbm C}}
\def\N{{\mathbbm N}}
\def\Z{{\mathbbm Z}}
\def\R{{\mathbbm R}}

\newcommand{\cG}{{\mathcal G}}
\newcommand{\bG}{{\partial\mathcal G}}

\newcommand{\cexG}{\mathcal G_{\text{color}}}

\newcommand{\cO}{{\mathcal O}}

\newcommand{\cL}{{\mathcal L}}
\newcommand{\cV}{{\mathcal V}}

\newcommand{\bdel}{ {\boldsymbol{\delta}} }

\newcommand{\bmu}{\boldsymbol{\mu}}

\newcommand{\Tr}{{\rm Tr}}

\newcommand{\Sym}{ {\rm Sym} }

\newtheorem{theorem}{Theorem}
\newtheorem{proposition}{Proposition}

\newcommand{\bea}{\begin{eqnarray}}
\newcommand{\eea}{\end{eqnarray}}
\newcommand{\beq}{\begin{equation}}
\newcommand{\eeq}{\end{equation}}

\newcommand{\be}{\begin{equation}}
\newcommand{\ee}{\end{equation}}

\makeatletter

\begin{document}

\begin{titlepage}
\begin{flushright}
\end{flushright}

\vspace{20pt}

\begin{center}

{\Large\bf One-loop beta-functions of \\ quartic enhanced  tensor field theories \\

\medskip

}
\vspace{15pt}

{\large Joseph Ben Geloun$^{a,c,\dag}$ and Reiko Toriumi$^{b,\ddag} $}

\vspace{15pt}

$^{a}${\sl Laboratoire d'Informatique de Paris Nord UMR CNRS 7030}\\
{\sl Universit\'e Paris 13, 99, avenue J.-B. Clement, 93430 Villetaneuse, France} \\

\vspace{5pt}

$^{b}${\sl Okinawa Institute of Science and Technology }\\
{\sl 1919-1 Tancha, Onna-son, Kunigami-gun, Okinawa, Japan 904-0495 }

\vspace{5pt}

$^{c}${\sl International Chair in Mathematical Physics 
and Applications\\ (ICMPA-UNESCO Chair), University of Abomey-Calavi,\\
072B.P.50, Cotonou, Rep. of Benin}\\

\vspace{5pt}

E-mails:  {\sl $^{\dag}$bengeloun@lipn.univ-paris13.fr,
$^\ddag$reiko.toriumi@oist.jp}

\vspace{10pt}

\begin{abstract}
Enhanced tensor field theories (eTFT) have dominant graphs that differ from the melonic diagrams of conventional tensor field theories.
 They therefore describe pertinent candidates to escape the 
so-called branched polymer phase, the universal geometry found for  tensor models. For generic order $d$ of the tensor field,  we compute the perturbative $\beta$-functions at one-loop of two just-renormalizable quartic eTFT coined by $+$ or $\times$,  depending on  their vertex weights.
The models $+$ has 
two quartic coupling constants
$(\lambda, \lambda_{+})$,
and two 2-point couplings 
(mass, $Z_a$). 
Meanwhile, 
the model $\times$ has two quartic coupling constants $(\lambda, \lambda_{\times})$
and three 2-point couplings  (mass, $Z_a$, $Z_{2a}$). 
At all orders,
both models have a constant wave function renormalization: 
$Z=1$ 
and therefore no anomalous dimension. Despite such  peculiar behavior, both models acquire nontrivial radiative corrections for the coupling constants.
The RG flow of the model $+$ exhibits a particular asymptotic
safety: 
$\lambda_{+}$ 
is marginal without corrections
thus is a fixed point of arbitrary constant value. 
All remaining couplings determine relevant directions and get
suppressed in the UV.   
Concerning the  model $\times$, 
$\lambda_{\times}$ is marginal and again a fixed point
(arbitrary constant value), 
$\lam$, $\mu$ and $Z_a$ 
are all relevant couplings and flow to 0. 
Meanwhile $Z_{2a}$ is a marginal
 coupling and becomes a linear function of the time scale. 
This model can neither be called asymptotically safe or free.
\end{abstract}

\today

\end{center}

\noindent  Pacs numbers:  11.10.Gh, 04.60.-m, 02.10.Ox
\\

\bigskip

\setcounter{footnote}{0}

\end{titlepage}

%%%%%%%%%%%%%%%%%%%%%%%

\tableofcontents

\section{Introduction}
\label{intro}

In the search for a gravitational theory resulting from the random generation of geometries, the most notable success to date restricts to two dimensions \cite{Di Francesco:1993nw}. The models at the base of this success are those of the celebrated random matrices. These statistical models generate simplicial complexes of dimension 2 and, therefore, discretized random surfaces. An impressive number of results pertains to  these models (KPZ equation, integrable models, string theory, ...) 
\cite{Knizhnik:1988ak} 
\cite{David:1988hj}
\cite{kawai} 
\cite{Morozov:1995pb}
\cite{thooft}.
More recently, it has been proven \cite{legallmiermont} \cite{Miller2015LiouvilleQG} that the continuous limit of such random surfaces provides a fluctuating spherical geometry, the Brownian sphere, endowed with the famous Liouville conformal field.
Building on the success of statistical matrix models, models of random tensors  
\cite{razvanbook} \cite{adrianbook}
generate discrete random geometries in higher dimensions and
aim at  extending these results  to higher dimensions \cite{ambj3dqg} \cite{mmgravity} \cite{sasa1} 
\cite{boul}.  
However, their continuum limit is singular and of (Hausdorff) dimension less than 2 
\cite{Bonzom:2011zz}.
The generated continuous spaces belong to the so-called universal branched polymer geometry which cannot
faithfully describe the geometry of our 4-dimensional space-time. 

Let us take a closer look at these tensor models and their universality classes. At large size $N$ of the tensor indices,  tensor statistical models are dominated by a class of graphs called melonic diagrams or melons. They 
mainly contribute to the construction of the continuum limit of tensor models. 
Melonic diagrams  define a subclass of topologically spheres and form a tree structure that characterizes the so-called branched polymer phase 
\cite{Gurau:2013cbh}. 
It is a striking feature of several tensor models that such melons  dominate universally and can be  studied analytically. Nevertheless, in the context of quantum gravity, this class of universal branched polymers is undesirable. We therefore aspire to escape from this branched polymer phase.

In order to improve  the critical behavior of tensor models, an interesting idea then emerged: one could adjust the scale $N$, considering coupling constants of non-melonic interactions in such a way that a broader class of graphs, including non-melonic ones, contributes to the critical behavior of tensor models \cite{Bonzom:2015axa}. 
Thus, making the graphs that were previously suppressed contribute to the large  limit $N$ defines the working hypothesis now put forward. We call the models of such a  program {\it enhanced tensor models}. Note that a model exhibiting a phase transition towards a two-dimensional geometry was found in \cite{Bonzom:2015axa}. 
Therefore, this already proves that enhanced tensor models produce different universality classes and phases from branched polymers.  
However, notably, the counterpart of these results in quantum field theory (QFT) still remains a vast territory to explore.

QFT is one of the most successful languages of modern physics. It describes with incredible precision complex systems with an infinity of degrees of freedom via the renormalization group (RG). 
In a different perspective from statistical models whose degrees of freedom are fixed,  QFT characterized by a nontrivial propagator unfolds a model flow from microscopic (ultraviolet)  to macroscopic (infrared) scales. 
We consider this point of view to be relevant for random geometry and quantum gravity. Indeed, if the degrees of freedom or modes of the tensor fields are geometric like spacetime quanta, 
and if the presence of the nontrivial kinetic term induces a flow from UV to IR governed by a given dynamics, we consider that going towards the infrared, these degrees  of freedom could agglomerate and form new degrees of freedom in a way similar to a condensation phenomenon 
\cite{oriti}
\cite{Oriti:2006ar} 
\cite{Marchetti:2022igl} \cite{Eichhorn:2017xhy}
\cite{BenGeloun:2011rc}.
The fact that some QFT of tensor fields called Tensor Field Theories (TFTs), are asymptotically free, and therefore act like Quantum Chromodynamics, brings one more stone to this building 
\cite{BenGeloun:2013vwi}
\cite{Rivasseau:2015ova}.

Studying TFTs requires adopting the powerful renormalization group methods of QFT to extract their critical behavior. The renormalization analysis of a TFT is complex because it is a nonlocal field theory 
\cite{BenGeloun:2011rc} \cite{Geloun:2014ema} 
\cite{Carrozza:2013mna} 
\cite{Geloun:2012fq}
\cite{BenGeloun:2012pu}
\cite{BenGeloun:2012yk}.
Additionally,  one  needs to  treat the dimensions  of coupling constants carefully in order to analyze their flows
\cite{Carrozza:2014rba} 
\cite{Carrozza:2014rya}
\cite{Benedetti:2014qsa}
\cite{BenGeloun:2015xrk}
\cite{BenGeloun:2016rqa}.

More recently, enhanced tensor field theories (eTFTs) have been introduced
as extensions of enhanced tensor models \cite{BenGeloun:2017xbd} 
\cite{Geloun:2015lta}. 
They are proven renormalizable 
at all orders of perturbation theory
and, for the first time, non-melonic graphs contribute to the renormalization group 
analysis of the coupling constants. 
Hence, they are crucial candidates for generating a geometric phase different from that of branched polymers and those of 2D geometries.

In this work, we take further the analysis of \cite{BenGeloun:2017xbd} and compute the $\beta$-functions of two eTFT models. One model is called $+$ and
the second model is labeled by $\times$. They were shown to be perturbatively just-renormalizable, at all orders, for given sets of parameters $(d,D,a,b)$, where $d$ is the tensor field order, 
$D$ is the background group dimension, 
$b$ parametrizes the kinetic term $\sim p^{2b}(\phi_{p})^2$, 
and $a$ determines the vertex weight $\sim p^{2a}(\phi_{p})^4$.
The $\beta$-function of model $+$ is studied for generic order
$d\ge 3$ of the tensor field and $(D=1, a=(d-2)/2, b=(d-3/2)/2)$. 
The second model $\times$ has fixed parameters
 $(D=1,d=3, a=1/2, b=1)$. 
Both models show exotic features compared to the ordinary TFT and ordinary scalar $\phi^4$ QFT. 

Our results are the following: 

(1) In both models,  the wave function renormalization is constant $Z=1$ and therefore the anomalous dimension is vanishing. Although this hints at a classical behavior,  there is still a RG flow; 
the system of $\beta$-functions is remarkably simple 
and explicitly integrable 
at one loop.
We give the solutions of all 
equations in terms of the so-called  time scale ($t =   \log (k/k_0)$). 
Based on the power counting theorem of each model and given the type of corrections that appear, we  conjecture that the system of $\beta$-functions
of each model can be explicitly 
solved and the models can be constructed 
in the way of \cite{Rivasseau:1991ub} at all orders of perturbation theory. This is a
genuine feature of the present quantum models.

(2) The model $+$
has two quartic couplings, 
$\lam$ and $\lam_+$, 
two 2-point couplings, 
a mass $\mu$ and another 2-point coupling $Z_a$
for a quadratic term $p^{2a}(\phi_{p})^2$ (introduced to ensure the  renormalizability of the model), and a wave function 
renormalization $Z$. 
The asymptotic UV-behaviors of all 
dimensionless couplings are given by (in loose notation that we will make precise)
\bea
&&
\lam_{\Lambda \to \infty} = 0 \crcr
&& 
\lam_{+;\Lambda \to \infty}  = \theta \crcr
&&
\mu_{\Lambda \to \infty}   = 0 \crcr
&& 
Z_{a; \Lambda \to \infty}  = c \, \theta \crcr
&&
Z = 1 
\eea
where $\theta$ and $c$ are real constants. 
The coupling  
$\lam_+$ is marginal and does not acquire any correction: it is a fixed point.  The second quartic coupling $\lam$ is relevant,  still running, 
tends to blow up in the IR and becomes 
exponentially suppressed in the UV.  
The mass and $Z_a$ are 
both relevant operators and 
get suppressed in the UV. 
In particular, $Z_a$ reaches a constant whereas the mass vanishes. 
We can call this model 
asymptotically safe with a free parameter $\lam_+ = \theta$ (a line of fixed points) and endowed with 3 relevant directions. However, one notes that $\lam_+$ does not 
flows to reach a fixed value, it 
is already set at a constant value. 
In a sense, this is a new kind of asymptotic safety. 
Strictly speaking,
this model is not asymptotically free
as two couplings, namely $\lam_+$
and $Z_a$ do not flow to 0, 
and therefore the Gaussian fixed
point is never reached unless 
$\lam_+$ is set to 0.  
The model is nearly super-renormalizable  because of the number
of relevant couplings 
that flow. 
However it cannot be stringently  
super-renormalizable 
because of the presence of a marginal coupling.

(3) The model $\times$ is endowed with two quartic couplings, $\lam_\times$ 
and $\lam$, three 2-point couplings
$\mu, Z_a$ and $Z_{2a}$, and a wave function renormalization $Z$. The UV-behavior of the dimensionless coupling is delivered
by the system (in loose notation again): 
\bea
&&
\lam_{\Lambda \to \infty} = 0 \crcr
&& 
\lam_{\times;\Lambda \to \infty}  = \theta \crcr
&&
\mu_{\Lambda \to \infty}   = 0 \crcr
&& 
Z_{a; \Lambda \to \infty}  =0\crcr
&&
Z_{2a; \Lambda \to \infty}  = \lim _{\Lambda \to \infty} (c\, \theta \log {\Lambda} + c' )=\infty
\crcr
&&
Z = 1 
\eea
where $\theta$, $c$ and $c'$ are constants.  
The coupling $\lam_\times$ is marginal and does not receive corrections. It becomes a fixed point: for any  $\lam_\times = \theta$ is a line of fixed points. The second coupling $\lam$ is relevant: it is suppressed in the UV and flows to 0. So behave the mass $\mu$ and the coupling $Z_a$. 
 Introduced
to make sense of  renormalizability,
the last 2-point coupling $Z_{2a}$
is marginal and grows linearly in the time scale. 
As we never reach the Gaussian 
fixed point (unless we set $\lam_\times = \theta =0$), this model, because of $\lam_+ = \theta \neq 0$, cannot be called asymptotically free. 
On the other hand, because $Z_{2a}$  runs 
to infinity, the model cannot be called
safe (unless once again $\lam_\times = \theta=0)$. This is again a noteworthy behavior different from conventional QFT and TFT models.

The paper is organized as follows.
In section \ref{rev}, we review the essence of \cite{BenGeloun:2017xbd} and present  two models named  model $+$ and  model $\times$.
In section \ref{betaf}, we compute the $\beta$-functions of the model 
$+$ at 1-loop and interpret the results. 
The next section \ref{betat} presents the same analysis for   
the model $\times$. 
The conclusion in section \ref{conc} provides a summary of our work 
and present some perspectives. 
In appendix \ref{app:SumtoInt}, we  detail the integral approximation techniques 
that we will use to tackle the spectral sums which appear in the calculation of the $\beta$-functions. In appendices \ref{app:2ndO+} and \ref{app:2ndOx}, we illustrate the divergent graphs at second order in perturbation theory to support our different conjectures.

\section{Review of the renormalizable enhanced quartic TFT}
\label{rev}

We review here the main 
results of  \cite{BenGeloun:2017xbd}
introducing a class of quartic eTFTs parametrized by 
$(D,d,a,b)\in \N\times \N\times\R_+\times\R_+$. 
In the following sections, we will specify the 4-tuple dealing only 
with specific models. 

\subsection{Enhanced TFT models}
\label{sec:etftmodels}

Consider a field theory defined by a complex function $\phi: (U(1)^{D})^{\times d} \to \C$. 
The Fourier transform of this complex field yields 
 a  order $d$ complex  tensor $\phi_{\bf P}$, with ${\bf P}=(p_1,p_2,\dots,p_d)$ a multi-index. 
$\bar\phi_{\bf P}$ denotes its complex conjugate. 
 Note that in the notation $\phi_{\bf P}$ the indices $p_s$ define themselves multi-indices: 
\be
p_{s} =(p_{s,1},p_{s,2},\dots,p_{s,D}) \,,\; \, p_{s,i}  \in \Z\,. 
\ee

A tensor field theory action  $S$ is built by a sum of convolutions of the tensors 
$\phi_{\bf P}$ and $\bar\phi_{\bf P}$:  
\bea\label{eq:actiond}
&&
S[\bar\phi,\phi]=\Tr_2 (\bar\phi \cdot 
{\bf K}
\cdot \phi) 
+ \mu
\, \Tr_2 (\phi^2) + S^{\inter}[\bar\phi,\phi]\,, 
\cr\cr
&&
\Tr_2 (\bar\phi \cdot 
{\bf K}
\cdot \phi) =
\sum_{{\bf P}, \, {\bf P}'} \bar\phi_{{\bf P}} \, 
{\bf K}({\bf P};{\bf P}')
\, \phi_{{\bf P}' } \,, 
\qquad 
\Tr_{2}(\phi^2) = \sum_{{\bf P}} \bar\phi_{{\bf P}}\phi_{{\bf P}}\,, 
\label{eq:generalaction}
\eea
and where $S^{\inter}[\bar\phi,\phi]$ is another convolution involving, 
in the case we are focusing on, 4 tensors.

 Hence, giving  ${\bf K}$ and   $S^{\inter}[\bar\phi,\phi]$ entirely determines 
 the models. For a real parameter $b\ge 0$, we  introduce the class of kernels  
 ${\bf K} ={\bf {K}}_b({ \bf P}; {\bf P'} ) 
=\bdel_{ { \bf P}; {\bf P'} } {\bf P}^{2b}$ where 
\bea
\label{delmom}
&&
\bdel_{ { \bf P}; {\bf P'} } =    \prod_{s=1}^d\prod_{i=1}^D  \delta_{ p_{s,i}, p'_{s,i}}  \,,\qquad
{\bf P}^{2b} = \sum_{s=1}^d |p_s|^{2b}\,,  \qquad  |p_{s}|^{2b} = \sum_{i=1}^D |p_{s,i}|^{2b} \,,
\eea
where $ \delta_{ p,q}$ denotes the usual Kronecker symbol on $\Z$. 
It is clear that ${\bf K}_b$ represents 
a power sum of eigenvalues of  $d$ Laplacian operators over the $d$ copies of $U(1)^D$
 ($b=1$ precisely  corresponds  to Laplacian 
eigenvalues on the torus). 
Seeking renormalizable theories \cite{BenGeloun:2017xbd}, more freedom
for $b$ values, allowing even values different from integers, leads to interesting models. 
In ordinary quantum field theory (QFT),   the restriction  
$b\leq 1$ ensures the Osterwalder-Schrader (OS)
positivity axiom \cite{Rivasseau:1991ub,Rivasseau:2011hm}.   
In any case,  $b$ will be set as a strictly positive real parameter
with no other restriction. 

We concentrate on the interaction part. Given a parameter $a\ge 0$,  
we distinguish the following quartic interactions: 
\bea
&& 
\Tr_{4;1}(\phi^4) = \sum_{p_{s},  p'_{s} \in  \Z^D}
\phi_{12\dots d} \,\bar\phi_{1'23\dots d} \,\phi_{1'2'3'\dots d'} \,\bar\phi_{12'3'\dots d'} 
\,, 
\label{phi4sim}\\
&&
\Tr_{4;1}([p^{2a}+p'^{2a}]\,\phi^4) = \sum_{p_{s}, p'_{s} \in \Z^D} 
\Big( |p_{1}|^{2a} + |{p'}_{1}|^{2a}\Big)\phi_{12\dots d} \,\bar\phi_{1'23\dots d} \,\phi_{1'2'3'\dots d'} \,\bar\phi_{12'3'\dots d'}
\,, \cr\cr
&& = 
2\sum_{p_{s}, p'_{s} \in \Z^D} 
|p_{1}|^{2a}\,\phi_{12\dots d} \,\bar\phi_{1'23\dots d} \,\phi_{1'2'3'\dots d'} \,\bar\phi_{12'3'\dots d'} = 2\,\Tr_{4;1}(p^{2a}\,\phi^4) 
 \,,
\label{intplus}  \\ 
&&
\Tr_{4;1}([p^{2a}p'^{2a}]\,\phi^4) = \sum_{p_{s}, p'_{s} \in \Z^D} 
\Big( |p_{1}|^{2a}  |{p'}_{1}|^{2a}\Big)\phi_{12\dots d} \,\bar\phi_{1'23\dots d} \,\phi_{1'2'3'\dots d'} \,\bar\phi_{12'3'\dots d'}
\,.
\label{intprod} 
 \eea
 where the notation $\phi_{12\dots d} $ stands for $\phi_{p_1,p_2,\dots, p_d} = \phi_{\bf P}$.
In the above equations \eqref{phi4sim}, \eqref{intplus} and \eqref{intprod}, the color index 1 plays
a special role.  The sum over all possible color configurations delivers  colored symmetric  interactions: 
\bea
&& 
 \Tr_{4}(\phi^4)
 :=  \Tr_{4;1} (\phi^4) + \Sym (1 \to 2 \to \dots \to d) 
\,, \label{eq:3dinter} \\
&& 
\Tr_{4}(p^{2a}\,\phi^4)
 :=  
 \Tr_{4;1} (p^{2a}\,\phi^4)
 + \Sym (1 \to 2 \to \dots \to d) \,, 
 \label{eq:ourinteraction+}\\
&& 
\Tr_{4}([p^{2a}p'^{2a}]\,\phi^4)
 :=  
 \Tr_{4;1} ([p^{2a}p'^{2a}]\,\phi^4)+ \Sym (1 \to 2 \to \dots \to d) 
 \,. 
\label{eq:ourinteractionx}
\eea
One might consider 
the momentum weights  in the above interactions as  derivative couplings
for particular choices of $a$. 
We look for just renormalizable models in our theory space, and then will constrain $a$ 
{\it a-posteriori} to some particular 
values. 
 It turns out that  the two interactions \eqref{eq:ourinteraction+}
and \eqref{eq:ourinteractionx} generate two new 2-point diverging graphs
that neither the mass nor the wave function renormalization can absorb. 
They are of the form: 
\bea
 \Tr_2 (p^{2\xi}\phi^2) = 
\Tr_2 (\bar\phi \cdot 
{\bf K}_{\xi}
\cdot \phi)\,, \qquad \xi = a,2a\,,
\eea
In \cite{BenGeloun:2017xbd}, the authors handled these
by adding them in kinetic term, in addition 
to the former term $\Tr_2 (p^{2b}\phi^2)$. 
The models that proves renormalizable have the following kinetic terms and  interactions: 
\bea
\text{model }+: 
&&
S^{\inter}_+[\bar\phi,\phi] =  \frac{ \lambda}{2}\,\Tr_{4}(\phi^4)  
 +\frac{\pet}{2}\, \Tr_{4}(p^{2a}\,\phi^4)
  + CT_{2}[\bar\phi,\phi] +\sum_{\xi=a,b}CT_{2;\xi}[\bar\phi,\phi] 
 \cr\cr
&& 
S^{\kin}_+[\bar\phi,\phi] =  
\sum_{\xi=a,b}  
Z_\xi   \Tr_2 (p^{2 \xi} \phi^2) 
 + \mu \Tr_2 (\phi^2) \,,
\label{model1} \\ 
\text{model }\times: &&
S^{\inter}_\times[\bar\phi,\phi] =   \frac{ \lambda}{2}\,\Tr_{4}(\phi^4)  
 +\frac{\lambda_\times}{2}\, \Tr_{4}([p^{2a}p'^{2a}]\,\phi^4)
 + CT_{2}[\bar\phi,\phi] +\sum_{\xi=a,2a,b}CT_{2;\xi}[\bar\phi,\phi]  \,,
 \cr\cr
&& 
S^{\kin}_\times[\bar\phi,\phi] =  
\sum_{\xi=a,2a,b} Z_\xi  \Tr_2 (p^{2 \xi} \phi^2) 
 + \mu \Tr_2 (\phi^2)
\label{model2}
\eea
where  $\lambda$, $\pet$, $\lambda_\times$ are coupling constants, 
$CT_{(\cdot)}$ are counterterms, $\mu$ is the mass coupling, $Z_a$ and $Z_{2a}$
are other kinetic term couplings,  and $Z_b$ will be called the wave function
renormalization. 
Note that we could be also interested 
in a model $\lam=\pet$, in which case these
couplings merge in a single one. 
We will address this after the extraction
of the $\beta$-function. 

Another important issue is the following: the choice of modifying
the covariance of the theory using a kinetic
term as $\sum_{\xi}   Z_\xi   \Tr_2 (p^{2 \xi} \phi^2) 
 + \mu \Tr_2 (\phi^2)$ make computations  lengthier. 
Indeed, the analysis of the amplitude could be made easier
by keeping a kinetic term as  $  Z_b   \Tr_2 (p^{2 b} \phi^2) 
 + \mu \Tr_2 (\phi^2)$ and let the extra term $ Z_a  \Tr_2 (p^{2 a} \phi^2) $ to be part of  the interaction. This makes the covariance much simpler. We will show that
both models share the same power counting and the renormalization analysis
applies equally well on both. The $\beta$-function will be computed with 
the second kind of models (with simpler covariance).

\subsection{Amplitudes and renormalizability}
\label{sect:perturb}
The models $+$ and $\times$ associated with the actions given 
by  \eqref{model1} and \eqref{model2}, respectively, give
the quantum models determined by the partition function
\be
Z_\bullet  = \int d\nu_{C_\bullet}(\bar\phi,\phi) \; e^{-S^{\inter}_{\bullet}[\bar\phi,\phi]}\,, 
\ee
where $\bullet =+,\times$, and $d\nu_{C_\bullet}(\bar\phi,\phi)$ is a field Gaussian measure with covariance $C_\bullet$
given by the inverse of the kinetic term:
\be
C_\bullet({\bf P};{\bf P'}) =\tilde{C}_\bullet({\bf P})\, \bdel_{{\bf P},{\bf P}'}\,,\qquad
\tilde{C}_\bullet({\bf P})\,= \frac{1}{\sum_\xi Z_\xi {\bf P}^{2 \xi}+ \mu }\,. 
\label{eq:cov}
\ee 
where, if $\bullet =+$, $\xi=a,b$ and if $\bullet =\times$,  $\xi=a,2a,b$.

\

\noindent 
{\bf Feynman graphs  in TFTs.}
Feynman graphs of TFTs and enhanced TFTs  have 
 two equivalent  representations. One is called ``stranded graph'' representation 
 and  incorporates more details of the structure
of the Feynman graph. The other representation of a Feynman
graph in this theory is  a bipartite colored graph \cite{Bonzom:2012hw,Gurau:2011xp}.
 
\begin{figure}[H]
 \centering
     \begin{minipage}{1\textwidth}
     \centering
\includegraphics[angle=0, width=14cm, height=6cm]{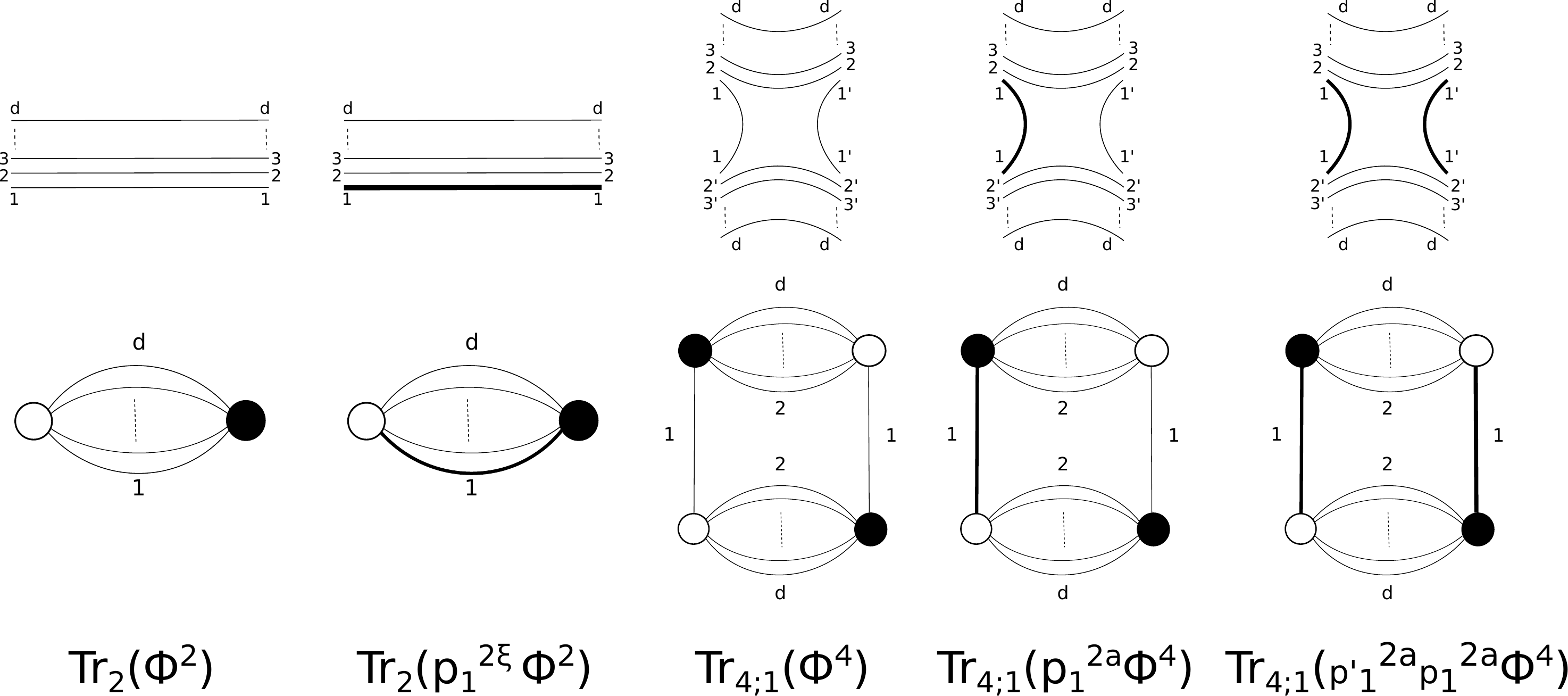}
\caption{ {\small  Order $d$  vertices of the mass, $\phi^2$- and $\phi^4$-terms.   
}} 
\label{fig:4vertex}
\end{minipage}
\end{figure}

\begin{figure}[H]
 \centering
     \begin{minipage}{1\textwidth}
     \centering
\includegraphics[angle=0, width=12cm, height=5cm]{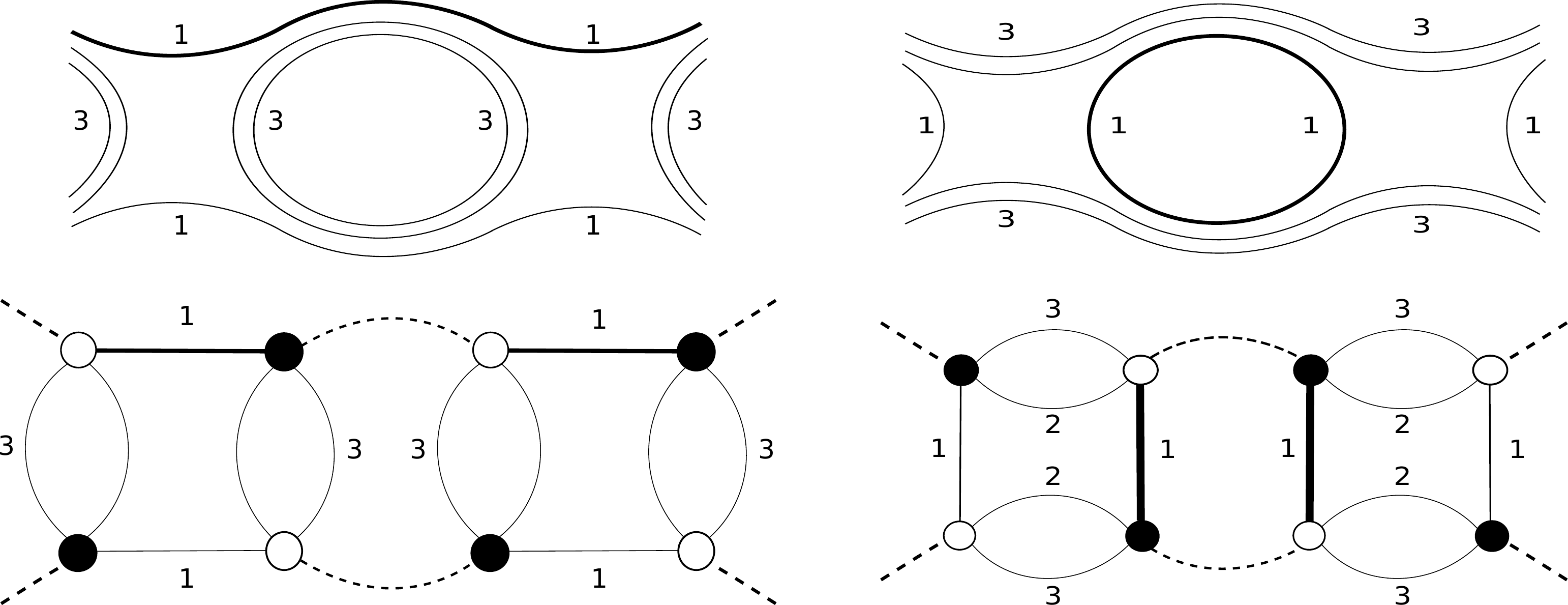} 
\caption{ {\small  Order $d=3$  Feynman graphs.   }} 
\label{fig:graphs}
\end{minipage}
\end{figure}

The propagator is drawn as a set of $d$ non-intersecting segments
called strands, or as a dotted line (see Figure \ref{fig:propacol}). 
\begin{figure}[H]
\centering
     \begin{minipage}[t]{0.7\textwidth}
\begin{tikzpicture}
\node (r1) at (-0.3,0.0)  {$p_{1}$};  
\node (r2) at (-0.3,0.3)  {$p_{2}$};
\node (r1) at (-0.3,0.8)  {$\vdots$};  
\node (r3) at (-0.3,1)  {$p_{d}$};  
\draw (0,0.3) -- (4,0.3) ;
\draw (0,1) -- (4,1) ;
\draw (0,0) -- (4,0) ;
\draw[dashed] (6,0.5) -- (10,0.5);
\end{tikzpicture}
\caption{\small  Two ways of representing the propagator of the theory.}
\label{fig:propacol}
\end{minipage}
\end{figure}
 In Figure \ref{fig:4vertex}, an interaction is pictured as a stranded vertex 
 (pictures above) or by a $d$-regular colored bipartite graph (pictures below).
We list therein all possible quadratic and quartic vertices
and a bold line represents a momentum weight of the term. 
In Figure \ref{fig:graphs}, we give some examples of two 4-point graphs. 

At the perturbative level, we compute the amplitude of a Feynman 
graph $\cG(\cV,\cL)$ with set of vertices $\cV$ and set of propagator 
lines $\cL$,  in the standard way: 
\bea
A_{\cG}(\{p_{\ext}\}) = \sum_{{\bf P}_v} \prod_{l \in \cL} C_{\bullet, l} ({\bf P}_{v},{\bf P}'_{v'} )
\prod_{v \in \cV} (-{\bf V}_v({\bf P}_v) ) 
\eea
where  $C_{\bullet, l}$ is a propagator with line index $l$, ${\bf V}_v({\bf P}_v)$
is a given vertex weight that contains a coupling constant but also a momentum 
weight if the vertex $v$ is enhanced.  The sum is performed over internal momenta 
and will follow the ordinary momentum routine whereas the set $\{p_{\ext}\}$ defines external momenta that are not summed over.

Some notation is needed to distinguish the different vertices
and their weight that we shall deal with:

 -  the set $\cV_{4;s}$ of vertices 
 $\Tr_{4;s}(\phi^4)  $ with color $s$ and with kernel  ${\bf V}_{4;s}$, $\cV_4 = \sqcup_{s=1}^d \cV_{4;s}$ (disjoint union notation); we denote
$V_4= |\cV_4|$. 

-  the set $\cV_{+; 4;s}$ of vertices 
$\Tr_{4;s}(p^{2a}\,\phi^4)$ with color $s$ and 
with vertex kernel  ${\bf V}_{+;4;s}$,
$\cV_{+;4} = \sqcup_{s =1}^d\cV_{+; 4;s}$;  we denote $V_{+;4}= |\cV_{+;4}|$; 

-  the set $\cV_{\times; 4;s}$ of vertices 
$\Tr_{4;s}(([p^{2a}p'^{2a}]\,\phi^4)$
with vertex kernel  ${\bf V}_{+;4;s}$,
$\cV_{\times;4}= \sqcup_{s =1}^d\cV_{\times; 4;s}$; 
we denote $V_{\times;4}= |\cV_{\times;4}|$; 

- the set $\cV_2$  of mass vertices with kernel 
${\bf V}_{2}$; we have 
$V_2 = |\cV_2| $; 

- 
 the set  $\cV_{2;\xi;s}$ of vertices 
$ Z_\xi   \Tr_2 (p^{2 \xi} \phi^2) $ each kind corresponding to 
a kernel ${\bf V}_{2;\xi;s}$,
 $\cV_{2;s} = \cup_{\xi} \cV_{2; \xi}$; $|\cV_{2;\xi;s}|=V_{2;\xi;s}$,  $V_{2;\xi}= \sum_s V_{2;\xi;s}$.
  
We denote

- the cardinalities
$|\cV_{4;s}|=V_{4;s}$,
$|\cV_{4}|=V_4$,
$|\cV_{\bullet;4;s}|=V_{\bullet;4;s}$, 
$|\cV_{\bullet;4}|=V_{\bullet;4}$, 
$\bullet=+,\times$; 

- $V_{(4)}=|\cV_{(4)}|=V_4 +  V_{\bullet;4}$, where
$ \cV_{(4)}= \cV_4  \sqcup 
\cV_{\bullet; 4}$ 
depending on the model $\bullet = +,\times$.

- $V_{(2)} = |\cV_{(2)}|= V_{2} +  \sum_{\xi} V_{\bullet;2}$ where $ 
\cV_{(2)} = \cV_2\sqcup \cup_{\xi} \cV_{2;\xi}$, where the value of
$\xi$
  depends on the model $\bullet = +,\times$.

Then, 
$\cV = \sqcup_{s =1}^d (\cV_{4;s} \cup \cV_{\bullet;4; s}   \cup \cV_{2;s}) \cup \cV_2 = \cV_{(4)} \sqcup \cV_{(2)}$,
and has cardinality 
$|\cV|=V =V_{(4)}+  V_{(2)}$.

\

\noindent{\bf Power counting theorems
and list of divergent graphs.}
We restrict now to 
%primitively 
divergent graphs. 
Given $a\le b$, the degree of divergence of 
a graph amplitude of the model $+$ is given by 
\bea
&&
\omp(\cG) =
\crcr
&& - {2 D \over (d-1)!} ( \omega(\cG_{\rm color}) - \omega (\partial \cG))  - D (C_{\partial \cG} - 1) 
-{1\over 2} \left[ ( D \,  (d-1) - 2 b) N_{\ext} - 2 D \,  (d-1) \right] 
\crcr
&&+ {1 \over 2} \left[ -2 D\,  (d-1) + (D \,  (d-1) - 2 b) n \right] \cdot V  + 2 a \rhop
+2a\rho_{2;a}+ 2b \rho_{2;b}  \,.
\eea
where $ \omega(\cG_{\rm color})$ is the Gurau degree \cite{Gur3,Gur4} of the extended 
colored graph $\cG_{\rm color}$ of $\cG$, $\partial \cG$ is the boundary graph 
of $\cG$ \cite{BenGeloun:2011rc}, $C_{\partial \cG}$ is the number of connected components of $\partial \cG$, $N_{\ext}  $ is the number of external legs of the diagram $\cG$, 
$ n \cdot V = \sum_{k} k V_{(k)}$ where $k$ is the valence of the vertex 
of type  $\cV_{(k)}$, 
$\rhop$ is the number of times that a momentum is enhanced passing 
through a strand of an enhanced vertex $\cV_{\bullet;4}$;
likewise $\rho_{2;\xi}$, $\xi=a,b$ is the number of times that a momentum gets an enhancement
passing through a vertex $\cV_{2;\xi;s}$. 

Using  bounds on $\rho_+$, $\rho_{2;a}$, and  
$\rho_{2;b}$, the same divergence degree finds the following useful bound:
\bea
&&
\omp(\cG) \le 
\crcr
&& - {2 D \over (d-1)!} ( \omega(\cG_{\rm color}) - \omega (\partial \cG))  - D (C_{\partial \cG} - 1) 
-{1\over 2} \left[ ( D \,  (d-1) - 2 b) N_{\ext} - 2 D \,  (d-1) \right] 
\crcr
&&- 2b V_{2} -2(b-a) V_{2;a}
+ \left[ D\,  (d-1)   - 4 b \right] V_4
+ \left[ D\,  (d-1) - 4b +2a\right]  V_{+;4}  \,.
\label{ompcG}
\eea
where the coefficient of $V_{2;b}$
vanished.

The following statement has been proved in \cite{BenGeloun:2017xbd}. 

\begin{proposition}[List of 
divergent graphs for the model $+$]
\label{prop:list+}
The $p^{2a}\phi^4$-model $+$ with parameters $a=D(d-2)/2, b=D(d-\frac32)/2$
for two integers $d>2$ and $D>0$, has 
divergent graphs  with $\Omega(\cG)=
 {2 D \over (d-1)!}(\omega(\cexG) - \omega(\bG))$:

\begin{table}[H]
\centering
\begin{tabular}{lcccccccccccccccc}
\hline\hline
$\cG$ && $N_{\ext}$ && $V_{2}$ &&  $V_{2;a}$  && $V_{4}$ && $\rhop$   && $C_{\bG}-1$ && $\Omega(\cG)$ && $\omp(\cG)$  \\
\hline\hline
 && 4 && 0 && 0 && 0 && $V_{+;4}$ && 0 && 1&& 0\\
I && 2 && 0  && 0  && 0 && $V_{+;4}$ && 0 && 1 && ${D \over 2}$  \\
II && 2 && 0  && 0 && 0 && $V_{+;4}-1$  && 0 && 0 && ${D \over 2}$ \\
III && 2 && 0  && 0 && 1 && $V_{+;4}$  && 0 && 0 && ${D \over 2}$ \\
IV && 2 && 0  && 1  && 0 && $V_{+;4}$ && 0 && 1 && $0$  \\
V && 2 && 0  && 1 && 0 && $V_{+;4}-1$  && 0 && 0 && $0$ \\
VI && 2 && 0  && 1 && 1 && $V_{+;4}$  && 0 && 0 && $0$ \\
\hline\hline
\end{tabular}
\caption{List of 
divergent graphs of the $p^{2a}\phi^4$-model $+$.} 
\label{tab:listprim1}
\end{table}

\end{proposition}

Having a look at Table \ref{tab:listprim1} and the row labeled by $\Omega(\cG)$, 
we see that the dominant graphs are not those labeled by $\Omega(\cG)=0$ which are the melonic diagrams, but are those that are $\Omega(\cG)=1$ and thus non-melonic graphs. 
This shows that the model $+$ delivers the expected output. 

\begin{theorem}
\label{theorem+}
The $p^{2a}\phi^4$ model $+$  with parameters $a=D(d-2)/2, b=D(d-\frac32)/2$ 
for arbitrary  order $d\ge 3$ and dimension $D>0$ with action defined by \eqref{eq:actiond} is just-renormalizable at all orders of perturbation
theory. 
\end{theorem}

Concerning the model $\times$, 
one obtains the degree of divergence in this model valid for $3a\leq 2b$,
\bea
&& 
\omt(\cG)
 = \crcr
 &&
 - {2 D \over (d-1)!} ( \omega(\cG_{\rm color}) - \omega (\partial \cG))  - D (C_{\partial \cG} - 1) 
-{1\over 2} \left[ ( D \, (d-1) - 2 b) N_{\ext} - 2 D \, (d-1) \right] 
\crcr
&&+ {1 \over 2} \left[ -2 D\, (d-1) + (D \, (d-1) - 2 b) n \right] \cdot V  + 2 a \rhot
+\sum_{\xi=a,2a,b}2\xi\rho_{2;\xi} \,.
\eea
We can bound the same divergence  as:
\bea
&&
\omt(\cG)
  \le 
\crcr
&&   - {2 D \over (d-1)!} ( \omega(\cG_{\rm color}) - \omega (\partial \cG))  - D (C_{\partial \cG} - 1) 
-{1\over 2} \left[ ( D \, (d-1) - 2 b) N_{\ext} - 2 D \, (d-1) \right] 
\crcr
&&- 2b V_{2} -2(b-a) V_{2;a}
 - 2(b-2a) V_{2;2a}
+ \left[ D\,  (d-1)   - 4 b \right] V_4
+ \left[ D\,  (d-1) - 4b +4a\right]  V_{\times;4}  \crcr
&& 
\label{omtcG}
\eea

\begin{proposition}[List of 
divergent graphs for  the model $\times$]
\label{prop:listx}
The $p^{2a}\phi^4$-model $\times$ with parameters $D = 1, d=3,a={1 \over2}, b=1$, 
has the following 
divergent  graphs which  obey
$\Omega(\cG)=  \omega(\cexG) - \omega(\bG)$:
\begin{table}[H]
\begin{center}
\begin{tabular}{lcccccccccccccccc}
\hline\hline
$\cG$ && $N_{\ext}$ && $V_{2}$  && $V_{2;a}$ && $ V_4$ && $\rhot$   && $C_{\bG}-1$ && $\Omega(\cG)$ && $\omt(\cG)$  \\
\hline\hline
I && 2 && 0  && 0 && 0 &&  $2 V_{\times;4} -1$ && 0 && 1 && ${0}$  \\
II && 2 && 0  && 0 && 0 && $2 V_{\times;4}-2$  && 0 && 0 && ${0}$ \\
III && 2 && 0  && 0 && 1 && $2 V_{\times;4}$  && 0 && 0 && ${0}$ \\
\hline\hline
\end{tabular}
{\caption{List of 
divergent graphs of the $p^{2a}\phi^4$-model  $\times$. \label{tab:listprim2}}}
\end{center}
\end{table}
\end{proposition}

\begin{theorem}
\label{theoremx}
The $p^{2a}\phi^4$ model $\times$  with parameters $D = 1, d=3,a={1 \over2}, b=1$,  
with the action defined by \eqref{eq:actiond}  is renormalizable at all orders of perturbation.
\label{theorenx}
\end{theorem}

This model $\times$ has an unexpected behavior: it does not have  divergent 4-point graphs,
only log-divergent 2-point graphs. One may ask if this is a super-renormalizable 
model with a finite number of divergent graphs. The answer is no because the model has  infinite  terms participating in the mass flow. The property can be 
regarded as a particularity entailed by both the nonlocality and the presence of enhanced
vertices  which make the mass behave like a marginal coupling.

\subsection{An alternative enhanced TFT model}

As stated previously, the presence of the enhanced vertices
generate new 2-point terms with external data following the pattern 
of $ \Tr_2 (p^{2 a} \phi^2) $ for the model $+$, and 
the patterns of
 $ \Tr_2 (p^{2a} \phi^2) $ and
$ \Tr_2 (p^{4a} \phi^2) $ for the model $\times$. 
Consequently, in \cite{BenGeloun:2017xbd}, the authors have modified 
the  
covariance and performed 
the
renormalization analysis. 
We present here an alternative way of dealing with 
these terms that make the whole simpler. 
We simply demand that the new terms are interactions and 
therefore propose the following models: 
(in the following, we use the same notation as in the previous sections
as no confusion may arise)
\bea
\text{model }+: 
&&
S^{\inter}_+[\bar\phi,\phi] =  \frac{ \lambda}{2}\,\Tr_{4}(\phi^4)  
 +\frac{\pet}{2}\, \Tr_{4}(p^{2a}\,\phi^4)
 + Z_a   \Tr_2 (p^{2a} \phi^2)  \crcr
 && 
  + CT_{2}[\bar\phi,\phi] +CT_{2;b}[\bar\phi,\phi] 
 \cr\cr
&& 
S^{\kin}_+[\bar\phi,\phi] =  
Z_b   \Tr_2 (p^{2b } \phi^2) 
 + \mu \Tr_2 (\phi^2) \,,
\label{qmodel1} \\ \cr
\text{model }\times: &&
S^{\inter}_\times[\bar\phi,\phi] =   \frac{ \lambda}{2}\,\Tr_{4}(\phi^4)  
 +\frac{\lambda_\times}{2}\, \Tr_{4}([p^{2a}p'^{2a}]\,\phi^4)
 + \sum_{\xi=a,2a} Z_\xi  \Tr_2 (p^{2 \xi} \phi^2) 
 \crcr
 &&
 + CT_{2}[\bar\phi,\phi] 
 +CT_{2;b}[\bar\phi,\phi]  \,,
 \cr\cr
&& 
S^{\kin}_\times[\bar\phi,\phi] =  
Z_b \Tr_2 (p^{2 b} \phi^2) 
 + \mu \Tr_2 (\phi^2)
\label{qmodel2}
\eea

Considering this proposal, the covariance of these models 
is unique and given by 
\be
C({\bf P};{\bf P'}) =\tilde{C}({\bf P})\, \bdel_{{\bf P},{\bf P}'}\,,\qquad
\tilde{C}({\bf P})\,= \frac{1}{
Z_b{\bf P}^{2 b}+ \mu }\,. 
\label{eq:newcov}
\ee 
The difference between this propagator \eqref{eq:newcov} and the former \eqref{eq:cov}
is that, for a given strand, the momentum which was previously
  $\sum_{\xi} p_s^{2\xi}$ becomes $p_s^{2b}$. 
The multiscale analysis of  \cite{BenGeloun:2017xbd} 
with appropriate parameter $M^{2b}$ ($M>1$)  can be mimicked 
with no difficulty to obtain the sliced propagator as
\bea
C_0 \le K \,, \qquad \quad 
C_i ({\bf P}) \le K M^{2bi } e^{\delta M^{-bi} ({\bf P}^{b}+ \mu/Z_b)} 
\eea
for $i>0$ a high slice  index, and for some constant $K$ and $\delta$. 

One way of considering that exchanging the covariance \eqref{eq:cov} for \eqref{eq:newcov} has no noticeable effect lies in the perturbation theory of large moment behavior: $p^{2b}\ge p^{2a}$, for $b>a$. For an IR analysis, it would have been more judicious to target
the smaller momentum $p^{2a}$ to define the theory covariance. That study and its implications are left for the future. 

To write the multiscale amplitude where each propagator
lives in an arbitrary slice is identical to the previous analysis 
\cite{BenGeloun:2017xbd} for both the model $+$ and the model $\times$. We need a convenient
bound on that quantity. One should arrive at the expression
of a sliced amplitude according to 
a momentum attribution $\bmu = (i_1, \dots i_{| \cL | })$: 
\bea
|A_{\cG; \bmu}| \leq  \tilde K  \prod_{l \in \cL} M^{-2bi_l }
\sum_{p_{f_s}} \prod_{f_s \in \ F_{\inter}} 
e^{-\delta (\sum_{l\in f_s} M^{-bi_l})  |p_{f_s}^{b}|}
\prod_{i=1}^d \prod_{v_s\in \cV_{\bullet;4;s} }
{\rm weight} (v_s)
\eea
where $\tilde K$ is a constant, 	and weight$(v_s)$ is self-explanatory and is identical 
to the former case. 

The rest of the analysis consists in performing the spectral sums
$\sum_{p_{f_s}} (\cdot)$. A close inspection of the method introduced
in \cite{BenGeloun:2017xbd} shows that one condition is imposed  
and gets rid of $a$ at leading order. The parameter $a$ does not contribute
to the sum. We have at $a\ge 0$:
\bea
\sum_{p_1; \dots p_D=1}^{\infty}
(\sum_{l=1}^{D}p_{l}^c )^n e^{-B (p^b + p^a)}
= k B ^{-\frac{cn + D}{b}} e^{-B^{1- \frac{a}{b}}} (1+ \cO(B^{\frac1b}))
\eea
Requesting $a\le b$ for the model $+$ and $3a\le 2b$ for the model $\times$
removes the parameter $a$ from the expansion at leading order in $B$. 
Then, the remaining part
of the momentum integration exactly performs in the same way. We  therefore reach the same power counting, the same list of divergent graphs, Propositions \ref{tab:listprim1} 
and \ref{tab:listprim2} are valid for the models $+$ and $\times$, respectively.
We perform the same subtraction procedure
making the models  \eqref{qmodel1} and \eqref{qmodel2} renormalizable 
at all orders of perturbation theory. Theorems \ref{theorem+} and \ref{theoremx}
hold for the models $+$ and $\times$, respectively. 

From this point onwards, we consider the models  \eqref{qmodel1}
and \eqref{qmodel2}. To simplify the  determination of the graph  combinatorial
factors, we will  distinguish all couplings by providing them with  colors. For instance, 
for the model $+$, we write:  
\bea
S'^{\inter}_+[\bar\phi,\phi] 
 := \sum_{c=1}^d \frac{\lac}{2} \; \Tr_{4;c}(\phi^4)
 + 
\sum_{c=1}^d \frac{\lapc}{2}\; \Tr_{4;c}(p^{2a}\,\phi^4) 
+\sum_{c=1}^d  Z_a^{(c)} \Tr_{2;c}(p^{2 a } \phi^2) \,.
\label{eq:ourinteraction}
\eea
In the end, the $\beta$-functions of 
$\lam$, $\pet$, and $Z_a$ will be directly inferred
by letting $\lac \to \lam$, $\lapc \to \pet$
and $ Z_a^{(c)} \to Z_a$. 
The same will be done for the model $\times$.

\section{
One-loop beta-functions  of the model $+$ }
\label{betaf}

This section now addresses a first set of new results in this contribution: we determine the RG flow at 1-loop of the model $+$. We compute the $\beta$-function of the coupling $\lambda$, the coupling $Z_a$ and the mass. 
Since, the action contains a vertex weight and an extra quadratic coupling, we carefully carry out the formalism of finding the effective action and write the resulting RG flow equations.  

\subsection{Effective action}

The presence of several coupling constants in our theory urges us to handle the effective action with care. 
We will compute the renormalization of coupling constants via multiscale analysis.
We start with performing a slice decomposition of the covariance.
We let $M >1$ a positive real number and we define the (sharp) cutoff functions as
\begin{equation}
    \chi^0(\alpha) = 
    \begin{cases}
      1 & \text{if  $1 \le \alpha $}\\
      0 & \text{if $1 > \alpha$}
    \end{cases}  \,,
\label{eq:chi0}
\end{equation}

$\forall i >0$,
\begin{equation}
\chi^i(\alpha) = 
    \begin{cases}
      0 & \text{if $\alpha \le M^{-2 bi}$}\\
      1 & \text{if  $M^{-2 bi} < \alpha \le M^{-2 b(i-1)} $}\\
      0 & \text{if $ M^{-2 b(i-1)} < \alpha $}
    \end{cases}  
\label{eq:chii}
\end{equation}
The presence of $2b$ in 
the cutoff function is
related to the propagator momentum power and
has a normalization 
effect for the degree of divergence $\omp(\cG)$. 
Then we write the covariance in \eqref{eq:cov} as expressed in Schwinger parametrization,
\be
C({\bf P};{\bf P'}) =\tilde{C}({\bf P})\, \bdel_{{\bf P},{\bf P}'}\,,\qquad
\tilde{C}({\bf P})\,= \frac{1}{{\bf P}^{2 b}+ \mu }
=
\sum_{i =0}^{\infty} \tilde{C}_{i} ({\bf P})
\,,
\label{eq:C}
\ee
with
\begin{equation}
\tilde{C}_i ({\bf P})
=
\int_{0}^{\infty} d \alpha \, 
e^{- \alpha ({\bf P}^{2 b} + \mu)} \chi^i(\alpha)\,.
\label{eq:Ctilde}
\end{equation}
We integrate out the fields at high scales greater than $i$ and include their effects in the effective action $W^i$. 
As written below,
\begin{equation}
Z = \int d \nu_{C_{\le i}} (\bar\phi_{\le i}, { {\phi}}_{\le i})
\, 
e^{- W^i (\bar\phi_{\le i},\, { {\phi}}_{\le i})}\,,
\end{equation}
where
\begin{equation}
C_{\le i} ({\bf P}; {\bf P}') = \bdel_{{\bf P}, {\bf P}'}\sum_{j \le i} {\tilde C}_j({\bf P})\,.
\end{equation}
Following Wilsonian renormalization group idea, after integrating up to the scale $i$, we continue integrating the slice $i$ in order to obtain an effective action at scale $i-1$.
We can do this by using the property of Gaussian measure; we can readily decompose $C_{\le i} =  C_i + C_{\le i-1} $ and the corresponding fields
$\phi_{\le i} = \psi_i + \phi_{\le i-1}$ (${\bar \phi}_{\le i} = {\bar \psi}_i + {\bar \phi}_{\le i-1}$). Then the partition function becomes
\begin{equation}
Z = \int 
d \nu_{C_{\le i-1}}(\bar\phi_{\le i-1}, {\phi}_{\le i-1})
e^{-W^{i-1}( \bar\phi_{\le i-1}, \, { \phi}_{\le i-1})}\,,
\end{equation}
where the effective action at  scale $i-1$ is given by
\begin{equation}
e^{-W^{i-1}( \bar\phi_{\le i-1}, \, { \phi}_{\le i-1})}
=
\int 
d \nu_{C_i}(\bar\psi_{i}, { \psi}_{i}) e^{-W^i(\bar \psi_i + \bar \phi_{\le i-1}, \, {\psi}_i + { \phi}_{\le i-1})}\,,
\end{equation}
Note that here, in the case that the theory is renormalizable, one can assert the effective action at any scale $i$ takes the same form as the interaction action,
\begin{equation}
W^{i}(\bar\phi_{\le i},\phi_{\le i}) = S^{\inter}
(\bar\phi_{\le i}, \phi_{\le i})\,,
\end{equation}
We can, then,  write formally: 
\bea
&& 
- W^{i-1}(\bar\phi_{\le i-1}, { \phi}_{\le i-1})
=
 \Tr_2 (\bar \phi_{\le i-1} \cdot \Sigma
 \cdot \phi_{\le i-1} )  
+
\frac{1}{2} \Tr_4({\phi^4_{\le i-1}} \cdot \Gamma_4)
+ R(\phi_{\le i-1})
\,,
\label{eq:effectiveW}
\eea
where $\Sigma(\{p\})$ is the sum over all amputated 1PI 2-point graphs, 
$\Gamma_4(\{p\})$ is the sum of 1PI 4-point graphs following the pattern of $\Tr_4(\phi^4)$, and 
$R(\phi_{\le i-1})$ is the rest of the terms containing 1PR graphs (they do not contribute to the iteration process) and the finite terms.  In the above equation, 
$\Sigma$ and $\Gamma_4$ are kernels
that are convoluted with the tensors
fields.

We separate the graph amplitudes into local and nonlocal parts, therefore,
\begin{equation}
\Sigma(\{p\}) 
= 
\Sigma (\{0\}) 
+ 
\sum_{c}
\vert p_{c}\vert^{2 b} \partial_{\vert p_{c}\vert^{2 b}} \Sigma \big\vert_{\{p\}=0}
+
\sum_{c}
\vert p_{c}\vert^{2 a} \partial_{\vert p_{c}\vert^{2 a}} \Sigma \big\vert_{\{p\}=0} 
+
\cdots
\label{eq:sigma} 
\end{equation} 
As a result of \cite{BenGeloun:2017xbd}, the renormalization analysis of the model dictated by the rows of I, III, IV, and VI in  Table \ref{tab:listprim1}  proved that 
$\partial_{\vert p_{c}\vert^{2 b}} \Sigma \big \vert_{\{p\}=0} =0 $ and the rest of the terms in $\cdots$ are finite, 
whereas the mass renormalization
$\Sigma (\{0\})  $
and
$\partial_{\vert p_{c}\vert^{2 a}} \Sigma\vert_{\{p\}=0} \equiv \Gamma^{(c)}_2 (\{0\})$ are divergent.
$\vert p_{c}\vert^{2 a}\Gamma^{(c)}_2(\{p\})$
is the sum of all  amputated 1PI (one-particle-irreducible) 2-point functions following the pattern of 
${\rm Tr}_2 ( p_{c}^{2a} \phi^2)$ 
on their boundary graphs as dictated by II and V rows of Table \ref{tab:listprim1}.
The boundary of the graphs contributing to this function are all melonic without the $\vert p\vert^{2a}$-enhancement (see Proposition \ref{prop:list+} and lines with $\Omega(\cG) = 0$).

Similarly, one can expand the contribution coming from $4$-point function
\begin{equation}
\Gamma_4(\{p\}) 
=
\sum_c 
\Big\{
\Gamma_4^{(c)}(\{0\})
+
\vert p_{c} \vert^{2 a}
\partial_{\vert p_{c} \vert^{2 a}} \Gamma_4^{(c)}\Big\vert_{\{p\} =0}
+
\vert p_{c} \vert^{2 b}
\partial_{\vert p_{c} \vert^{2 b}} \Gamma_4^{(c)}\Big\vert_{\{p\} =0}
\Big \}
+\cdots\,,
\label{eq:4ptfcn} 
\end{equation}
where $ \sum_c \Gamma_4^{(c)}(\{0\})
\equiv \Gamma_4(\{0\}) 
$ is the sum of all amputated 1PI  4-point functions 
following the pattern of ${\rm Tr}_4 (\phi^4)$ on their boundary graphs  as dictated by the first row of Table \ref{tab:listprim1}.
We define
$
\partial_{\vert p_{c} \vert^{2 a}} \Gamma^{(c)}_{4}\big\vert_{\{p\} =0}
\equiv
\Gamma_{4;+}^{(c)}(\{0\})
$ which are all amputated 1PI 4-point functions following the pattern of $ \Tr_{4;c}( p^{2a} \phi^4)$, with characteristics given by the first row of Table \ref{tab:listprim1} and having a boundary with external $\vert p\vert^{2a}$-enhancement (see Proposition \ref{prop:list+}). 
In fact, from Proposition \ref{prop:list+}, there is only the leading order ${\mathcal O}(\lambda_+)$ contribution in $\Gamma_{4;+}^{(c)}(\{0\})$ and there are no contributions from higher orders in perturbation theory
in $\lambda_+$.
Also, from the Proposition \ref{prop:list+}, the rest denoted by $\cdots$ and
 $\partial_{\vert p_{c} \vert^{2 b}} \Gamma_4^{(c)} \big \vert_{\{p\} =0}$ are finite.

Now, reorganizing and absorbing all the finite parts into ${\tilde R}({\phi_{\le i-1}})$,
however intentionally leaving the term with
$\partial_{\vert p_{c}\vert^{2 b}} \Sigma \big \vert_{\{p\}=0} $
even though it is zero,
\bea
- W^{i-1}(\phi_{\le i-1}, {\bar \phi}_{\le i-1})
&=&
\Sigma_{i-1} (\{0\})
\Tr_2 (\phi^2_{\le i-1})  
+
(\partial_{\vert{p_{c}\vert}^{2 b}} \Sigma \big \vert_{\{p\}=0})_{i-1}
\Tr_{2} (p^{2 b}\phi^2_{\le i-1}) 
\nonumber
\\
&&+
\sum_c
\Gamma^{(c)}_{2, \; i-1} (\{0\})
\Tr_{2;c}(p^{2 a}\phi^2_{\le i-1})
+
\sum_c
\frac{\Gamma^{(c)}_{4, \; i-1} (\{0\})}{ 2} \Tr_{4;c}({\phi^4_{\le i-1}})
\nonumber
\\
&&+ 
\sum_c
\frac{\Gamma^{(c)}_{4;+, \; i-1} (\{0\})}{ 2} \Tr_{4;c}( p^{2a}\phi^4_{\le i-1})
+
{\tilde R}(\phi_{\le i-1})\,.
\label{eq:effectiveW2}
\eea

The effective theory is then defined by a new measure given by
\bea
d \nu_{\tilde{C}^{i-1}(\phi_{\le i-1})}
\exp \Big[
{\Sigma_{i-1}(\{0\}) {\rm Tr}_2 (\phi_{\le i-1}^2) 
+ 
\sum_c 
(\partial_{\vert p_c\vert^{2 b}} \Sigma  \vert_{\{p\}=0})_{i-1}
{\rm Tr}_2( p_c^{2b} \phi_{\le i-1}^{2})}
\Big]
\,,
\eea
which is still a Gaussian measure.
Let us compute the new covariance for the above Gaussian measure, 
\bea 
\frac{1}
{Z_{b,\, i-1}}
\int_0^\infty
d \alpha \,
e^{- \alpha (\vert p \vert^{2b} + \mu_{{\rm ren}, i-1})}
\chi^{i-1}(\alpha)
=
\frac{1}
{Z_{b,\, i-1}}
{\tilde{C}}^{i-1}(p)
\,,
\eea
where we defined the renormalized mass $\mu_{{\rm ren}, i-1}$ to be
\beq
\mu_{{\rm ren}, i-1} = \frac{\mu_{i-1} - \Sigma_{i-1}(\{0\})}
{Z_{b,\, i-1}}
\,,
\eeq
and the wave function renormalization to be
\beq
Z_{b,\, i-1} \equiv 1 + 
(\partial_{\vert p_c\vert^{2 b}} \Sigma \big \vert_{\{p\}=0})_{i-1}
\,.
\eeq
Note that the color dependence on   $(\partial_{\vert p_{c}\vert^{2 b}} \Sigma \big \vert_{\{p\}=0})$
should be actually absent.  
Then, the effective theory for $\phi_{\le i-1}$ can be written as
\bea
d \nu_{\frac{1}{Z_{b,\, i-1}}{\tilde{C}^{i-1}}}(\phi_{\le i-1})
&\exp&
\Big[
\sum_c \Gamma^{(c)}_{2, \, i-1}(\{0\}) \Tr_2 ( p_c^{2a} \phi_{\le i-1}^2) 
+
\sum_c 
\frac{\Gamma^{(c)}_{4, \, i-1} (\{0\})}{2} \Tr_4 (\phi_{\le i-1}^4) 
\crcr
&&
+
\sum_c 
\frac{\Gamma^{(c)}_{4;+, \, i-1} (\{0\})}{2} \Tr_4 ( p_c^{2a} \phi_{\le i-1}^4) 
+
{\tilde R} (\phi_{\le i-1})
\Big]
\,.
\eea
With a field rescaling $\phi_{\le i-1} \rightarrow \sqrt{Z_{b,\, i-1}} \phi_{\le i-1}$ (which in our specific case, there is no actual rescaling because $Z_{b,\, i-1} = 1$ and trivial), the effective theory 
for $\phi_{\le i-1}$ can be recast:
\bea
d \nu_{{\tilde C}_{i-1}}(\phi_{\le i-1})
&{\rm exp}&
\Big[
\sum_c\frac{\Gamma^{(c)}_{2, i-1}(\{0\})}{Z_{b, i-1}}
\Tr_{2;c}( p^{2 a}
\phi_{\le i-1}^2)
+
\sum_c
\frac{\Gamma^{(c)}_{4, i-1}(\{0\})}{{2}Z_{b, i-1}^2}
\Tr_{4;c}(\phi_{\le i-1}^4)
\nonumber 
\\
&&
+
\sum_c
\frac{\Gamma^{(c)}_{4;+, i-1}(\{0\})}{{ 2} Z_{b, i-1}^2}
\Tr_{4;c}( p^{2a} \phi_{\le i-1}^4)
+
{\tilde R} (\sqrt{Z_{b,\, i-1}}\phi_{\le i-1})
\Big]
\,.
\eea
Now we can identify the effective couplings at scale $i-1$, 
\bea
Z_{a, i-1} 
&=&
-\frac{\Gamma^{(c)}_{2, i-1}(\{0\})}{Z_{b, i-1}} 
\,,
\nonumber
\\
\lambda^{(c)}_{i-1}
&=&
-\frac{\Gamma^{(c)}_{4, i-1}(\{0\})}{Z_{b, i-1}^2}
\,,
\nonumber 
\\
\lambda^{(c)}_{+; \, i-1}
&=&
-\frac{\Gamma^{(c)}_{4;+, i-1}(\{0\})}{Z_{b, i-1}^2}
\,.
\eea
Note that in our case, 
$(\partial_{\vert p_{c}\vert^{2 b}} \Sigma \big \vert_{\{p\}=0})_{i-1} = 0$
therefore, throughout, we actually had 
\bea
Z_{b,\, i-1} &=& 1\,,
\crcr
\mu_{{\rm ren}, i-1} &=& \mu_{i-1} - 
\Sigma_{i-1}(\{0\})
\,,
\crcr
Z_{a, i-1} 
&=&
-\Gamma^{(c)}_{2, i-1}(\{0\})\,,
\crcr
\lambda^{(c)}_{i-1}
&=&
-\Gamma^{(c)}_{4, i-1}(\{0\})\,,
\crcr
\lambda^{(c)}_{+; \, i-1}
&=&
-\Gamma^{(c)}_{4;+, i-1}(\{0\})
\,.
\label{eq:rgeqns}
\eea

In the following sections, we will compute these perturbative renormalization group
flow equations restricting to 1-loop 
corrections.  
For the model $+$, 
the $\beta$-functions can be computed for generic parameters
$a=(d-2)/2$, and $b=(d-3/2)/2$ and $d>2$ but with 
fixed group dimension $D=1$. 
More general equations for $D>1$ can be also computed 
with a bit more work \ref{theorem+}.

\subsection{4-point function and   beta-function}
\label{sect:}

We will restrict the analysis by only considering up to one-loop corrections to the $\beta$-function in the perturbation theory.
Note that, from Proposition \ref{prop:list+}, 
$\Gamma^{(c)}_{4;+}(\{p\})$
 only contains the first order zero-loop graph with one $\lambda_+$ coupling with four external legs. 
Next, we focus on $\Gamma^{(c)}_{4}(\{p\})$
and write 
\bea
\Gamma^{(c)}_{4}(\{p\})
 = \sum_{\cG_{4,\iota}^{(c)}} K_{\cG_{4,\iota}^{(c)}} \;  S_{\cG_{4,\iota}^{(c)}} (\{p\})\,,
\eea
where the sum over $\cG_{4,\iota}^{(c)}$
runs over a list of 4-point graphs obeying the first row of Table \ref{tab:listprim1}, 
$K_{\cG_{4,\iota}^{(c)}}$ is a combinatorial factor and $  S_{\cG_{4,\iota}^{(c)}} (\{p\})$ is a formal amplitude sum. 
At zero-loop, the first order graph is made of one $\lambda$ interaction bubble with four external legs.
At one-loop 
a single graph that we call $n_4^{(c)}$ shown in Fig. \ref{fig:V4singleloop} contributes. 
If one is further interested in two-loop contribution, see Appendix \ref{app:4pt2+}.

\begin{figure}[H]
\centering
     \begin{minipage}[t]{0.8\textwidth}
      \centering
\includegraphics[angle=0, width=6cm, height=4cm]{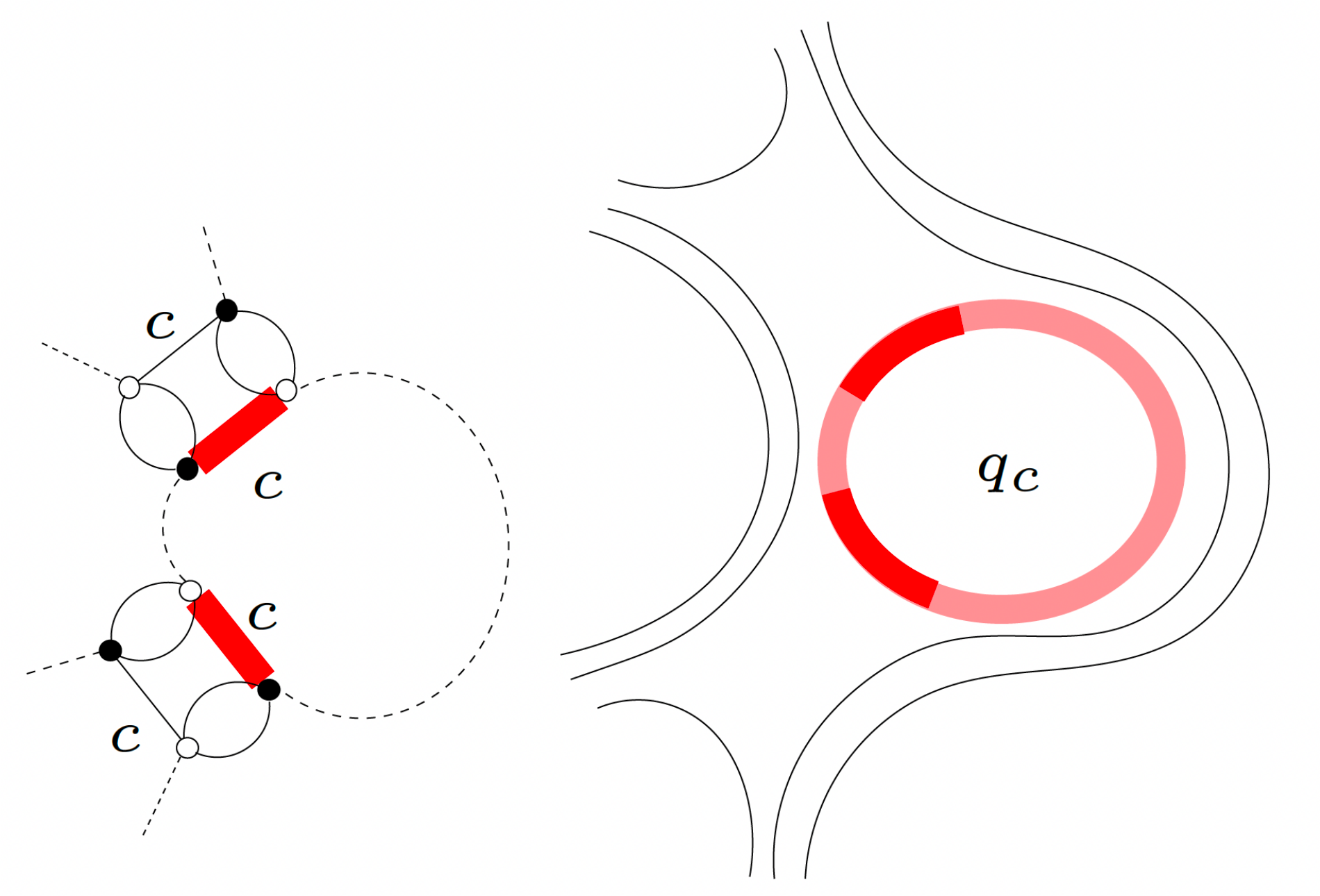}
\caption{ {\small  
One-loop divergent graph, $n_4^{(c)}$
at $d=3$ contributing to the flow of $\lambda$.  
}} 
\label{fig:V4singleloop}
\end{minipage}
\end{figure}

Explicitly, 
at one-loop,
\bea
&&
 K_{n_4^{(c)} } = 2 \,,
 \crcr
 &&
  S_{n_4^{(c)}} (\{\mathbf{p}, \mathbf{p}'\}) = \frac{1}{2!}
  \Big(\frac{-\lapc}{2}\Big)^2 
  \sum_{q_{c}} 
  \frac{\vert q_{c}\vert^{2a}}{(|\mathbf{p}_{\check c}|^{2b} + \vert q_{c}\vert^{2b} + \mu)}
    \frac{\vert q_{c}\vert^{2a}}{( |\mathbf{p}'_{\check c}|^{2b} + \vert q_{c}\vert^{2b} + \mu)}
    \,,
\eea
where $\mathbf{p}_{\check c} = (p_{1},\dots, p_{c-1},p_{c+1}, \dots, p_{d})$, 
and 
$|\mathbf{p}_{\check c}|^{2b} = 
\sum_{l=1| l \ne c}^d p_{l}^{2b}$.

We compile these equations and deliver the $\beta$-functions of the couplings up to one-loop as 
\bea
&&
\lacren =  \lam^{(c)}  -  \frac{1}{4} (\lambda^{(c)}_+)^2 S_0 
\,,
\qquad \quad 
S_0 = \sum_{q}\frac{\vert q \vert^{4a}}{(\vert q\vert^{2b} + \mu_i)^2}
\,,
  \\
  &&
\lapcren  =  \lapc
\,,
\eea 
where $\lam_{\rm ren} $ and $\lam_{+, {\rm ren}} $ are the obvious corresponding renormalized coupling constant. 
We set all couplings to 
$\lac = \lam$, and $\lapc= \lam_{+} $ to simplify 
these 
\bea
&&
\laren =  \lam   -  \frac{1}{4} (\lambda_+)^2 S_0 
\,,
  \\
  &&
\lapren  =  \lam_{+} \; . 
\eea 
These RG flow equations carry already some information. The second equation displays the fact
that the marginal coupling $ \lam_{+}$ does
not run and defines a fixed point  at all orders of perturbation. 
These equations also reflect that the two couplings $ \lam$ and $\lam_{+} $ could not be set the same (assuming equal value of the couplings is inconsistent with
these equations). 

We want to understand the
qualitative feature of RG flow
given by the above coupled system. 
In the multiscale analysis with discrete scale $i$, the system can be written as
\bea
&&
\lam_{i-1} =  \lam_i-  \frac{1}{4} \lam_{+,i}^2 S_{0,i} 
\,,
\label{eq:lambda+S0i}
  \\
  &&
\lam_{+,i-1} =  \lam_{+,i}  \,,
\eea
where $S_{0,i}$ stands for the cut-off
amplitude of  the formal sum $S_0$
and  formulates as
\bea
S_{0,i} &=& \sum_{q}\vert q\vert^{4a}
\int_0^\infty 
d\alpha \chi^{i} (\alpha) e^{-\alpha 
(\vert q\vert^{2b} + \mu_i)}
\int_0^\infty 
d\alpha' \chi^{i} (\alpha') e^{-\alpha' 
(\vert q\vert^{2b} + \mu_i)}
\crcr
&= &
\int_0^\infty 
d\alpha \,\chi^{i} (\alpha) 
\int_0^\infty 
d\alpha' \chi^{i} (\alpha')
e^{-(\alpha+  \alpha')
\mu_i}
\sum_{q}\vert q\vert^{4a}
e^{-(\alpha+  \alpha')
\vert q\vert^{2b} }\,,
\label{s0idebut}
\eea
where the sharp cutoff function $\chi^i(\alpha)$ was defined in \eqref{eq:chii}.

\

\noindent{\bf On scaling dimensions.} 
Computing $\beta$-functions of our different couplings at first order of perturbation and solving them,  
we must deliver these equations using dimensionless couplings.  

Using Peskin and Schroeder's argument \cite{Peskin:1995ev}, the scaling dimensions of a coupling $g_{\bullet}$ can be read from the degree of divergence as ($-1$ times) the coefficient of the corresponding vertex number $V_{\bullet}$. The same argument has been extended to the scaling of TFT couplings  according to \cite{BenGeloun:2016rqa}. 
This agrees also with the scaling dimension of field and coupling 
using integrability arguments, 
see for instance 
\cite{Ferdinand}, for applications
involving both stochastic analysis
in the TFT setting. 

 Denoting the scaling dimension of a coupling $g$ by $\{g\}$, and using \eqref{ompcG}, we have 
 
\bea
&&
\{\lam_+\} = -(d-1-4b +2a) = 0\;,  \qquad 
\{\lam\} = -( d-1-4b) = 2a=d-2 \;, 
\crcr
&&
\{\mu\} = 2b = d-\frac32 \;,  \qquad  \{Z_a\} = 2(b-a)= \frac12\; , 
\label{scal+}
\eea
where the vanishing of the first scaling dimension is due to the marginality of $\lam_+$. 
The dimensions also show that both $\lam$ and $Z_a$ define
relevant directions.

\

\noindent{\bf Perturbative $\beta$-function and integration.} 
First, we give an approximation of 
$S_{0,i}$ \eqref{s0idebut}. 
The sum over $q$ 
will be handled
using an integral via Euler-Maclaurin approximation (Appendix \ref{app:SumtoInt} details the
following result): 
\bea
S_{0,i} =
 \frac{1}{b}
 \log 
  \frac{(M^{2b} +1)^2}{
  4 M^{2b }}
 + {\mathcal O}(  M^{-2b i}
 \log(M^{-2b i})) 
 \,.
 \label{s0iapprox}
\eea
We write the $\beta$-function for a given coupling $g$ as $\beta_g (k) = k\partial_k g (k)$, where $k$ is a momentum scale. In our present setting dealing 
with discrete slices (multiscale analysis), 
we have finite difference equations that we will turn into differential equation. 
The momentum scale must be compared
to the slice range as
 $k/k_0 \sim M^{i}$, given in terms of a given momentum unit $k_0 $. 
Using the so-called time scale
 $t= \log (k/k_0) \sim i \log M$,
given that the coupling 
$\lambda_{+} = \lambda_{+, i}$ does
not run and that, at leading order
 $S_{0, i}= c \log M $,
where $c>0$,  
the difference $\lambda_{i-1} - \lambda_i$ takes the form: 
\bea
-(\lambda_{i-1} - \lambda_i) = 
\frac{\partial \lambda_{i}}{   \partial i}
&=&
\frac{1}{4} \lambda_{+,i}^2 \, S_{0, i}
\,,
\eea
which can be translated into 
the following first order ODE
\bea
\frac{\partial \lambda_{i}}{\partial ( (\log M)  i) } = \partial_t \lambda (t)
&=&
- \beta_\lambda  \lambda_{+}^2 
\crcr
\beta_\lambda 
&=& 
- 
\frac{1}{4b}
\frac{
\log 
  \frac{(M^{2b} +1)^2}{
  4 M^{2b }} 
  }
 {\log(M)} <0 
\,.
\eea
One may wonder why $M$, the propagator slice parameter,  appears
in the perturbative expansion and if the present
equation does not depend on the slicing scheme. 
The multi-scale analysis justifies this
entirely: we are computing a flow
between two scales $\sim M^{-2bi}$, therefore
$M$ becomes an input of our equation.

Using the scaling dimensions 
\eqref{scal+}, we have $S_{0,i} = \widetilde S_{0,i}$, $\lambda = k^{d-2} \widetilde \lambda$ and $\lambda_{+} = \widetilde \lambda_{+}$, where $\; \widetilde{}\; $ indicates a dimensionless quantity{\footnote{Dimensionless here means that  the quantity with $\, \widetilde{}\, $  does not have scaling behavior in $t$ nor $k$.}}. 
Then, we obtain $\partial_t \lambda  
= (d-2)k^{d-2} \widetilde \lambda +  k^{d-2}\partial_t \widetilde \lambda$
and 
\beq
\label{eq:lambdat+}
\partial_t \widetilde \lambda = -(d-2) \widetilde \lambda + k^{-(d-2)} \vert \beta_\lambda \vert \lambda_+^2\,.
\eeq
We use $k = k_0 e^{t}$, and any equation dependence in the unit of momentum $k_0$ will be confined to a single constant $c_0 $
(though, this constant may vary from equation to equation). 
Massaging \eqref{eq:lambdat+},
\beq
\partial_t (e^{(d-2)t} \widetilde \lambda) = 
c_0  \vert \beta_\lambda \vert  \lambda_+^2\,.
\eeq
Now, we integrate the equation to obtain
\bea
\widetilde \lambda (t)
&=& c_0 
\vert \beta_\lambda \vert  \lambda_+^2\, t \, e^{-(d-2)t} + const. \, e^{-(d-2)t}
\crcr
&=&  c_0
\vert \beta_\lambda \vert  \lambda_+^2 (t-t_0) \, e^{-(d-2)t}
+ \widetilde \lambda (t_0) \, e^{- (d-2)(t-t_0)} 
\label{eq:lambdarunlinear+}
\eea
where the initial condition
was set at some IR scale 
$ e^{t_0} \ll \Lambda/k_0 = e^{t}$.

 As opposed to the usual $\phi^4_4$ model, 
there is no pole in the solution at first order. This proves that 
there is no Landau ghost and therefore this model 
is not similar to the ordinary $\phi^4_4$ model. 
Furthermore, 
at large $t\ge t_0$,
$\widetilde\lam(t)$ becomes suppressed
by the exponential factors.  
This is of course the ordinary behavior of relevant couplings. 
One may be tempted to conclude to  asymptotic freedom, however, we will show that another coupling does not run to 0. 
In the IR, $t\to -\infty$, $\lam$ grows 
and such that we may expect phase transition in this regime. 
Note also that the resulting model differs 
from the usual 
$T^4$-TFT model
with only $\lam$ coupling. 
Indeed, 
the latter describes 
a different class of dominant graphs (melonic ones) yielding  asymptotic 
freedom via a marginal coupling. 
Finally, at $d=2$, something special happens. Of course, this is an (enhanced) matrix model which enhances some planar graphs. This deserves a 
full-fledged investigation.

\subsection{Computing $\Gamma_2$
and $Z_{a}$ RG equation}
\label{compGamma2+}

Now we compute the renormalization of the $2$-point coupling, $\Gamma_2^{(c)}(\{0 \})$, at fixed color $c$: 
\bea
 \vert p_{c}\vert^{2 a}
\Gamma_2^{(c)}(\{p \})
& =& \sum_{{\cal G}_{2;a;\iota}^{(c)} }
K_{{\cal G}_{2;a;\iota}^{(c)}} S_{{\cal G}_{2;a;\iota}^{(c)}} (\{p\})\,,
\label{Gamma2Za}
\eea
where the sum is over all amputated $1$PI $2$-point graphs at $1$-loop whose boundaries to be in the form of ${\rm Tr}_{2;(c)} ({ p}^{2 a} \phi^2)$.

Up to the first order in perturbation theory, we have 
${\cal G}_{2;a; \iota}^{(c)} \in \{z_a^{(c)},  m_{e}^{(c)} \}$, where $z_a^{(c)}$ is the leading order (zero-loop) graph with one $Z_a^{(c)}$ interaction with two external legs. (Appendix \ref{app:2ndO+} shows additional graphs up to the second order in perturbation theory.) 
In  \eqref{eq:sigma}, we have identified $\partial_{\vert p_{c}\vert^{2 a}} \Sigma\vert_{\{p\}=0} \equiv \Gamma^{(c)}_2 (\{0\})$ as divergent.
The other  contributing graphs, namely $m_{e}^{(c)}$, $c=1,2,3$,  are divergent at first order in perturbation theory  (row II of Table \ref{tab:listprim1}) 
 with  $\omega_{m_{e}^{(c)}} = \frac{D}{2}$. 

\begin{figure}[H]
\centering
     \begin{minipage}[t]{0.7\textwidth}
      \centering
\includegraphics[angle=0, width=5cm, height=2.5cm]{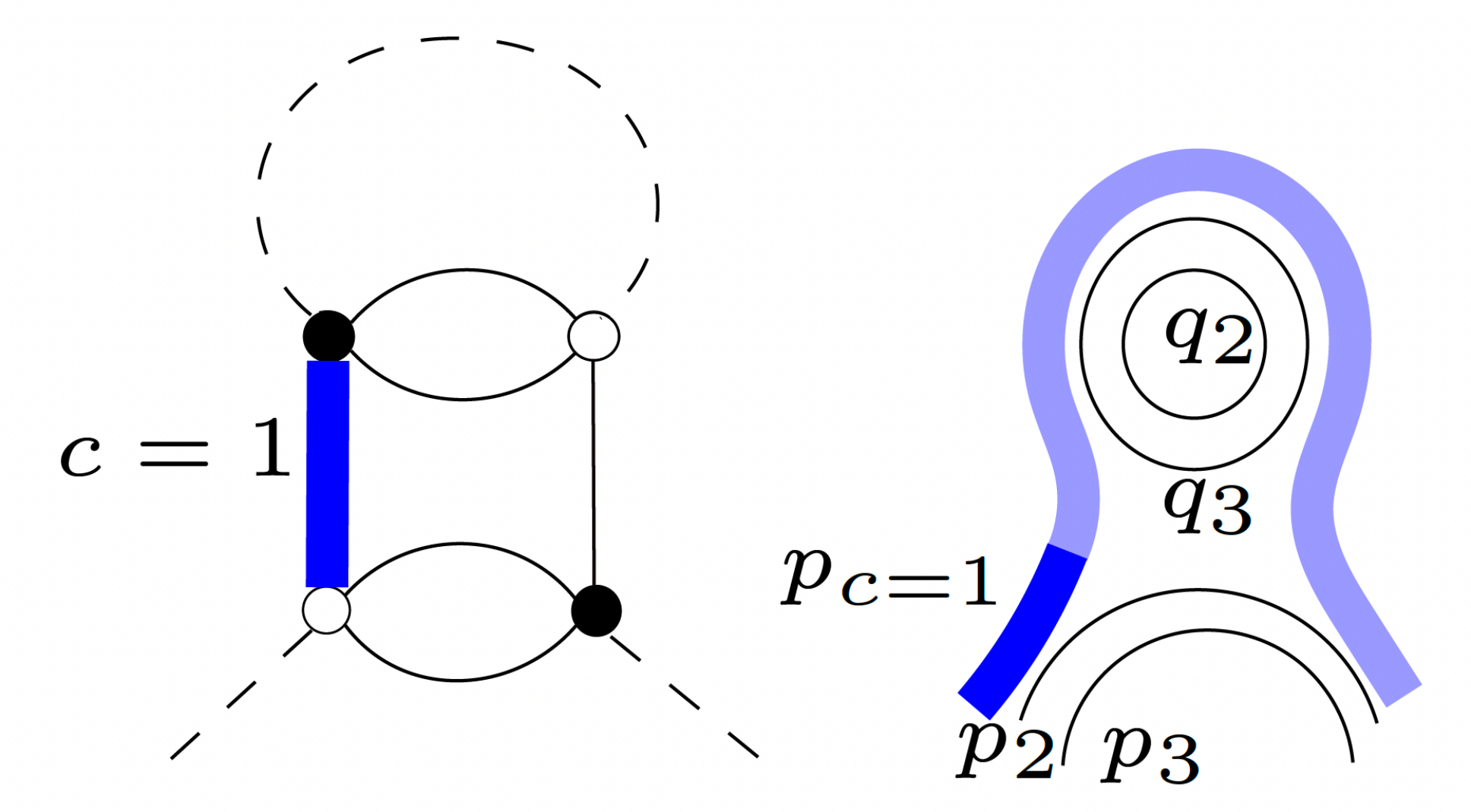}
\caption{ {\small  
Illustration of $m_{e}^{(c=1)}$ for $d=3$ in colored (left) and stranded (right) representations. $\{q_{\check c}\} = \{q_{2}, \; q_{3}\}$, ${|{\bf q}_{\check c}|}^{2 b} = {|q_{2}|}^{2 b} + {|q_{3}|}^{2 b}$.
}} 
\label{fig:m1e}
\end{minipage}
\end{figure}
\bea
K_{m_{e}^{(c)}} S_{m_{ e}^{(c)}} (\{p\}) 
&=&
K_{m_{e}^{(c)}}
\Big( - \frac{{\lambda_{+}}^{(c)}}{2}\Big)
\sum_{\{q_{\check c}\}} \frac{\vert p_{c}\vert^{2 a}}{( {|{\bf q}_{\check c}|}^{2 b} + \vert p_{c}\vert ^{2 b} + \mu )}
\,,
\eea
where
$K_{m_{e}^{(c)}} = 2$.
Putting all together, up to first order in perturbation theory, 
\bea
 \vert p_{c}\vert^{2 a}
\Gamma_2^{(c)}(\{p\})
=
 - \vert p_{c}\vert^{2 a}
Z_a^{(c)} 
 - \lambda_+^{(c)} \vert p_{c}\vert^{2 a}
\sum_{\{q_{\check c}\}} \frac{1}{( {|{\bf q}_{\check c}|}^{2 b} + {\vert p_{c}\vert}^{2 b} + \mu )}
\,.
\eea
$\Gamma_2^{(c)}(\{p\})$ is therefore
identified easily. 
For general $d$, recalling the renormalization group equations \eqref{eq:rgeqns} and setting all the external momenta $\{p\} =0$, we obtain: 
\bea
Z^{(c)}_{a,{\rm ren}} &=&
- \Gamma^{(c)}_2(\{0\})
=
Z^{(c)}_{a}
+
\lambda_+^{(c)} 
\sum_{\{q_{\check c}\}} \frac{1}{( {|{\bf q}_{\check c}|}^{2 b} + \mu )}
\,.
\eea
Note that we only keep the first order in Taylor expansion in couplings.
Setting the couplings independent of colors, 
\bea
Z_{a,{\rm ren}} 
&=&
Z_{a}
+
\lambda_+ 
\sum_{\{q_1,\dots, q_{d-1}\}} 
\frac{1}{({|{\bf q}|}^{2 b}  + \mu )}
\,.
\eea
In the multiscale analysis language,
\bea
Z_{a, i-1} 
=
Z_{a, i}
+
\lambda_{+,i} S_{1,i}
\,,
\eea
where $S_{1,i}$ is given in \eqref{s1idebut} with explicit expression given in \eqref{eq:S1itildeS1i}, where we fixed $D=1$.
We recall that $\lambda_+ = \lambda_{+,i}$ does not run.
Making explicit the dimensions, we obtain the renormalization group equation for $Z_a$ as
\bea
Z_{a, i-1} 
=
Z_{a, i} + 
k^{1/2}
\lambda_{+,i} 
\,
\widetilde S_{1,i}
\,,
\eea
which,  
using $\widetilde S_{1,i}$ given in \eqref{eq:runningmass2}, 
 gives  the $\beta$-function for $Z_a$: 
\bea
- ( Z_{a, i-1} -  Z_{a, i}) 
&=&
\frac{\partial Z_{a, i}}{\partial i}
=
- k^{1/2} \lambda_{+,i} \, \widetilde S_{1,i}
\,,
\crcr
\frac{\partial  Z_{a, i}}{\partial ( (\log M)  i) } 
&=&
\partial_t  Z_{a} (t)
= 
- k^{1/2}  \beta_{Z_a} \, \lambda_{+}
\,,
\crcr
\beta_{Z_a} 
&=&
\frac{\widetilde S_{1,i}}{\log(M)} > 0 
\,.
\eea
Introducing dimensionless quantities, 
$Z_a(t)= k^{1/2}\widetilde Z_a(t)$, 
the dimensionless RG equation can be written 
\bea
\partial_t  \widetilde Z_a(t)
&=& - \frac{1}{2} \widetilde Z_a(t)+ k^{-1/2} \partial_t Z_a(t) \crcr
& = &
- \frac{1}{2} \widetilde Z_a(t)
-  \beta_{Z_a} \, \lambda_{+}
\,.
\eea
This integrates easily with respect to $t$ and
gives
\bea
 \widetilde Z_a(t) = 
 c_1 \, e^{- t/2}   - 2 \beta_{Z_a}  \, \lambda_{+} 
 \,,
\eea
where $c_1$ is  an integration constant.
This equation just expresses the fact that
 $\widetilde Z_a(t)$ is a relevant coupling and
decreases exponentially in the UV ($t \rightarrow \infty$) and suppressed up to reach a constant $ - 2 \beta_{Z_a}  \, \lambda_{+} $. 
This and the fact that $\lam_+$ is a constant make  this model 
not asymptotically free
although the coupling $\widetilde\lam$ flows to 0. 
On the other hand, in the IR ($t \rightarrow -\infty$),  $ \widetilde {Z}_a(t)$ blows up as any relevant coupling.

\

\subsection{Self energy and mass RG equation}
\label{sect:selfenergmass}

Our following task is to compute the so-called self energy, which we denote by $\Sigma_b(\{ p \})$: 
\bea
\Sigma_b(\{p \})
=
\sum_{c=1}^d 
\sum_{{\cal G}^{(c)}_{2,\iota}} {K_{{\cal G}^{(c)}_{2,\iota}}} {S_{{\cal G}^{(c)}_{2,\iota}}} (\{p\})\,,
\label{selfnrg}
\eea
where the first sum 
is broken by color $c$
and the second sum 
is performed over 
all  amputated 1PI 2-point graphs 
${\cal G}^{(c)}_{2,\iota}$
(with color label $c$) at 1-loop 
 with  boundary to be in  the form of ${\rm Tr}_2 (p^{2 b} \phi^2)$
 (hence the index $b$
 in subscript in $\Sigma_b(\{p \})$). 
 
It is noteworthy that 
$\Sigma_b(\{p \})$ corresponds to the part 
$ \Sigma (\{0\}) 
+  \sum_{c}
\vert p_{c}\vert^{2 b} \partial_{\vert p_{c}\vert^{2 b}} \Sigma \big\vert_{\{p\}=0}$
of total self-energy   function $\Sigma(\{ p \})$ in \eqref{eq:sigma}. 
Noting that the second term $\partial_{\vert p_{c}\vert^{2 b}} \Sigma \big\vert_{\{p\}=0}=0$, we only focus on the contribution $\Sigma (\{0\}) 
$, namely the contribution to the mass renormalization. 

The graphs we are interested in are denoted
${{\cal G}^{(c)}_{2,\iota}} \in \{ m^{(c)}, n^{(c)}\}_{c=1,2,\dots, d}$, see Figures \ref{figu:m1} and
\ref{fig:n1}. 

\begin{itemize}
\item
For the graph $m^{(c)}$, $c=1,\dots, d$,
the degree of divergence
$\omega_{d;+}(m^{(c)})$ reaches $\frac{D}{2}$,
as shown in the class III of Table \ref{tab:listprim1}. 
In Figure \ref{figu:m1}, 
we display $m^{(c)}$, when the vertical line is color $c$. 

\begin{figure}[H]
\centering
     \begin{minipage}[t]{0.6\textwidth}
      \centering
\includegraphics[angle=0, width=5cm, height=2.5cm]{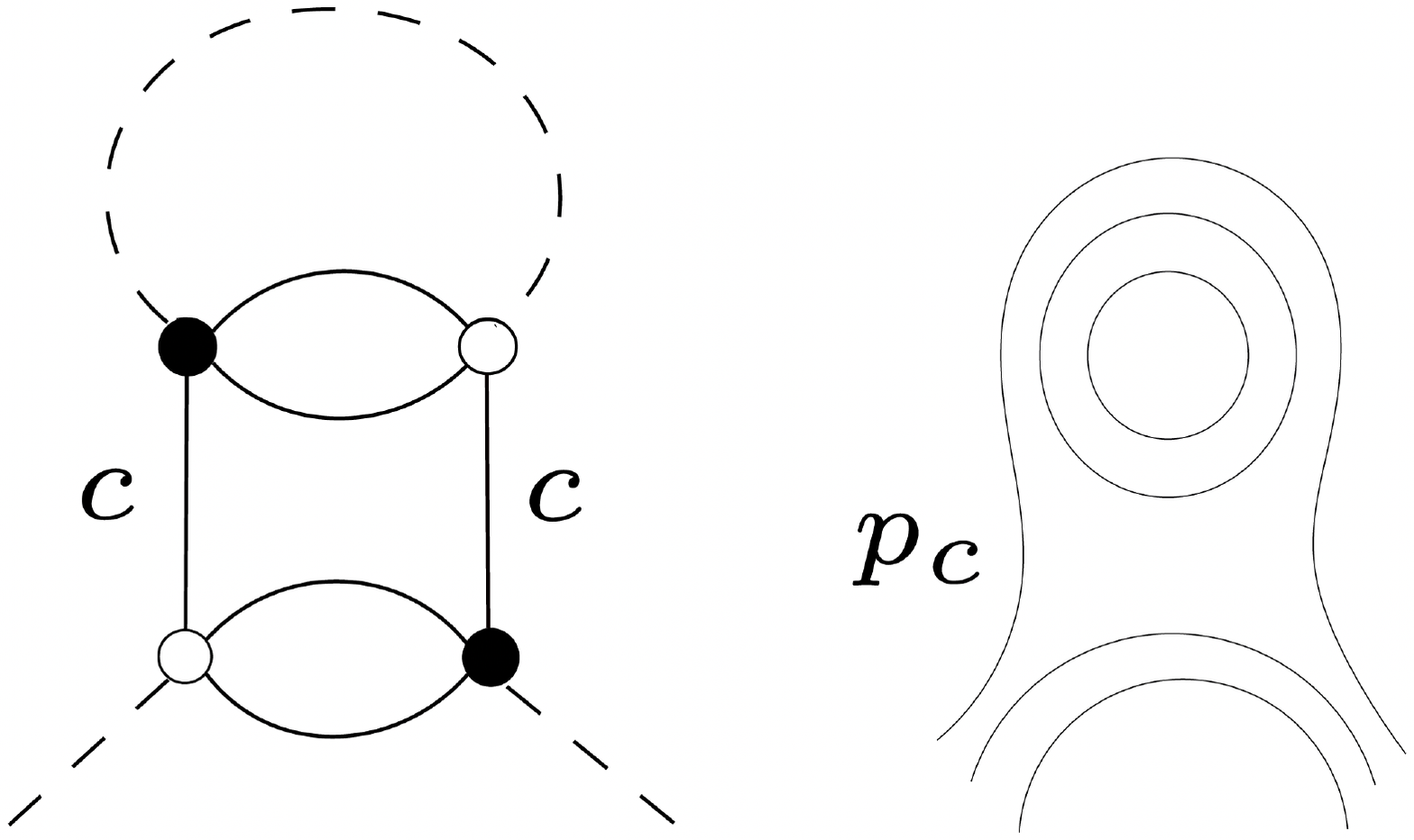}
\caption{ {\small {The graph $m^{(c)}$ in the case $d=3$ in colored (left)
and stranded (right) representations. 
}}} 
\label{figu:m1}
\end{minipage}
\end{figure}

The contribution  to the amplitude brought by the graphs $m^{(c)}$ is 
\bea
\sum_{c=1}^{d}
K_{m^{(c)}} S_{m^{(c)}}  ( \{p\})
&=&
\sum_{c=1}^{d} 2 \big(- \frac{\lambda^{(c)}}{2} \big) \sum_{\{q_{\check c}\}} 
\frac{1}{({|{\bf q}_{\check c}|}^{2 b} + \vert p_{c}\vert^{2 b} + \mu )} \crcr
&=& 
\sum_{c=1}^{d} \sum_{\{q_{\check c}\}} 
(- \lambda^{(c)})
\frac{1}{({|{\bf q}_{\check c}|}^{2 b} + \vert p_{c}\vert^{2 b} + \mu )}
\,,
\eea
where $K_{m^{(c)}} = 2$,
for any $c$.

\item
In the second class of graphs denoted each $n^{(c)}$, the amplitude has divergence degree $\omega_{d;+}(n^{(c)}) = \frac{D}{2}$.
This graph belongs to the class I in Table \ref{tab:listprim1}.
Figure \ref{fig:n1} shows $n^{(c)}$
at order $d=3$ with horizontal line colored by index $c$. 

\begin{figure}[H]
\centering
     \begin{minipage}[t]{0.5\textwidth}
      \centering
\includegraphics[angle=0, width=5cm, height=2.5cm]{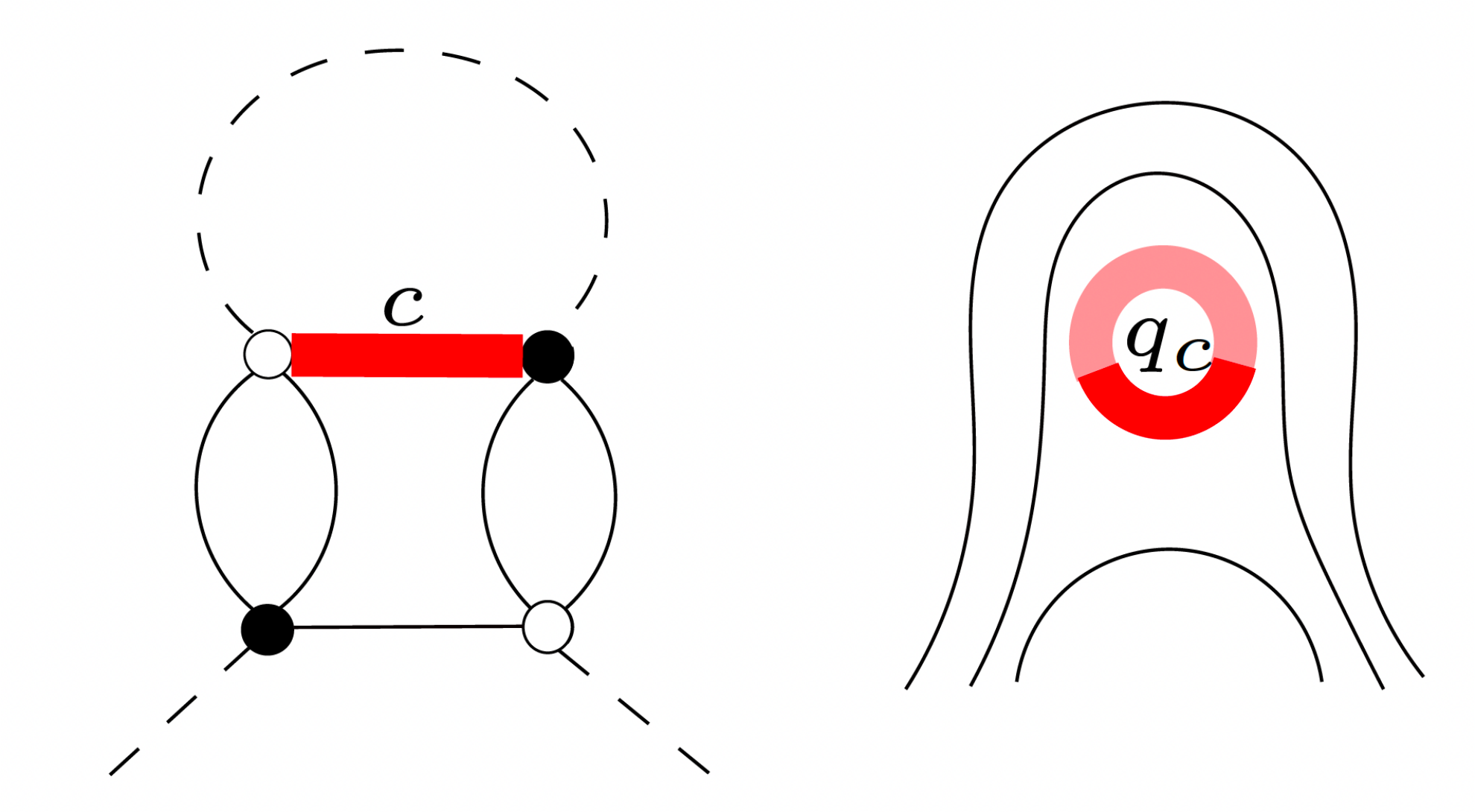}
\caption{ {\small  $n^{(c)}$ for $d=3$ in colored (left) and stranded (right) representations. 
}} 
\label{fig:n1}
\end{minipage}
\end{figure}

The sum of Feynman amplitude associated with the graph $n{(c)}$ yields
\bea
\sum_{c=1}^{d}
K_{n^{(c)}} S_{n^{(c)}} ( \{p\})
&=&
\frac12
\sum_{c=1}^{d}
(-\lambda_{+}^{(c)} )
\sum_{q_{c}}
\frac{\vert q_{c}\vert^{2 a}}{(\vert q_{c}\vert ^{2 b} + {|{\bf p}_{\check c}|}^{2 b} + \mu)}\,,
\eea
where $K_{n^{(c)}} = 1$ for any $c$. 
\end{itemize}

Up to the 
first order in perturbation theory and 
evaluating at $0$ external momenta, we obtain 
\bea
&&
\Sigma_b(\{0\})
=
\sum_{c=1}^d  
\Big( K_{m^{(c)}} S_{m^{(c)}} (\{p\}) 
+  K_{n^{(c)}} S_{n^{(c)}} (\{p\}) 
\Big)  \Big|_{p=0}
\crcr
&& = 
\sum_{c=1}^{d}
\Big[(- \lambda^{(c)}) \sum_{\{q_{\check c}\}} 
\frac{1}{({|{\bf q}_{\check c}|}^{2 b} + \mu )}
+
\frac{1}{2}(- 
{\lambda_{+}}^{(c)})
\sum_{q_{c}}
\frac{\vert q_{c}\vert^{2 a}}{(\vert q_{c}\vert^{2 b} + \mu)} \Big]
\crcr
&&
\eea
Therefore, for general $d$, the renormalized mass equation up to the first order in perturbation is given by:
\bea
{\mu_{\rm ren}} 
= 
\mu +
\sum_{c=1}^{d}
\Bigg\{
\lambda^{(c)} \sum_{\{q_{\check c}\}} 
\frac{1}{({|{\bf q}_{\check c}|}^{2 b}  + \mu )}
+
\frac{1}{2}
{\lambda_{+}}^{(c)} 
\sum_{q_{c}}
\frac{\vert q_{c}\vert^{2 a}}{(\vert q_{c}\vert^{2 b}  + \mu)}
\Bigg\}
\,.
\eea
 Now assert color independence, namely $\lambda^{(c)} = \lambda$, ${\lambda_{+}}^{(c)} = \lambda_{+}$, one gets 
\bea
&&
\mu_{\rm ren}
=
\mu +
d \Big(
\lambda S_1
\: + \;   \frac{1}{2}
{\lambda_{+}}
S_2 
\Big) 
\crcr
&&
S_1 = \sum_{\{q_1,\dots, q_{d-1}\}} 
\frac{1}{({|{\bf q}|}^{2 b}  + \mu )}
\qquad 
S_2 = 
\sum_{q}
\frac{\vert q \vert^{2 a}}{(\vert q \vert^{2 b}  + \mu)}
\,,
\label{eq:S1S2}
\eea
where we restrict ourselves to $D=1$.

We cut off the propagators, switch to dimensionful quantities, and write the sums
\bea
&&
S_{1,i} = 
\sum_{{\bf q} \in k \Z^{d-1}}
\int_{0}^{\infty}
d\alpha \; \chi^i(\alpha)\, 
e^{ - \alpha( |{\bf q}|^{2 b} + \mu_i )}
\,,
\label{s1idebut}
\\
&&
S_{2,i} = 
\sum_{q\in  k \Z} 
\int_{0}^{\infty}
d\alpha \; \chi^i(\alpha)\, 
\vert q \vert^{2 a}
e^{ - \alpha( |q|^{2 b}  + \mu_i )}
\,.
\label{s2idebut}
\eea
In writing explicitly the dimensions of $q$ and $\alpha$ in terms of  momentum  scale $k$, \eqref{eq:S1S2dimfuldimless} in Appendix \ref{app:SumtoInt} proves that
we can approximate those
as
\bea
&&
S_{1,i} = k^{1/2} 
\Bigg(
\Big( \frac{1}{b} \Gamma\left(\frac{1}{2b} \right) \Big)^{d-1} 
(2d-3)
\Big(  1 - M^{-1/2} \Big)
M^{i/2}
+{\mathcal O}(M^{-i/2}) 
\Bigg)
\,,
\label{eq:S1itildeS1i}
\\
&&
S_{2,i} = k^{1/2} 
\Bigg(4 \, \Gamma\Big(\frac{2(d-1)}{2d-3}\Big)
(1 - M^{-1/2} ) M^{i/2}
 + 
 {\mathcal O}\big(  M^{-i (d -2)}  \big) 
 \Bigg)
 \label{eq:S2itildeS2i}
 \,,
\eea
where we have set $D=1$ at any order $d$  and   $b = \frac{1}{2} (d - \frac{3}{2})$ given in Proposition \ref{prop:list+}.
Compiling this, 
we obtain the following
equation at leading order,
\bea
\mu_{i-1} 
&=&
 \mu_i 
+
 k^{1/2} 
d\Big(
\lambda_i \widetilde S_{1,i}
+
\frac{1}{2}
\lambda_{+, i}
\widetilde S_{2,i}
\Big) 
\,,
\crcr
\widetilde S_{1,i} & = & k^{-1/2}S_{1,i}
\,,
\crcr
\widetilde S_{2,i} &=&  k^{-1/2}S_{2,i}
\,.
\label{eq:murenorm}
\eea
We obtain the $\beta$-function for the
mass 
\bea
- ( \mu_{i-1}- \mu_i)
&=&
\frac{\partial  \mu_i}{\partial i}
= - k^{1/2} d \Big( \widetilde S_{1,i} \, \lambda_i 
+ \frac{1}{2} \widetilde S_{2,i} \, \lambda_{+,i}\Big) 
\,,
\crcr
\frac{\partial \mu_i}{\partial ((\log M) i)}
&=&
\partial_t  \mu 
= 
- k^{1/2} (\beta_{\mu, 1}\,  \lambda
+  \beta_{\mu, 2}\, \lambda_{+})
\,,
\crcr
\beta_{\mu, 1} 
&=&
\frac{d}{\log M}  \, \widetilde S_{1,i} >0
\,,
\crcr
\beta_{\mu, 2}
&=&
\frac{d}{2 \, \log M} \,  \widetilde S_{2,i} >0
\,.
\eea
According to \eqref{scal+}, 
we recall the scaling dimensions $\{\mu\} = 2 b = d- 3/2$, $\{\lam_+\} = 0$
and $\{\lam\} = 2a = d-2$. 
We therefore switch to dimensionless quantities, 
$\mu = k^{d- \frac32} \widetilde \mu$, 
therefore, $\partial_t  \mu = k^{d- \frac32} ( (d-\frac{3}{2}) \widetilde \mu  + \partial_t  \widetilde \mu)$.
This equation becomes
\bea
\partial_t  \widetilde \mu(t)=- (d- \frac32)\widetilde \mu(t)+ k^{-(d- \frac32)} \partial_t \mu (t)
\,.
\eea
Given that the coupling 
$\lambda_{+} =  \lambda_{+, i}$ does
not run and that,
 $ \widetilde \lambda$ runs according to \eqref{eq:lambdarunlinear+}, we make the $t$ dependence explicit and 
 \bea
 \label{eq:dtmu+}
 \partial_t \widetilde \mu (t)
 &=& -  (d- \frac32) \widetilde \mu (t)
- 
( \beta_{\mu, 1}\,\tilde \lambda(t)
+ c_0 e^{- (d-2) t} \, \beta_{\mu, 2}\,   \lambda_{+}) \cr\cr
&=& 
- (d- \frac32)\widetilde \mu(t)  \, 
-\left( \beta_1   (t-t_0)+ \beta_{2} \right) e^{-(d-2)t} \,, 
 \eea
 where $\beta_1 = c_0\beta_{\mu, 1} \,  \vert\beta_\lambda\vert \, \lambda_{+}^2$, $\beta_2 = \beta_{\mu, 1} \tilde \lambda (t_0)  e^{(d-2)t_0} +  c_0 \beta_{\mu, 2}\,    \lambda_{+}$. 
The above differential equation \eqref{eq:dtmu+} takes the form: 
 \beq
  \partial_t (e^{(d- \frac32)t} \widetilde \mu (t))  =  - e^{ \frac12 t}\left( \beta_1   (t-t_0)+ \beta_{2} \right)
  \,.
 \eeq
 We integrate this and obtain 
 \bea
  \widetilde \mu (t)
    &=&
  -  2 e^{-(d- 2)t} \Big(
   \beta_1 (t- t_0) 
    + ( - 2\beta_1 + \beta_{2}) \Big)  + const. e^{-(d- \frac32)t}  
    \,,
 \eea
 with 
 \bea
const. =  e^{(d- \frac32)t_0} \widetilde \mu (t_0)  +  2 e^{ \frac12 t_0} \Big( 
    - 2\beta_1 + \beta_{2}  \Big) 
    \,.
 \eea
In the UV, $\widetilde\mu$ runs to $0$ whereas in the IR the mass exponentially increases. This behavior is common to any relevant mass coupling.

\ 

\noindent{\bf Summary --} 
We list our dimensionless 1-loop RG flow equations for the
model $+$ and their solutions below.
The constants $\kappa_i$, $i=1,2,3,4$, are integration 
constants. 
\begin{table}[H]
\centering
\begin{tabular}{l|l}
\hline 
\hline 
\\
$\partial_t \,\widetilde\lambda (t) =
-(d-2)\widetilde\lambda (t)  + c_0|\beta_\lambda|  \lambda_{+}^2 e^{-(d-2)t} $\,
% \qquad $\beta_{\lambda} <0$  
 &  $\widetilde\lambda(t) 
= e^{-(d-2)t}(c_0|\beta_\lambda| \, \lambda_{+}^2\,t + \kappa_1 )$ \\ 
\\
$\partial_t\,  \lambda_{+} = 0 $  & $\lambda_{+} =\kappa_2$\\
\\
$\partial_t \,\widetilde \mu (t)
=- (d-\frac{3}{2}) \widetilde \mu (t)$  
&
$\widetilde \mu (t)= 2e^{-(d-2)t}
\big[
 -\beta_1 \,t +  \gamma \big]+\kappa_3 e^{-(d-\frac32)t}$ \\
\qquad\qquad $
-  (\beta_{\mu, 1}\,  \widetilde\lambda(t)
+ c_0 e^{-(d-2)t}\beta_{\mu, 2}\,  \lambda_{+})$ 
& $\beta_1 = c_0\beta_{\mu, 1} |\beta_\lambda| \lambda_{+}^2 > 0 $\,,
 \\ 
$\beta_{\mu, 1} > 0\,, \;\, c_0\beta_{\mu, 2} > 0 $& 
$\gamma =   \beta_1(t_0+2) - \beta_2 $
\\
& $ \beta_2 = \beta_{\mu, 1} \tilde \lambda (t_0)  e^{(d-2)t_0} +  c_0 \beta_{\mu, 2}\,    \lambda_{+}$
 \\
 \\
$\partial_t \,  \widetilde Z_a(t)
 = 
- \frac{1}{2} \widetilde Z_a(t)
-  \beta_{Z_a} \, \lambda_{+} $\,,
\quad 
  $\beta_{Z_a} > 0$   &
$\widetilde Z_a(t) = 
 \kappa_4 \, e^{- t/2}   - 2 \beta_{Z_a}  \, \lambda_{+} $
 \\ 
 \hline 
 \hline 
\end{tabular}
\caption{Summary of the RG flow equations for the couplings in the model $+$.}
\label{tab:RG+}
\end{table}

\subsection{Integration at arbitrary loops}
\label{sec:allloops+}

The power counting of the model $+$, provided by Proposition \ref{prop:list+},  gives us a lot of information about the $\beta$-functions of the couplings even in at arbitrarily high order of perturbation theory. 
In this section, we investigate general forms of the $\beta$-functions of the couplings in the model $+$ at all orders of perturbation theory.

\

\noindent{\bf 4-point couplings  $\lambda$ and $\lambda_+$  RG equation --}
 Proposition \ref{prop:list+} dictates that there are no diverging amplitudes contributing to the renormalization of $\lambda_+$ at all orders in perturbation, therefore, {\it $\lambda_+$ is constant at all orders}.
Furthermore, from the first row of
Table \ref{tab:listprim1} of Proposition \ref{prop:list+}, which governs the renormalization of the coupling $\lambda$, we know that in order for the amplitude to be divergent, Feynman graph must only contain $\lambda_+$ couplings, and no other couplings. Then, we can readily conclude that,
the subleading corrections to $\lambda$ at  $n$-loops, $n$ being arbitrary,
can be written as a  polynomial $P_n(\lambda_+)$ in the variable $\lambda_+$ that is fixed to a constant. 
At arbitrary $n$-th order in perturbation, the $\beta$-function of the coupling $\lambda$ assumes the form 
\bea
\partial_t \lambda (t)
= P_n(\lambda_{+}) 
\eea
and that can be integrated in terms of dimensionless coupling as 
 \bea
\widetilde\lambda(t) 
= 
e^{-(d-2)t}( P_n(\lambda_{+})  t + const. )
\,.
\label{eq:lambda+allorders}
\eea
The particular form of $P_n(\lambda_{+}) = -\beta_{\lambda}\lambda_{+}^2 + \dots $ is left for future  investigation.  
However whatever form  $P_n(\lambda_{+})$ may have, the asymptotic behavior of $\widetilde \lambda$ remains unchanged: it vanishes in the UV.

\

\noindent{\bf 2-point coupling $Z_a$ RG equation --}
The classes II and V of Proposition \ref{prop:list+} contribute to the renormalization of the $2$-point coupling $Z_a$.
All class II amplitudes involve only the coupling $\lambda_+$ 
yielding a similar result as to the one-loop computation above.  
The class V contains  contributions with exactly one $Z_a$ and 
several $\lambda_+$ (an example is given in Fig. \ref{fig:m2e} Appendix \ref{app:2ndO+}).
Therefore, at an arbitrary $n$-th order in perturbation theory, 
we expect
\bea
\partial_t Z_a(t) 
&=&
e^{t/2} Q_{1;n} (\lambda_+) + t\,  Q_{2;n} (\lambda_+)\, Z_a(t)
\,,
\eea
where $Q_{i;n}  (\lambda_+) $, $i=1,2$ are polynomials in $\lambda_+$. 
This equation can be integrated 
but we refrain to display the solution.

\

\noindent{\bf Mass RG equation --}
Looking at Proposition \ref{prop:list+}, the classes that contribute to the renormalization of the mass are I, III, IV, and VI.
The class I only contains $\lambda_+$, which is held constant.
The class III contains only exactly one $\lambda$ and the rests are all $\lambda_+$.
The class IV contains only exactly one $Z_a$ coupling and the rests are all $\lambda_+$.
Finally, the class VI contains exactly only one $Z_a$, and exactly only one $\lambda$, and the rests are $\lambda_+$.
Noting that $\lambda_+$ does not run at all orders, the most complicated one could have is the class VI where $\lambda$ and $Z_a$ are coupled (whose example is given in Fig.{\ref{fig:m2}}).
Therefore, we expect 
\bea
\partial_t \mu(t)
&=&
e^{t/2} \Big(R_{1;n}(\lambda_+) \, \lambda (t)
+ R_{2;n}(\lambda_+) \Big)
+ 
\, t\, Z_a (t)\Big(
R_{3;n}(\lambda_+) \lambda (t) \, 
+ 
 R_{4;n} (\lambda_+)   \Big)
\,,
\qquad 
\label{eq:mass+allorders}
\eea
to an arbitrary $n$-th order in perturbation theory, and $R_{i;n}$, with $i = 1, 2, 3, 4$ are polynomials in $\lambda_+$.

\

In summary, we 
expect that the coupled system of RG equations of the model $+$ to an arbitrary $n$-th order is given in terms of dimensionful couplings as
\bea
\partial_t \lambda_+ 
&=&
0
\,,
\crcr
\partial_t \lambda (t)
&=&
 P_n(\lambda_{+}) 
 \,,
 \crcr
\partial_t Z_a(t) 
&=&
 e^{t/2} Q_{1;n} (\lambda_+) + t\, Z_a(t) Q_{2;n} (\lambda_+)
\,,
\cr
\partial_t  \mu(t)
&=& 
e^{t/2} \Big(R_{1;n}(\lambda_+) \, \lambda (t)
+ R_{2;n}(\lambda_+) \Big)
+ 
\, t\, Z_a (t)\Big(
R_{3;n}(\lambda_+) \lambda (t) \, 
+ 
 R_{4;n} (\lambda_+) \,   \Big)
\,.
\qquad
\eea
Apart from $\widetilde\lam$, 
the solution of these
equations requires more asumptions before interpreting the UV asymptotic behavior of the model. For instance,
the behavior of 
$\widetilde Z_a$ strongly depends on $Q_{2;n}$ that is yet unknown.

\

\section{One-loop beta-functions of
the model $\times$}
\label{betat}

Just as performed for the model $+$, we will compute the 
1-loop RG flow equations of coupling constants via multiscale analysis
for the model $\times$. 
Because the scheme and proofs are nearly identical, the derivations and explanations are given in a streamlined analysis. 
We stress the following fact: although the notation and expressions 
are similar to previous section, they actually
refer to different quantities.  As the model is different, there is no confusion 
and no need to introduce new notation. 

\subsection{Effective   coupling equations}

The integration of high modes yields a formal effective action of the form \eqref{eq:effectiveW}
where notation keeps its meaning but adapts to the present model. 
Therein, the 2-point  amplitudes expand in local and nonlocal parts, obtaining,
a self-energy of the same form as
given in  \eqref{eq:sigma} in addition with the following term: 
\bea
 \sum_{c}
\vert p_{c}\vert^{4 a} \partial_{\vert p_{c}\vert^{4 a}} \Sigma \big\vert_{\{p\}=0} 
\,.
\label{eq:sigmax}
\eea
According to \cite{BenGeloun:2017xbd},  Table \ref{tab:listprim2} dictates  the renormalization analysis of the model. One shows that 
\begin{align}
 \partial_{\vert p_{c}\vert^{2 b}} \Sigma \big \vert_{\{p\}=0} =0& 
  \qquad \text{implies }\;    Z_b = 1 \quad  \text{wave function renormalization} \crcr
 \Sigma (\{0\})  \sim   \log-{\rm divergent}&   \qquad    \text{(row III) \quad mass renormalization} \crcr
\partial_{\vert p_{c}\vert^{2 a}} \Sigma\vert_{ \{p\}=0 } \equiv \Gamma^{(c)}_{2;a} (\{0\})  \sim   \log-{\rm divergent} &    \qquad    \text{(row I)  \quad}  Z_a \text{\; renormalization} \crcr
\partial_{\vert p_{c}\vert^{4 a}} \Sigma\vert_{\{p\}=0} \equiv \Gamma^{(c)}_{2;2a} (\{0\})  \sim   \log-{\rm divergent}&    \qquad     \text{(row II)  \quad}  Z_{2a} \text{\; renormalization}
\end{align}
where $\vert p_{c}\vert^{2 a}\Gamma^{(c)}_{2;a}(\{p\})$
and $\vert p_{c}\vert^{4 a}\Gamma^{(c)}_{2;2a}(\{p\})$
are the sum of all  amputated 1PI 2-point functions following the patterns of 
${\rm Tr}_{2;c} ({ p}^{2a} \phi^2)$ and ${\rm Tr}_{2;c} ( {p}^{4a} \phi^2)$, respectively, on their boundary graphs. 
Because $4$-point functions converge we do not need to report them. 

We reorganize the effective action as 
\bea
- W^{i-1}(\phi_{\le i-1}, {\bar \phi}_{\le i-1})
&=&
\Sigma_{i-1} (\{0\})
\Tr_2 (\phi^2_{\le i-1})  
+
\sum_c
\Gamma^{(c)}_{2;a, \; i-1} (\{0\})
\Tr_{2;c}({p}^{2 a}\phi^2_{\le i-1})
\crcr
&&+
\sum_c
\Gamma^{(c)}_{2;2a, \; i-1} (\{0\})
\Tr_{2;c}( {p}^{4 a}\phi^2_{\le i-1})
+
{\tilde R}(\phi_{\le i-1})\,,
\eea
where ${\tilde R}(\phi_{\le i-1})$ contains all finite contributions. 

Following step by step, the previous analysis, we deliver the 
the effective couplings at scale $i-1$: 
\bea
Z_{b,\, i-1} &=& 1\,,
\\
\mu_{{\rm ren}, i-1} &=& \mu_{i-1} - 
\Sigma_{i-1} (\{0\})
\,,
\label{eq:renormmassxi}
\\
Z_{a, i-1} 
&=&
-\Gamma^{(c)}_{2;a, i-1}(\{0\})\,,
\\
Z_{2a, i-1} 
&=&
-\Gamma^{(c)}_{2;2a, i-1}(\{0\})\,,
\label{eq:rgeqnsx}
\eea

Following Proposition {\ref{prop:listx}}, we work with the set of parameters
$d=3$, $D=1$, $a = \frac{1}{2}$, and $b= 1$ so that the model is just-renormalizable.

\subsection{Self energy and mass  RG equation}
\label{sect:}

 For the model $\times$, we calculate the self energy which we denote again by 
$\Sigma_{b}(\{ p \})$ and which has an expression similar to \eqref{selfnrg}, 
where the graphs ${\cal G}^{(c)}_{2,\iota}$ should be chosen 
among all  amputated 1PI 2-point graphs 
with  boundary  of the form ${\rm Tr}_{2; c} (p^{2 b} \phi^2)$. 

Up to the first order in perturbation theory, 
${\cal G}^{(c)}_{2,\iota} \in \{ m^{(1)}, m^{(2)}, m^{(3)}\}$. 
(At second order in perturbation theory, the interested reader 
may look at Fig. \ref{fig:mmee} in Appendix \ref{app:2ndOx}.)

Recall that $\Sigma_{b}(\{p \})$ corresponds to
$  \Sigma (\{0\}) 
+  \sum_{c}
\vert p_{c}\vert^{2 b} \partial_{\vert p_{c}\vert^{2 b}} \Sigma \big\vert_{\{p\}=0}$. 
Considering  that the second term vanishes, i.e. $\partial_{\vert p_{c}\vert^{2 b}} \Sigma \big\vert_{\{p\}=0}=0$, only the contribution $\Sigma (\{0\}) $,
 namely the mass renormalization, survives.

For $m^{(c)}$,  $c=1,2,3$, 
$\omega_{d; \times}({m^{(c)}}) = 0$.
This graph belongs to the class III in Table \ref{tab:listprim2}, also, it appears at first order. 
\begin{figure}[H]
\centering
     \begin{minipage}[t]{0.7\textwidth}
      \centering
\includegraphics[angle=0, width=4cm, height=2.5cm]{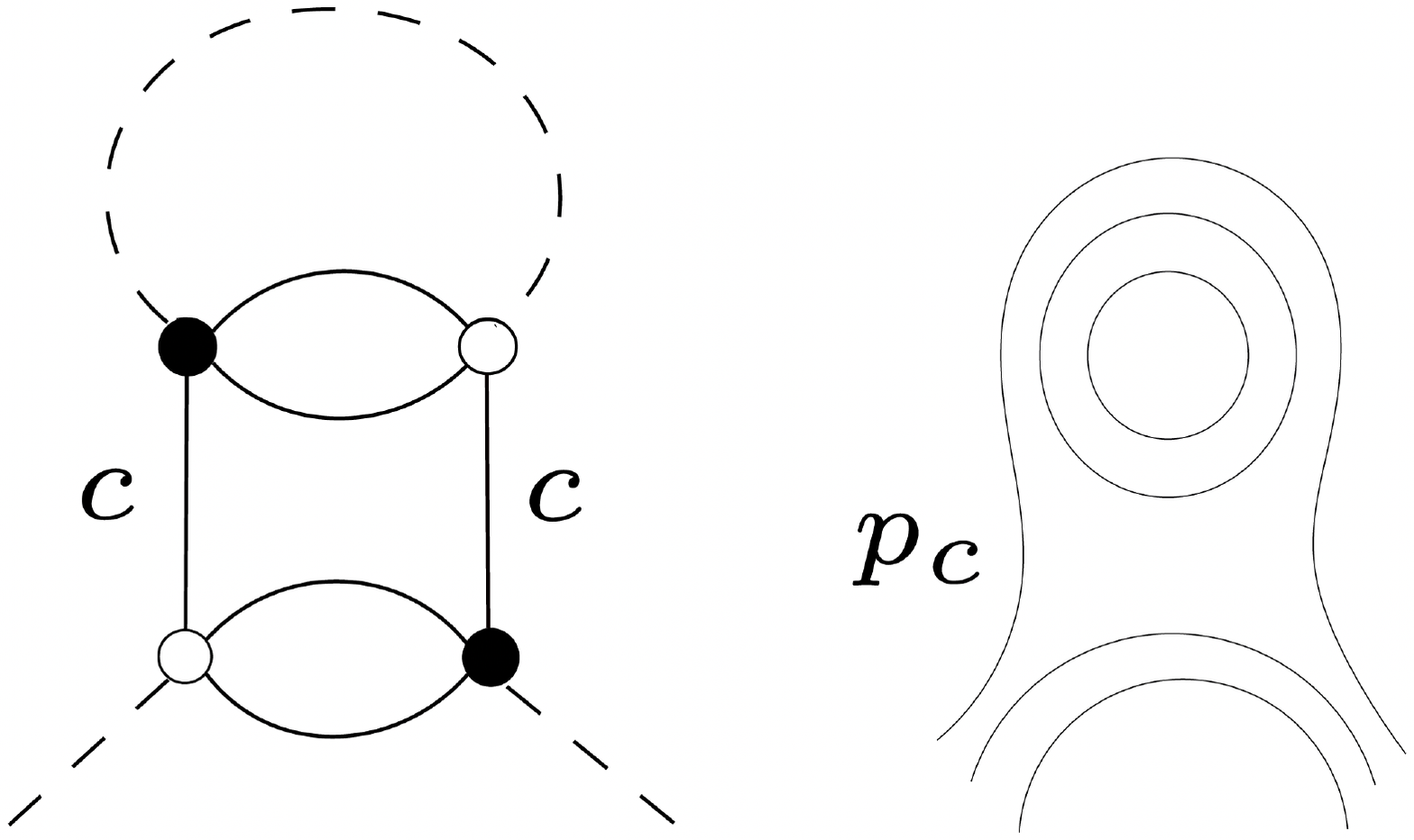}
\caption{{\small {The graph $m^{(c)}$. In the case $d=3$.  
${|{\bf q}_{\check c}|}^{2 b} = |q_{1}|^{2 b} + |q_{2}|^{ 2b}$. Here the parameters should follow that of 
Proposition {\ref{prop:listx}}.
}}}
\label{fig:m1}
\end{minipage}
\end{figure}
We sum the contributions of the graphs $m^{(c)}$ 
and write: 
\bea
\Sigma_{b}(\{p \}) = 
\sum_{c=1}^{d}
K_{m^{(c)}} S_{m^{(c)}} ( \{p\})
&=&
\sum_{c=1}^{d}
2 \big(- \frac{\lambda^{(c)}}{2} \big) \sum_{\{q_{\check c}\}} 
\frac{1}{({|{\bf q}_{\check c}|}^{2 b} + p^{2 b}_{c} + \mu )}
\,,
\eea
where $K_{m^{(c)}} = 2$.

The renormalized mass  \eqref{eq:renormmassxi} finds the expression: 
\beq
{\mu_{\rm ren}} 
=  \mu - \Sigma_b(\{p\})\Big|_{ \{p\}=\{0\} } 
=  \mu + \sum_{c=1}^{d} \lambda^{(c)}  \sum_{\{q_{\check c}\}} 
\frac{1}{({|{\bf q}_{\check c}|}^{2 b}  + \mu )}
\,.
\eeq
Now assert color independence, namely $\lambda^{(c)} = \lambda$, and ${\lambda_{\times}}^{(c)} = \lambda_{\times}$, 
the renormalized mass can be expressed as 
\bea
\mu_{\rm ren}
&=&
\mu +
d  \lambda \sum_{\{q_{\check c}\}} 
\frac{1}{({|{\bf q}_{\check c}|}^{2 b}  + \mu )}
=
\mu + d \, \lambda \, S_{1}
\,,
\eea
where $S_{1}$ is defined earlier  \eqref{eq:S1S2}.

\noindent
{\bf On scaling dimensions.}
 In the same vein as explained before, the scaling dimensions are obtained using this time \eqref{omtcG}. We have, 
 fixing $b=1,a=1/2,D=1,d=3$, 
\bea
&&
\{\lam_\times\} = -(D(d-1)-4b +4a) = 0 \; ,\qquad 
\{\lam\} = -( D(d-1)-4b) = 4a = 2
\crcr
&&
\{\mu\} = 2b= 2 \;, \qquad  \{Z_a\} = 2(b-a)= 1 \; , 
\qquad  \{Z_{2a}\} = 2(b-2a) = 0
\eea

The couplings $\lam$ and $\lam_\times$ have only finite corrections. We are led to 
the equations and easily reached solutions
\bea
&&
\partial_t \widetilde \lam = 
-2\widetilde \lam \;,  \qquad 
\widetilde \lam(t) = c_1 e^{-2t} \crcr
&&
\partial_t  \lam_\times  = 0
\;, 
\qquad 
 \lam_\times(t) = c_2\,, 
\eea 
with $c_1,$ and $c_2$  integration constants. 
Fixing an initial condition
at $e^{t_0} \ll e^{t}$, we get
\bea
\widetilde \lam(t) = 
\widetilde \lam(t_0) e^{-2(t-t_0)} \,. 
\label{lamsimp}
\eea
As expected, $\widetilde\lam$ is suppressed in the UV.

Let us address the mass equation. 
We perform a similar analysis as done in section \ref{sect:selfenergmass}, 
use \eqref{eq:runningmass2x} for $\widetilde S_{1,i}$ and  \eqref{eq:S1S2dimfuldimless}
to obtain 
 the running of the mass at leading order,
\bea
\mu_{i-1} = \mu_{i} + d \, \widetilde S_{1,i} \, \lambda_i 
\,.
\eea
For the mass  has  scaling dimension $\{\mu\}=2$, and
 that $\lambda$ does not run  in this model (see \eqref{eq:rgeqnsx}), then 
we write the RG flow equation for $\mu$: 
\bea
- (\mu_{i-1}- \mu_i)
&=&
\frac{\partial \mu_i}{\partial i}
= - d \, \widetilde S_{1,i} \, \lambda_i 
\,,
\crcr
\frac{ \partial\mu_i}{\partial ((\log M) i)}
&=&
\partial_t   \mu (t) 
= 
- \beta_{\mu, 1} \, \lambda
\,,
\crcr
\beta_{\mu, 1} 
&=&
\frac{d}{\log M}  \, \widetilde S_{1,i} = 2d \pi >0
\,.
\label{eq:massRGfirstorderx}
\eea

We obtain in terms of dimensionless 
quantities: 
\bea
\partial_t  \widetilde \mu (t) 
= 
- 2 \widetilde \mu (t) 
- \beta_{\mu, 1} \,  \widetilde \lambda
\,.
\eea
Inserting the solution \eqref{lamsimp}, we solve this equation and obtain: 
\bea
\partial_t  \widetilde \mu (t) 
&=& 
- 2 \widetilde \mu (t) 
- \beta_{\mu, 1} \, c_1e^{-2t}  
\,,
\crcr
\widetilde \mu (t) 
&=&
( -  \beta_{\mu, 1} c_1 t
 + c_3 ) e^{-2t}
 \,.
\eea
Fixing an initial condition  at $t_0$, $c_3 =  \widetilde \mu (t_0)e^{2t_0} + \beta_{\mu, 1} c_1 t_0$ we finally get 
\bea
 \widetilde \mu (t)
= 
\Big( 
\widetilde \mu (t_0)  
-\beta_{\mu, 1}\widetilde\lam(t_0)(t-t_0) \Big) e^{-2(t-t_0)}
\,.
\label{eq:massfirstorderx}
\eea
Therefore, as expected from a relevant coupling, the mass decays exponentially fast up to a constant value in the UV.

\subsection{Computing $\Gamma_{2;a}$ and  $Z_a$  RG equation}

We address here the flow of the 2-point coupling $Z_a$. 
The 2-point diagram sum $\Gamma^{(c)}_{2;a} (\{p\})$ follows again
an equation similar to \eqref{Gamma2Za} of the previous section \ref{compGamma2+}. 
The sum performs over all amputated 1PI 2-point graphs at 1-loop 
of the form ${\rm Tr}_{2;c} ({ p}^{2 a} \phi^2)$.

\begin{figure}[H]
\centering
     \begin{minipage}[t]{0.7\textwidth}
      \centering
\includegraphics[angle=0, width=4.8cm, height=2.5cm]{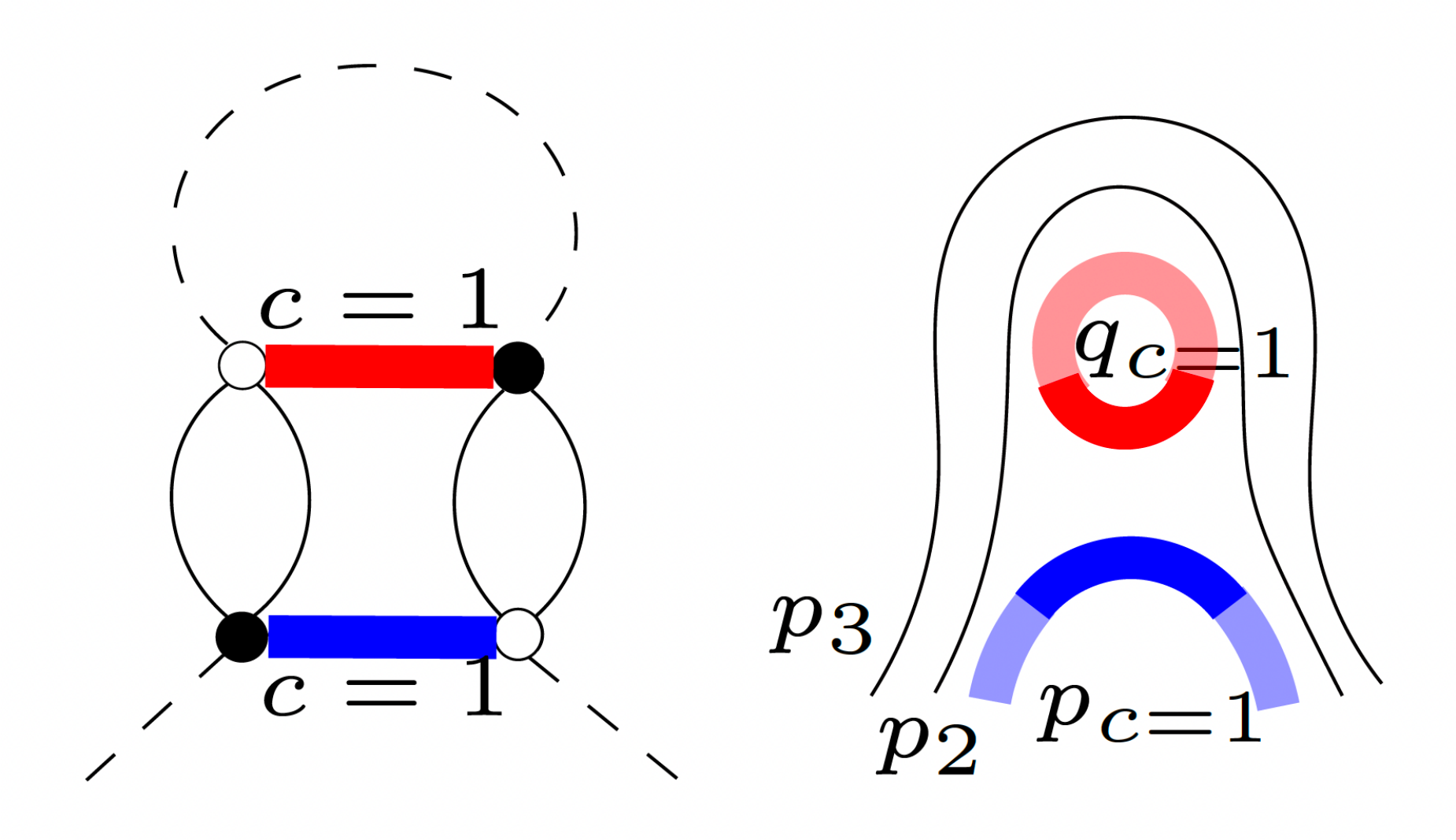}
\caption{ {\small  For $d=3$, $n_{e}^{(c=1)}$ is shown above in colored (left) and stranded (right) representations. $\{p\} = \{p_{1}, p_{2}, p_{3}\}$ and ${|{\bf p}_{\check c}|}^{2 b}={|p_{}|}^{2 b}+{|p_{2}|}^{2 b}$.
}} 
\label{fig:nee}
\end{minipage}
\end{figure}

At first order, the diagrams $n_e^{(c)}$, $c=1,2,3$, see Fig. \ref{fig:nee}, contribute to the flow. 
(The next order in perturbation theory will have the additional graphs 
of Appendix \ref{app:2ndOx}.) 
In  \eqref{eq:sigma}, we have defined $\partial_{\vert p_{c}\vert^{2 a}} \Sigma\vert_{\{p\}=0} \equiv \Gamma^{(c)}_{2;a} (\{0\})$ which is divergent.

The graphs $n_{e}^{(c)}$ satisfy  $\omega_{d;\times}(n_{e}^{(c)}) = 0$
and   belong  to the class I in Table \ref{tab:listprim2}. 

The Feynman amplitude associated with the graph $n_{e}^{(c)}$ shown in Fig. {\ref{fig:nee}} is
\bea
K_{n_{e}^{(c)}} S_{n_{e}^{(c)}} ( \{p\})
&=&
2 
\Big(- \frac{{\lambda_{\times}}^{(c)}}{2} \Big)
\vert p_{c}\vert^{2 a}
\sum_{q_{c}}
\frac{\vert q_{c}\vert^{2 a}}{(\vert q_{c}\vert ^{2 b} + {|{\bf p}_{\check c}|}^{2 b} + \mu)}\,,
\eea
where  $K_{n_{e}^{(c)}} = 2$.

Let us now compute the renormalization of $Z_a$, up to first order in perturbation theory 
\bea
 \vert p_{c}\vert^{2 a} 
\Gamma_{2;a}(\{p\})
&=&
- \vert p_{c}\vert^{2 a}  Z_{a}^{(c)} +
K_{n_{e}^{(c)}} S_{n_{e}^{(c)}} ( \{p\})
\nonumber
\\
&=&
- \vert p_{c}\vert^{2 a}  Z_{a}^{(c)} 
- \lambda_{\times}^{(c)}
\vert p_{c}\vert^{2 a}
\sum_{q_{c}}
\frac{\vert q_{c}\vert^{2 a}}{(\vert q_{c}\vert ^{2 b} + {|{\bf p}_{\check c}|}^{2 b} + \mu)}
\,.
\eea
Therefore, to the first order in perturbation, 
\bea
Z_{a, {\rm ren}}^{(c)} 
&=& 
- \Gamma_{2;a}^{(c)} (\{p\}) \big \vert_{\{p\}=0}
=
Z_{ a}^{(c)} 
+ \lambda_{\times}^{(c)}
\sum_{q }
\frac{\vert q \vert^{2 a}}{\vert q \vert ^{2 b} 
+ \mu}
\,.
\eea
By imposing the color independence,  we write
\bea
Z_{a, {\rm ren}}
=
Z_{ a}
+ \lambda_{\times}\,
S_2
\,,
\eea
where $S_2$ is given in \eqref{eq:S1S2}.
In multiscale analysis, we write
\bea
Z_{a, i-1}
=
Z_{ a, i}
+ \lambda_{\times, i}\,
S_{2, i}
\,,
\eea
where $S_{2, i}$ is given in \eqref{s2idebut}. 

We approximate $S_{2, i}$ in \eqref{eq:S2iexacttimes}. 
Now, we use the fact that  $Z_a$  has  scaling dimension $\{Z_a\}=1$, 
and that $\lambda_{\times}$  does not run, 
and express the $\beta$-function of the coupling $Z_a$ as: 
\bea
- (Z_{a, i-1}- Z_{a, i})
&=&
\frac{\partial Z_{a, i}}{\partial i}
= - \widetilde S_{2,i} \, \lambda_{\times,i} 
\,,
\crcr
\frac{\partial  Z_{a, i}}{\partial ((\log M) i)}
&=&
\partial_t Z_a(t)  = 
- \beta_{Z_a} \, \lambda_\times
\,,
\crcr
\beta_{Z_a} 
&=&
\frac{\widetilde S_{2,i}}{\log M} = 2 >0
\,,
\label{eq:ZaRGfirstorderx}
\eea 
where $\widetilde S_{2,i}$ is given in \eqref{eq:S2iexacttimes}.
Given the scaling dimension 
$\{Z_a\} = 1$, and after fixing an initial condition at some $t_0$, 
the dimensionless solution of the above equation
is straightforward: 
 \bea
\widetilde Z_{a}(t) 
&=& c_0 e^{-t}( -\beta_{Z_a} \, \lambda_\times t + c_4  )  \crcr
&=& 
 -  c_0 \beta_{Z_a} \, \lambda_\times  (t-t_0) e^{-t}
 +  \widetilde Z_{a}(t_0)e^{-(t-t_0)}
\,.
\label{eq:Zafirstorderx}
\eea 
In the UV, $t\to \infty$, the 2-point coupling $Z_a$ in the model $\times$
flows to 0.

\subsection{Computing of $\Gamma_{2;2a}$ and 
$Z_{2a}$  RG equation}

We focus on the RG flow for $Z_{2a}$. 
We denote by $\vert p_{c}\vert^{4 a}  \Gamma_{2;2 a}^{(c)}(\{p \}) $
the sum is over all amputated 1PI 2-point graphs at 1-loop 
whose boundaries  follow the pattern  of ${\rm Tr}_{2;c} ({ p}^{4 a} \phi^2)$.

The diagrams that will contribute at 1-loop are denoted $m_{e e}^{(c)}$
and  depicted in Fig.  \ref{fig:mee}. 
(Higher order diagrams are listed in Appendix \ref{app:2ndOx})
For $m_{e e}^{(c)}$, $\omega_{d; \times}(m_{e e}^{(c)}) = 0$ 
and belongs to the class II in Table \ref{tab:listprim2}. 
\begin{figure}[H]
\centering
     \begin{minipage}[t]{0.7\textwidth}
      \centering
\includegraphics[angle=0, width=5cm, height=2.5cm]{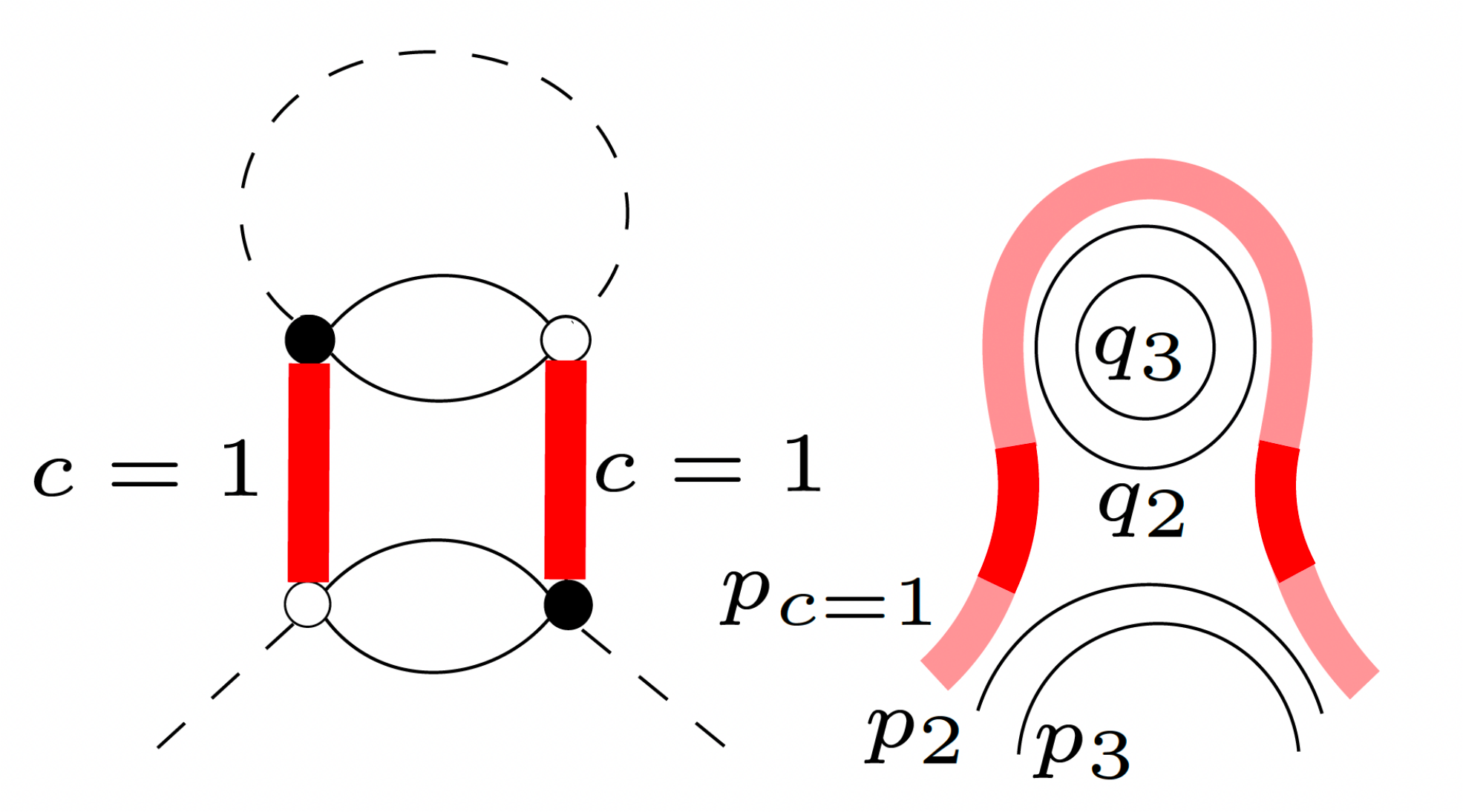}
\caption{{\small  $m_{e e}^{(c=1)}$ for $d=3$ is illustrated in colored (left) and stranded (right) representations. $\{q_{\check c}\} = \{q_{2}, \; q_{3}\}$, ${|{\bf q}_{\check c}|}^{2 b} = {|q_{2}|}^{2 b} + {|q_{3}|}^{2 b}$.}}
\label{fig:mee}
\end{minipage}
\end{figure}
\bea
K_{m_{ee}^{(c)}} S_{m_{e e}^{(c)}} (\{p\}) 
&=&
2
\Big( - \frac{{\lambda_{\times}}^{(c)}}{2}\Big) 
\vert p_{c}\vert^{4 a}
\sum_{\{q_{\check c}\}} \frac{1}{( {|{\bf q}_{\check c}|}^{2 b} + \vert p_{c}\vert ^{2 b} + \mu )}
\,,
\eea
where $K_{m_{e e}^{(c)}} = 2$.
Given this,  we compute the renormalization of $Z_{2 a}$: 
where
\bea
\vert p_{c}\vert^{4 a}
\Gamma_{2; 2 a}^{(c)}\big (\{p\}) 
&=&
- \vert p_{c}\vert^{4 a}   Z_{2 a}^{(c)} + 
K_{m_{ee}^{(c)}} S_{m_{e e}^{(c)}} (\{p\}) 
\nonumber
\\
&=&
-\vert p_{c}\vert^{4 a} Z_{2 a}^{(c)} 
- {\lambda_{\times}}^{(c)}
\vert p_{c}\vert^{4 a}
\sum_{\{q_{\check c}\}} \frac{1}{( {|{\bf q}_{\check c}|}^{2 b} + \vert p_{c}\vert ^{2 b} + \mu )}
\,.
\eea
After a small manipulation,  we obtain
\bea
Z_{2 a, {\rm ren}}^{(c)} 
&=&
Z_{2 a}^{(c)} 
+ {\lambda_{\times}}^{(c)}
\sum_{\{q_{\check c}\}} \frac{1}{( {|{\bf q}_{\check c}|}^{2 b} 
+ \mu )}
\,.
\eea
The equation for $Z_{ 2 a}$ is reached after requiring color 
independence: 
\bea
Z_{2 a, {\rm ren}}
=
Z_{ 2 a}
+ \lambda_{\times}\,
S_1
\,,
\eea
where we recognize  $S_1$  \eqref{eq:S1S2}.
Passing through the same steps via multiscale
regularization, the coupling equation can be written  as: 
\bea
Z_{2 a, i-1}
&=&
Z_{2 a, i}
+ \lambda_{\times, i}\,
S_{1, i}
\,, \crcr
- (Z_{2a, i-1}- Z_{2a, i})
&=&
\frac{\partial Z_{2a, i}}{\partial i}
= - \, \widetilde S_{1,i} \, \lambda_{\times, i} 
\,,
\crcr
\frac{\partial  Z_{2 a, i}}{\partial ((\log M) i)}
&=&
\partial _t Z_{2 a}(t) 
= 
- \beta_{Z_{2a}} \, \lambda_\times
\,,
\crcr
\beta_{Z_{2a}} 
&=&
\frac{\widetilde S_{1,i}}{\log M} = 2 \pi >0
\,,
\label{eq:Z2aRGfirstorderx}
\eea
where $\widetilde S_{1,i}$ is computed in \eqref{eq:runningmass2x}. 
Then, because both couplings
$Z_{2a}$ and $\lam_\times$ are dimensionless, at this order of perturbation, 
$Z_{2a}$ yields also a linear function in the time scale $t$.  
\bea
Z_{2 a}(t) 
&=& - (t-t_0) \, \beta_{Z_{2a}} \, \lambda_\times + Z_{a}(t_0)
\,.
\label{eq:Z2afirstorderx}
\eea
The 2-point coupling $Z_{2a}$ in the model $\times$ grows linearly in $t$ in its magnitude.
This behavior prevents one to conclude that the model is asymptotically safe.  As commented before, the model is neither asymptotically 
free nor asymptotically safe which makes it interesting to understand better.

\

\noindent{\bf Summary --} 
We  give a summary of the 1-loop RG flow equations for the
model $\times$ and their solutions, 
up to integration constants: 
\begin{table}[H]
\centering
\begin{tabular}{l|l}
\hline 
\hline 
&\\
$\partial_t \widetilde\lambda  =-2\widetilde\lambda    $  &   $\widetilde\lambda(t)  = c_1 e^{-2t}$ \\ 
%%%
$\partial_t\,  \lambda_{\times}  = 0 $  &  $\lambda_{\times} (t)=c_2$\\
$\partial_t \, \widetilde\mu =-2\widetilde\mu - \beta_{\mu,1}  \widetilde\lambda  $\,,
\qquad   
$\beta_{\mu,1}= 2d\pi >0$
 & $ \widetilde\mu (t)= 
 (- \beta_{\mu,1} c_1 t + c_3)e^{-2t}$ \\ 
%%% 
$\partial_t \, \widetilde  Z_a
 =  - \widetilde  Z_a 
-  c_0e^{-t}\beta_{Z_a} \, \lambda_{\times} $\,,
\qquad 
  $\beta_{Z_a} = 2 > 0$   &
$\widetilde Z_{a}(t) 
= c_0( -\beta_{Z_a} \, \lambda_\times t + c_4  )  e^{-t}$\\ 
 %%% 
 $\partial_t \,  Z_{2 a}
 = 
- \beta_{Z_{2a}} \, \lambda_\times $ 
\,,
\qquad 
  $\beta_{Z_{2 a}} = 2\pi > 0$   &
$Z_{2 a}(t) 
= - \beta_{Z_{2a}} \, \lambda_\times t+ c_5$\\ 
 \hline 
 \hline 
\end{tabular}
\caption{Summary of the RG flow equations for the couplings in the model $\times$.}
\label{tab:RGx}
\end{table}

\subsection{Integration at arbitrary loops}
\label{sec:allloopsx}

The power counting of the model $\times$ is provided by Proposition \ref{prop:listx}. 
This gives us more information on the $\beta$-functions of the couplings at arbitrary order of perturbation theory. Indeed, the information is stringent enough to let us present explicit general forms of $\beta$-functions of the couplings.

\

\noindent{\bf 4-point couplings  $\lambda$ and $\lambda_\times$  RG equation --}
The power counting theorem of the model $\times$ with Proposition \ref{prop:listx} determines that at all orders in perturbation theory, there are no amplitudes which are divergent contributing to the renormalization of 4-point couplings $\lambda$ and $\lambda_\times$. Hence, $\lambda$ and $\lambda_\times$ of the model $\times$ are constant
and therefore 
\bea
\partial_t \lambda = 0 \;, 
\qquad \;  
\partial_t \lambda_{\times} = 0
\,, 
\eea
which trivially yield, for the dimensionless coupling:
\bea
\widetilde\lambda (t) = c_1 e^{-2t}
\;, 
\qquad \;  
\lambda_{\times}(t) = c_2
\,.
\label{eq:4ptallordersx}
\eea

\

\noindent{\bf Mass,  2-point couplings $Z_a$ and $Z_{2a}$ RG equations --}
Observation of Proposition \ref{prop:listx} tells us that the mass renormalization is decided by the class III, where only exactly one $\lambda$ and a number of $\lambda_\times$ contribute.
The $Z_a$ renormalization is decided by the class I, where only $\lambda_\times$ contributes.
We also notice that only $\lambda_\times$ contributes to the renormalization of $Z_{2a}$, as class II dictates.

With \eqref{eq:4ptallordersx}, one immediately conclude that the coupled differential RG equations for the couplings of the model $\times$, which generalize trivially the equations for the first order RG equations \eqref{eq:massRGfirstorderx}, \eqref{eq:ZaRGfirstorderx}, and \eqref{eq:Z2aRGfirstorderx}, to arbitrary $n$-th orders, by introducing polynomials $P_n (\lambda_\times)$, $Q_n ( \lambda_\times)$, and $R_n ( \lambda_\times)$,  are
\bea
&& 
\partial _t \mu  
= 
\lambda \, 
P_n( \lambda_\times) \,, 
\qquad 
\partial _t \widetilde \mu  
= -2\widetilde \mu  
+ 
\widetilde \lambda \, 
P_n( \lambda_\times) 
\,,
\crcr
&&
\partial _t Z_{a}
= 
Q_n( \lambda_\times)
\,,
\qquad 
\partial _t\widetilde  Z_{a}
= - \widetilde  Z_{a} + 
 e^{-t}Q_n( \lambda_\times)
 \,,
\crcr
&&
\partial _t Z_{2 a}
= 
R_n( \lambda_\times)
\,,
\eea
which can be integrated easily. 
Thus each coupling will keep 
its behavior at arbitrary order
of perturbation.

\section{Conclusion}
\label{conc}

We have explicitly  computed 
the one-loop $\beta$-functions of the couplings of two eTFTs, the model $+$ and the model $\times$, at first order of perturbation theory. 
The system of RG flow equations can be explicitly solved. 
Both models $+$ and $\times$ do have a constant wave function renormalization ($Z_b=1$). Nevertheless, we have obtained some nontrivial RG flows of the couplings. 
Table \ref{tab:RG+} and Table \ref{tab:RGx}
 summarize the upshot of this analysis.

For the model $+$, the enhanced $4$-point coupling $\lambda_+$
is marginal but without corrections;  therefore it is a fixed point $\lam_+ = \theta$. Meanwhile, the ordinary $4$-point coupling $\lambda$ is
relevant and therefore exponentially suppressed at large 
time $t= \log k/k_0$, where $k$ is momentum scale.
This statement is true for all orders in perturbation theory. 
The 2-point coupling $Z_a$ and 
mass coupling 
have  exponential behavior in $t$:  
they are suppressed in the UV and reach a constant.  
This is common for relevant operators. 
As a result, the current  eTFT model
$+$ behaves like an asymptotically 
safe model: one marginal direction $\lam_+ = \theta$ is kept fixed 
and there are three relevant operators $(\lam, \mu, Z_a)$  with  dimensionless counterparts
 $(\widetilde \lam, \widetilde \mu, \widetilde Z_a)$
flowing to $(0,0,c\,  \theta)$. This a one-dimensional line
of fixed points 
that makes such a QFT interesting and special.

Concerning the model $\times$, 
 the $4$-point coupling
$\lambda_\times$ is marginal with no corrections: it becomes constant at all orders of perturbation and a fixed point, $\lam_\times = \theta$. The second $4$-point coupling
$\lambda$ is relevant but it is without corrections: it is exponentially suppressed towards the UV.  
On the other hand, the 2-point couplings, the mass and $Z_a$, are relevant and UV suppressed (they flow to 0). 
The last coupling $Z_{2a}$ is marginal and grows
 in its magnitude linearly in the scale $t$ with coefficient depending on $\lam_{\times}=\theta$. We obtain the UV-behavior the 
 dimensionless couplings 
 $(\widetilde \lam, \widetilde \mu, \widetilde Z_a, Z_{2a})$ leads us 
 to $(0,0,0, \infty)$. 
 This behavior is not common of ordinary QFT and TFT as it cannot be   associated neither with asymptotic freedom 
 nor with asymptotic safety. 
 Finally, comparing the RG flow equations between the conventional models and these eTFT, $+$ or $\times$ models, shows drastic
 differences. In the
 present context, they are simple enough to exhibit explicit solutions.

With the explicit forms of the running of the couplings at first order, together with the close observations of the power counting theorems for each model, $+$ or $\times$, in \cite{BenGeloun:2017xbd}, we have deduced the generic form of the 
coupling $\beta$-functions at all orders of perturbation. 
The determination of the polynomial coefficients at all-loops will be left for future investigations.  
We conjecture that the systems could be explicitly integrated at arbitrary and given order. 
More generally, the  amplitudes might be simple enough 
to be re-summed at all orders. 
Proving this property will require more work.

Concerning the quantum gravity side, the  model $+$ can be called  
asymptotically safe at first order of perturbation for generic $ \lambda_+$ and 
makes the UV completion of the quartic eTFTs likely. However, much less is known about its IR behavior.
  The computation of higher order perturbations of the models 
$+$ and $\times$  may reveal IR fixed points. Flowing backwards in the IR direction, the system of $\beta$-function collects polynomial coefficients in the coupling $\lambda_+$. The root of these polynomials may lead to vanishing $\beta$-functions and hence may produce non trivial IR fixed points. The IR study of eTFTs will require the use of different tools (like the Functional Renormalization Group Equation) and even different covariance with 
$p^{2a}$ weight. This could be also a following-up study from our work.

\section{Acknowledgements}
We thank anonymous referees for their reading
and remarks that has led to corrections
and radical improvement of our original work. 
We would like to thank Dario Benedetti, Sylvain Carrozza, Riccardo Martini, and Fabien Vignes-Tourneret for the insightful discussions.
The authors would also like to thank the thematic program ``Quantum Gravity, Random Geometry, and Holography'' 9 January - 17 February 2023 at Institut Henri Poincar\'e, Paris, France
for the platform for discussions and the collaboration and letting us progress further on this project.
The authors acknowledge support of the Institut Henri Poincar\'e (UAR 839 CNRS-Sorbonne Universit\'e) and LabEx CARMIN (ANR-10-LABX-59-01).

\section*{ Appendix}
\label{app}

\appendix

\renewcommand{\theequation}{\Alph{section}.\arabic{equation}}
\setcounter{equation}{0}

\section{Euler-Maclaurin formula 
and Feynman amplitude approximations}
\label{app:SumtoInt}

We briefly review 
the approximation of 
a discrete spectral sum by an integral
using Euler-Maclaurin formula \cite{BenGeloun:2013vwi}. In perturbation theory, this approximation is well controlled and sufficient to achieve our calculations. 

Consider $h_n(x) = x^n e^{-A x^a}$, with $x \ge 0$, $n \in \N$, and the sum $\sum_{p=0}^{\infty} h_n(p)$.
We use Euler-Maclaurin formula to obtain for a finite integer $q \le 1$,
\beq
\sum_{p=1}^q h_n(p) = \int_1^q h_n(p) dp + R(q)
\,, 
\eeq
where
\beq
R(q) = - B_1\big(h_n(1) + h_n(q) \big)
+
\sum_{k=1}^\infty 
\frac{B_{2k}}{2 k!} 
\big(
h_n^{(2 k -1)} (q)
-
h_n^{(2k-)} (1)
\big)\,,
\eeq
where $B_k$ are Bernoulli numbers and $h^{(2k-1)}$ denotes $2k-1^{\rm th} $ derivative of $h_n(p)$ with respect to $p$.

One can show that 
\beq
{\rm lim}_{q \rightarrow \infty} R(q) 
=
- B_1 
- \sum_{k=1}^{\infty} \frac{B_{2 k}}{2 k}
\binom{n}{2k-1} 
+ \cO(A)
=
\cO(1) 
+ \cO(A)
\,.
\eeq

For any $A$, the integral below is exact,
\bea
{\rm lim}_{q \rightarrow \infty}\int_1^q h_n(p) dp 
=
\frac{1}{a}
A^{- \frac{1+n}{a}}
\Gamma\Big[ \frac{1 + n}{a}, A \Big] 
=
\frac{1}{a}
A^{- \frac{1+n}{a}}
\Gamma\Big[ \frac{1 + n}{a}\Big] 
-
\frac{1}{1+n}
+
\cO(A)
\,,
\eea
with the incomplete Gamma function $\Gamma [ \cdot, \cdot]$, and Euler Gamma function $\Gamma[\cdot]$.
This then gives us
\bea
\sum_{p=1}^{\infty} h_n(p)
=
{\rm lim}_{q \rightarrow \infty}
\sum_{p=1}^q h_n(p)
=
\frac{1}{a}
A^{- \frac{1+n}{a}}
\Gamma\Big[ \frac{1 + n}{a}\Big] 
-
\frac{1}{1+n}
+\cO(1)
+
\cO(A)\,.
\label{sumh}
\eea

\noindent{\bf Expanding $\widetilde{S}_{0,i}$.}  
The goal is to approximate $S_{0,i}$ \eqref{s0idebut}
using the above developments. 
Use Euler-Maclaurin expansion, we compute the sum 
\bea
\tilde \tau 
&=&
\sum_{q\in \Z}\vert q\vert^{4a}
e^{-(\alpha+  \alpha') 
\vert q\vert^{2b} }
= 2\sum_{q= 1}^\infty   q^{4a} 
e^{-(\alpha+  \alpha') \vert q\vert^{2b} }
\crcr 
&& 
= 2\int_{1}^{\infty}
dq\,
 q^{4a}
e^{-(\alpha+  \alpha') 
\vert q\vert^{2b} }
+
R
 = 2 \times 
\frac{1}{2b (\alpha+  \alpha')^{\frac{4a+1}{2b}}}
\Gamma\left(\frac{4a+1}{2b},\alpha+  \alpha'\right)
+R
\crcr
&&
= 
\frac{(\alpha+  \alpha')^{\frac{-(4a+1)}{2b}}}{b}
\Gamma\left(\frac{4a+1}{2b},\alpha+  \alpha'\right)
+R \,,
\label{sumoverq}
\eea
where $\Gamma(s,x)= \int_{x}^\infty t^{s-1} e^{-t} dt$ 
denotes the upper incomplete Gamma function. $R$ is the Euler-Maclaurin remainder which behaves, according to \eqref{sumh}, as $R = {\mathcal O}(1) + {\mathcal O}(\alpha + \alpha')$.  $\Gamma(s,x)$ admits a known asymptotic expansion when $x\to 0^+$ given by 
\bea
\Gamma(s,x)= 
\Gamma(s) - \frac{x^s}{s} + {\mathcal O}(x^{s+1})
\label{eq:incompletegammaexpand0}
\eea
with the obstruction $s\notin \{0,-1,-2,\dots\}$. 
Now, for simplicity, 
we constrain our 
model and use $D=1$, 
but keep $a$ and $b$
accordingly as 
given by Theorem \ref{theorem+}. 
We use the above expansion \eqref{eq:incompletegammaexpand0}
in \eqref{sumoverq} as $ \alpha+  \alpha'$ is a small parameter (recalling that $\alpha$ and $\alpha'$ are small in the UV)
\bea
&&
\sum_{q\in \Z}\vert q\vert^{4a}
e^{-(\alpha+  \alpha') 
\vert q\vert^{2b} } 
\crcr
&&
=
\frac{(\alpha+  \alpha')^{\frac{-(4a+1)}{2b}}}{b}
\left( 
\Gamma\left(\frac{4a+1}{2b}\right) 
- \frac{(\alpha+  \alpha')^{\frac{4a+1}{2b}} }{\frac{4a+1}{2b}} + \cO\big((\alpha+  \alpha')^{\frac{4a+1}{2b}+1}\big) \right)  
+ R
\crcr
&&
= 
\frac{1}{b}
\Gamma\left(\frac{4a+1}{2b}\right) (\alpha+  \alpha')^{\frac{-(4a+1)}{2b}} 
 + {\mathcal O}(1) + {\mathcal O}(\alpha+  \alpha') \,.
\eea
Having a look at \eqref{s0idebut}, 
using  $e^{-(\alpha+  \alpha')\mu} = 1  
+ \cO(\alpha+  \alpha')$,
we integrate this expression above over $\alpha$ and $\alpha'$, 
with $\frac{1}{b}
\Gamma\left(\frac{4a+1}{2b}\right)= \frac1b \Gamma(2) = \frac1b$, and 
write: 
\bea
\widetilde S_{0,i}&=&
\int_{M^{-2 b i}}^{M^{-2 b(i-1)}}
d\alpha 
\int_{M^{-2b i}}^{M^{-2 b(i-1)}}
d\alpha' 
\Big[ 
 \frac{1}{b}(\alpha+  \alpha')^{-2} 
+ 
 {\mathcal O}\big( (\alpha+  \alpha')^{-1}
 \big)
 \Big] 
 \crcr
 &=&
\int_{M^{-2 bi}}^{M^{-2 b(i-1)}}
d\alpha 
\Big[ -\frac{1}{b} (\alpha +\alpha ')^{ -1}
 +
 {\mathcal O}\big( 
\log(\alpha+  \alpha')\big)
\Big]_{\alpha'=M^{-2 bi}}^{\alpha'=M^{-2 b(i-1)}}
 \crcr
 &=&
\frac{1}{b}
 \Big[
 \log(\alpha +M^{-2b i})
  - 
  \log(\alpha +M^{-2b(i-1)})
 \crcr
 &&
+ 
\Big(
  \;
  {\mathcal O}\big( \alpha)
 +
 {\mathcal O}\big( \alpha \log(\alpha+  M^{-2b i})\big)
+ 
  {\mathcal O}\big( M^{-2b i} \log(\alpha+  M^{-2b i})\big)
\Big)
 \Big]_{\alpha = M^{-2b i}}^{\alpha =M^{-2b (i-1)}}
 \crcr
 &=&
 \frac{1}{b}
 \Big[
 -
 \log(2 M^{-2b i})
+ 2
 \log(M^{-2b (i-1)} +M^{-2b i})
  - \log(2 M^{-2b (i-1)} )
 \crcr
 &&
 + \;
  {\mathcal O}\big( M^{-2b i})
 +
 {\mathcal O}\big( M^{-2b i} \log( M^{-2b i})\big)
 \Big]
 \crcr 
&=&
 \frac{1}{b}
 \Big[
2 \log [M^{-2b i}(M^{2b} +1)]
 - \log(2 M^{-2b i})
  - 
  \log(2M^{-2b i+2b })
 \Big]
 \crcr
 &&
 + {\mathcal O}(M^{-2b i})
+  {\mathcal O}\big( M^{-2b i} \log (M^{-2b i})\big)
 \cr\cr 
&=&
 \frac{1}{b}
 \Big[
2 \log (M^{2b} +1)
 - \log(4)
  - 
  \log(M^{2b })
 \Big]
 + {\mathcal O}( M^{-2b i})
+  {\mathcal O}\big(  M^{-2b i} \log (M^{-2b i})\big) \crcr
& = & 
\frac{1}{b}
 \Big[
  \log 
  \frac{(M^{2b} +1)^2}{
  4 M^{2b }}
 \Big]
 + {\mathcal O}( M^{-2b i})
+  {\mathcal O}\big(  M^{-2b i} \log (M^{-2b i})\big)
\crcr
&&
\eea
Thus $\widetilde  S_{0,i}$ is approximated by 
\bea
\widetilde  S_{0,i} =
\frac{1}{b}
 \log 
  \frac{(M^{2b} +1)^2}{
  4 M^{2b }}
 + {\mathcal O}( M^{-2b i}
 \log(M^{-2b i})) 
\eea
As $M\ge 1$, 
$\frac{(M^{2b} +1)^2}{
  (4 M^{2b })}>1$, 
  then the leading coefficient in the above expanding is positive.

\

We can put the above in 
a standard $\log M$ approximation by expanding at large enough
$M\gg 1$: 
\bea
&&
\frac{1}{b}
 \log 
  \frac{(M^{2b} +1)^2}{
  (4 M^{2b })}
 = 
\frac{1}{b}
 \Big[
\log  (M^{2b})  +  2 \log (1+1/ M^{2b} )
 - \log(4)
 \Big] 
 \crcr
 &&
 = \frac{1}{b}\log M^{2b}
 + \frac{2}{b}\left (\frac{1}{M^{2b}}
 - \log(2) \right) 
 + \cO(M^{-4b})
\eea 
At large $M$, the first term will correspond to 
the standard $\log \Lambda $ divergence for marginal coupling.

  \

\noindent{\bf Dimensionful computations for $S_{0,i}$.}
We address now the crucial question of the dimension 
in our computation. The previous calculations
of were performed without taking care of that aspect
that we now restore. 
The couplings and fields have scaling dimensions.
Expressing the propagator in Schwinger parameterization as in  \eqref{eq:Ctilde}, we notice that $\alpha$ should have 
a dimension of $-2b$ in units of momentum scale $k$. In other words, we can write
$\alpha = k^{-2b} \, {\tilde {\alpha}}$
where ${\tilde {\alpha}}$ is dimensionless.
There is a dimensionful quantity 
$\tau$ associated with $\tilde \tau$ \eqref{sumoverq},
in which $q$ acquires dimension of $1$ in the units of the momentum scale. 
The integral approximates the discrete sum  over $q$ as in the second line of \eqref{sumoverq}, but gains one more dimension
in the computation of $\tau$. 
A similar fact concerning discrete sums having scaling dimensions 
was advocated and used in \cite{Benedetti:2014qsa}. 
This also can be understood by the power counting theorem: 
in order to make the couplings marginal, so $\log$-divergent in 
the cut-off, discrete sums must carry scale dimension. 
We perform the following change of variables to let the dimensions be explicit in terms of a momentum scale $k$:
\bea
&&
q  = k \tilde q \, , \qquad  \tilde q \in  \Z\crcr
&& \alpha = k^{-2b} \, {\tilde {\alpha}} 
\label{eq:dimfuldimless}
\eea
We obtain in terms of dimensionless $\tilde \tau$ 
and $\widetilde S_{0,i}$: 
\bea
&&
\tau  =   k^{4 a +1} \tilde \tau 
\crcr 
&&
S_{0,i} =  k^{4 a +1} 
 k^{-4b}
\widetilde S_{0,i}  = \widetilde S_{0,i}
\,,
\eea
where in the last equality, we notice $4 a +1 - 4b = 0$, and therefore
we get $S_{0,i}$ \eqref{s0iapprox}.

\

\noindent{\bf Expanding $\widetilde S_{1,i}$ and  $\widetilde S_{2,i}$ for model $+$.} 
We use a similar technique to  provide an approximation of  $S_{1,i}$ 
 \eqref{s1idebut} and  $S_{2,i}$  \eqref{s2idebut}. 

Express the propagators in Schwinger representation and carefully converting the above sums into integrals and get 
\bea
\widetilde
S_{2,i} 
= 
\int_0^\infty 
d\alpha \chi^{i} (\alpha) 
e^{-\alpha
\mu_i}
\sum_{q \in \Z}\vert q\vert^{2 a}
e^{-\alpha
\vert q\vert^{2 b} }\,.
\label{eq:masslambda+}
\eea
This integral is similar to \eqref{sumoverq}, changing only  $\alpha+\alpha'$ to $\alpha$ 
and for a particular choice of $2a$. 
We perform an approximation in an  analogous
manner as before and get: 
\bea
\sum_{q\in \Z}\vert q\vert^{2a}
e^{-\alpha 
\vert q\vert^{2b} }
= 
\frac{\alpha^{\frac{-(2a+1)}{2b}}}{b}
\Gamma\left(\frac{2a+1}{2b},\alpha\right)
+R 
 = \frac{1}{b}
\Gamma\left(\frac{2a+1}{2b}\right) \alpha^{\frac{-(2a+1)}{2b}} 
 + {\mathcal O}(1) 
\eea
We insert this expression in \eqref{eq:masslambda+} and obtain 
\bea
\widetilde
S_{2,i}
&=&
\int_{M^{-2 bi}}^{M^{-2b(i-1)}}
d\alpha 
e^{-\alpha 
\mu_i}
\Big[ c_{1;a,b} \, \alpha^{\frac{-(2a+1)}{2b}} 
 +{\mathcal O}(1)
 \Big]
\crcr
&=&
\int_{M^{-2bi}}^{M^{-2b(i-1)}}
d\alpha   
(1   + {\mathcal O}(  \alpha) ) 
\Big[ c_{1;a,b} \, \alpha^{\frac{-(2a+1)}{2b}} 
 +{\mathcal O}(1) 
 \Big]
 \label{eq:S2icommon}
\crcr
&=&
\int_{M^{-2bi}}^{M^{-2b(i-1)}}
d\alpha 
\Big[ c_{1;a,b} \, \alpha^{\frac{-(2a+1)}{2b}} 
 + 
 {\mathcal O}\big(  \alpha^{1- \frac{(2a+1)}{2b} }\big)
 \Big] \crcr
&=&
\Big[ 
\frac{1}{ 1 - \frac{(2a+1)}{2b}} \, c_{1;a,b}\, \alpha^{1-\frac{(2a+1)}{2b}} 
+ 
{\mathcal O} \big(  \alpha^{2-\frac{(2a+1)}{2b}}\big)
 \Big]_{M^{-2bi}}^{M^{-2b(i-1)}}
 \,,
\eea
where $c_{1; a,b}= \frac{1}{b}
\Gamma\left(\frac{2a+1}{2b}\right)$.

For the model $+$, we use 
$ a = \frac{1}{2} D (d - 2)$ and
$ b = \frac{1}{2} D (d - \frac{3}{2})$
for just-renormalizability (see Proposition \ref{prop:list+}) with $D = 1$, then,
$\frac{2a+1}{2b} = \frac{2(d-1)}{2d-3}= 1 + \frac{1}{2d-3}$.  
Thus 
\bea
\widetilde
S_{2,i}
&=&
\Big[ 
- \frac{(2d-3)}{(2d-3)/4} \, \Gamma\left(\frac{2(d-1)}{2d-3}\right) \alpha^{-1/(2d-3)} 
 + 
{\mathcal O}\big(  \alpha ^{1- 1/(2d-3)} \big) 
 \Big]_{M^{-2 bi}}^{M^{-2b(i-1)}} 
 \crcr
 &=&
 4 \, \Gamma\Big(\frac{2(d-1)}{2d-3}\Big)
 (M^{-2bi(-1/4b) }
 - M^{-2b(i-1)(-1/4b) } )
 + 
  {\mathcal O}\big(  M^{-i (d -2)}  \big) 
 \crcr
 &=&
 4 \, \Gamma\Big(\frac{2(d-1)}{2d-3}\Big)
 M^{i/2}(
1 - M^{-1/2} )
 + 
 {\mathcal O}\big(  M^{-i (d -2)}  \big) 
\,.
 \label{eq:runningmass1}
\eea

Now we compute $\widetilde S_{1,i}$
 namely,
\bea
\widetilde
S_{1,i}
 = 
\sum_{{\bf q} \in \Z^{d-1}}
\int_{0}^{\infty}
d\alpha \; \chi^i(\alpha)\, 
e^{ - \alpha( |{\bf q}|^{2 b}  + \mu_i )}
 = 
\int_0^\infty 
d\alpha \chi^{i} (\alpha) 
e^{-\alpha \mu_i}
\left(\sum_{q\in \Z}
e^{-\alpha
\vert q\vert^{2b} }
\right)^{d-1}\,.
\label{eq:masslambdaintegral}
\eea
Let us focus on 
a single sum over $q$: 
\bea
&&
\sum_{q\in \Z}
e^{-\alpha 
\vert q\vert^{2b}}
= 2\sum_{q= 1}^\infty
e^{-\alpha q^{2b}}
+
1
= 
2
\int_{1}^{\infty} dq
e^{-\alpha q^{2b} }
+
R
+
1
\crcr
&&
 = 2 
\Bigg(\frac{1}{2b} \alpha^{-\frac{1}{2b}}
\Gamma\left(\frac{1}{2b},\alpha \right)\Bigg)
+R+1\,,
\label{eq:sumtoint}
\eea
where $R = {\mathcal O}(1) + {\mathcal O}(\alpha)$. 
We use \eqref{eq:incompletegammaexpand0}
in \eqref{eq:sumtoint} as $ \alpha$ is a small parameter in the UV, 
\beq
\sum_{q\in \Z}
e^{-\alpha 
\vert q\vert^{2b}}
=
\frac{1}{b} \alpha^{-\frac{1}{2b}}
\Big[
\Gamma\left(\frac{1}{2b}\right)
- 2b \alpha^{\frac{1}{2b}}
+ \cO(\alpha^{\frac{1}{2b}+1})
\Big] +R+1
 = 
\frac{1}{b} 
\Gamma\left(\frac{1}{2b}\right)
\alpha^{-\frac{1}{2b}} +{\mathcal O}(1) 
\,.
\eeq
We insert this expression in  \eqref{eq:masslambdaintegral},
\bea
&&
\widetilde
S_{1,i} = 
\int_0^\infty 
d\alpha \chi^{i} (\alpha) 
e^{-\alpha \mu_i}
\left(\frac{1}{b} 
\Gamma\left(\frac{1}{2b}\right)
\alpha^{-\frac{1}{2b}} +{\mathcal O}(1) \right)^{d-1}
\crcr
&=&
\int_{M^{-2 b i}}^{M^{-2 b (i-1)}}
d\alpha  \,
e^{-\alpha \mu_i}
\Big[
\Bigg(\frac{1}{b} \Gamma\left(\frac{1}{2b} \right) \Bigg)^{d-1} \alpha^{-\frac{d-1}{2b}}
+{\mathcal O}(\alpha^{-\frac{d-2}{2b}}) 
 \Big]
\crcr
&=&
\int_{M^{-2 b i}}^{M^{-2 b (i-1)}}
d\alpha 
(1   + {\mathcal O}(\alpha) 
)
\Big[
\Bigg(\frac{1}{b} \Gamma\left(\frac{1}{2b} \right) \Bigg)^{d-1} \alpha^{-\frac{d-1}{2b}}
+{\mathcal O}(\alpha^{-\frac{d-2}{2b}}) 
 \Big]
\crcr
&=&
\Big[
\Bigg(\frac{1}{b} \Gamma\left(\frac{1}{2b} \right) \Bigg)^{d-1} 
\frac{1}{1-\frac{d-1}{2b}}\alpha^{1-\frac{d-1}{2b}}
+{\mathcal O}(\alpha^{1-\frac{d-2}{2b}}) 
 \Big]_{M^{-2 b i}}^{M^{-2 b (i-1)}}
 \crcr
&=&
\Bigg(\frac{1}{b} \Gamma\left(\frac{1}{2b} \right) \Bigg)^{d-1} 
\frac{2b}{2b-d+1}
M^{-i(2b-d+1)}
\Big( 
M^{(2b-d+1)} - 1
\Big)
+{\mathcal O}(M^{-i(2b-d+2)})
 \,,
 \qquad
 \label{eq:interm}
\eea
For the model $+$, we use 
$ b = \frac{1}{2} D (d - \frac{3}{2})$
for just-renormalizability (see Proposition \ref{prop:list+}) with $D = 1$, then:
\beq
\widetilde
S_{1,i}
=
\Bigg( \frac{1}{b} \Gamma\left(\frac{1}{2b} \right) \Bigg)^{d-1} 
(2d-3)
M^{i/2}
\Big(  1 - M^{-1/2} \Big)
+{\mathcal O}(M^{-i/2}) 
\,.
\label{eq:runningmass2}
\eeq

\noindent{\bf  $\widetilde S_{1,i}$ and  $\widetilde S_{2,i}$ for the model $\times$.} 
For the model $\times$, from the Proposition \ref{prop:listx}, we have a specific set of parameters, i.e., $D=1$, $d=3$, $a=\frac{1}{2}$, $b=1$, and $\frac{2a+1}{2b}=1$. 
We then specialize the previous computation.  Starting from \eqref{eq:S2icommon},
we have 
\bea
\widetilde S_{2,i} &=&
\int_{M^{-2i}}^{M^{-2(i-1)}}
d\alpha  
(1   + {\mathcal O}(  \alpha) ) 
\Big[ \alpha^{-1}   +{\mathcal O}(1)   \Big]
 \crcr
 &=&
 \Big[\log \alpha + { \cO (\alpha) } 
 \Big]_{M^{-2i}}^{M^{-2(i-1)}}
 \crcr
 &=&
2 \log M  +   \cO(M^{-2(i-1)})
\,.
\label{eq:S2iexacttimes}
\eea

For the model $\times$, from the Proposition \ref{prop:listx}, we have a specific set of parameters, i.e., $D=1$, $d=3$, $a=\frac{1}{2}$, $b=1$, and $\frac{d-1}{2b}=1$.
So, recomputing \eqref{eq:interm}, we get: 
\bea
\widetilde
S_{1,i}
&=&
\int_{M^{-2  i}}^{M^{-2  (i-1)}}
d\alpha 
(1   + {\mathcal O}( \alpha) 
)
\Big[
\Big( \Gamma\left(\frac{1}{2} \right) \Big)^{2}   \alpha^{-1}
+{\mathcal O}(\alpha^{-\frac{1}{2}}) 
 \Big]
\crcr
&=&
\Big[ \pi  \log \alpha + \cO (\alpha^{\frac{1}{2}})
\Big]^{M^{-2  (i-1)}}_{M^{-2  i}}
\crcr
&=&
2\pi \log M 
+ \cO \big( {M^{-  i}}\big)
\,,
\label{eq:runningmass2x}
\eea
where we substitute $ \Gamma\left(\frac{1}{2} \right) = \sqrt {\pi}$.

\ 

\noindent{\bf
Dimensionful computations of $S_{1,i}$ \eqref{s1idebut} and $S_{2,i}$ \eqref{s2idebut}.} 
Following the similar argument as for $S_{0,i}$,
we perform the changes of variables given in \eqref{eq:dimfuldimless} so that the dimensions are explicit in computations.
Then, we obtain in terms of dimensionless $\widetilde S_{1,i}$ and $\widetilde S_{2,i}$: 
\bea
&&
S_{1,i}   
=
k^{- 2b} 
k^{d-1}
\widetilde S_{1,i}  
\crcr
&&
S_{2,i} 
=
k^{-2 b}\, 
k^{2 a}\, 
k
\widetilde S_{2,i} 
\,,
\label{eq:S1S2dimfuldimless}
\eea
where in the last equalities, we note 
from Propositions \ref{prop:list+} and \ref{prop:listx} that
\bea
    d-1- 2 b = 2 a - 2b +1
    = 
\begin{cases}
    \frac{1}{2} 
    ,& \text{for the model } +\\
    0 
    ,              & \text{for the model } \times
\end{cases}
\eea

\section{Higher order perturbation of model $+$}
\label{app:2ndO+}

This appendix illustrates the higher order corrections of the different couplings RG flow.
These manifestly highlight the structure of the RG flow  at higher loops.

\subsection{4-point function at third order}
\label{app:4pt2+}

\begin{itemize}
    \item The graph $n_{42}^{(c)}$ has degree of divergence $\omega_{n_{42}^{(c)}} = 0$, following the first row of Table \ref{tab:listprim1}, and its amplitude appears at the third order in perturbation theory.

\begin{figure}[H]
\centering
     \begin{minipage}[t]{0.8\textwidth}
      \centering
\includegraphics[angle=0, width=12cm, height=4cm]{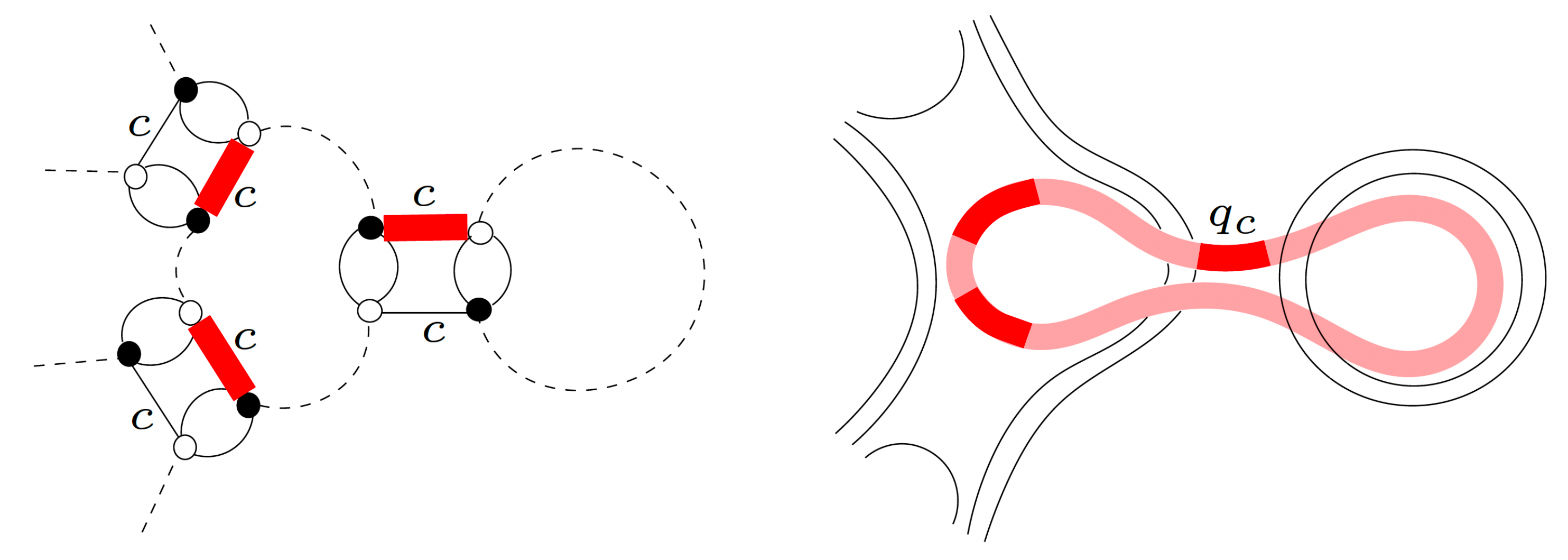}
\caption{ {\small  
A two-loop divergent graph, $n_{42}^{(c)}$  contributing to $\Gamma_4^{(c)}(\{p\})$, at $d=3$.  
}} 
\label{fig:V4twoloop}
\end{minipage}
\end{figure}

Explicitly, 
at two-loop,
\bea
&&
 K_{n_{42}^{(c)} } = 12 \,, 
 \crcr
 &&
  S_{n_{42}^{(c)}} (\{\mathbf{p}, \mathbf{p}'\}) = \frac{1}{3!}
  \Big(\frac{-\lapc}{2}\Big)^3 
  \sum_{\{q\}} 
  \frac
  {
  \big(\vert q_{c}\vert^{2a}\big)^{3}
  }
  {
  (|\mathbf{p}_{\check c}|^{2b} + \vert q_{c}\vert^{2b} + \mu)
    ( |\mathbf{p}'_{\check c}|^{2b} + \vert q_{c}\vert^{2b} + \mu)^2
    ( \vert \mathbf{q}\vert^{2b} + \mu)
    }
    \,.
    \nonumber
\eea
The contribution to the Feynman amplitude will be of the form 
$\sum_c K_{n_{42}^{(c)}} S_{n_{42}^{(c)}}$.

\end{itemize}

\subsection{$\Gamma_2$ at second order}
\label{app:Gamma2+}
The following graphs
will contribute to the flow of the $Z_a$ coupling. Computing it, we distinguish the colors and introduce $Z_a^{(c)}$
which makes us 
work at fixed color $c$.

\begin{itemize}

\item
For $m_{2 e}^{(c \, c')}$,
$\omega_{m_{2 e}^{(c \, c')}} = 0$.
This graph belongs to the class V in Table \ref{tab:listprim1}. 
\begin{figure}[H]
\centering
     \begin{minipage}[t]{0.7\textwidth}
      \centering
\includegraphics[angle=0, width=5.5cm, height=2.5cm]{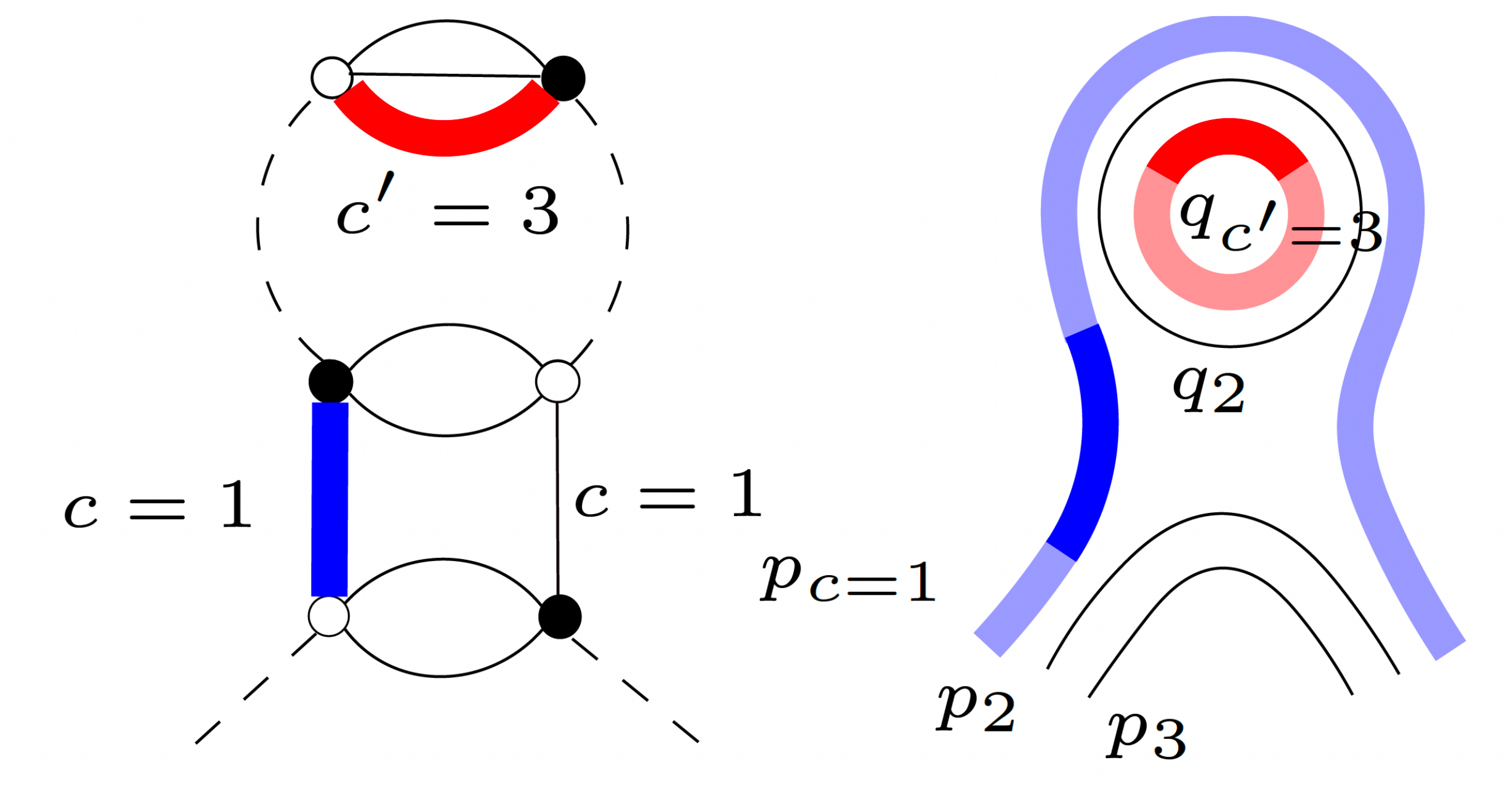}
\caption{{\small  The graph $m_{2 e}^{(c=1 \, c'=2)}$, for $d=3$ is illustrated. $\{q_{\check c= \check{1} }\} = \{q_{c'=2}, \; q_{3}\}$, and ${|{\bf q}_{\check{1}}|}^{2 b} = {|q_{2}|}^{2b} + {|q_{3}|}^{2 b} $.}}
\label{fig:m2e}
\end{minipage}
\end{figure}
The contribution to the Feynman amplitude is then,
\bea
\sum_{c'}
K_{m_{2 e}^{(c \, c')}} S_{m_{2 e}^{(c \, c')}} (\{p \})
& =& 
\sum_{c' \ne c}^{(d-1) \; {\rm terms}}
2 
\Big[
\Big(- \frac{{\lambda_{+}}^{(c)}}{2}\Big) (- Z_a^{(c')}) 
\Big]
\sum_{\{q_{\hat c}\}}
\frac{{\vert q_{c'} \vert}^{2 a} {\vert p_{c}\vert}^{2 a}} {( {|{\bf q}_{{(\check c)}}|}^{2 b}   + \vert{ p_{c}\vert}^{2 b} + \mu)^2}
\nonumber 
\\
&=&
\lambda_{+}^{(c)}  
{\vert p_{c}\vert}^{2 a}
\sum_{c' \ne c}^{(d-1) \; {\rm terms}}
 Z_a^{(c')}
\sum_{\{q_{\check c}\}}
\frac{{\vert q_{c'}\vert}^{2 a}} {( {|{\bf q}_{{(\check c)}}|}^{2 b}   + {\vert p_{c}\vert}^{2 b} + \mu)^2}
\,,
\eea
where $K_{m_{2 e}^{(c \, c')}} = 2$.

\item 
The graph $meme_2^{(c \, c')}$,
is such that 
$\omega_{meme_2^{(c \, c')}} = \frac{D}{2}$ and it 
appears in the class II in Table \ref{tab:listprim1}. 
\begin{figure}[H]
\centering
     \begin{minipage}[t]{0.7\textwidth}
      \centering
\includegraphics[angle=0, width=5cm, height=5cm]{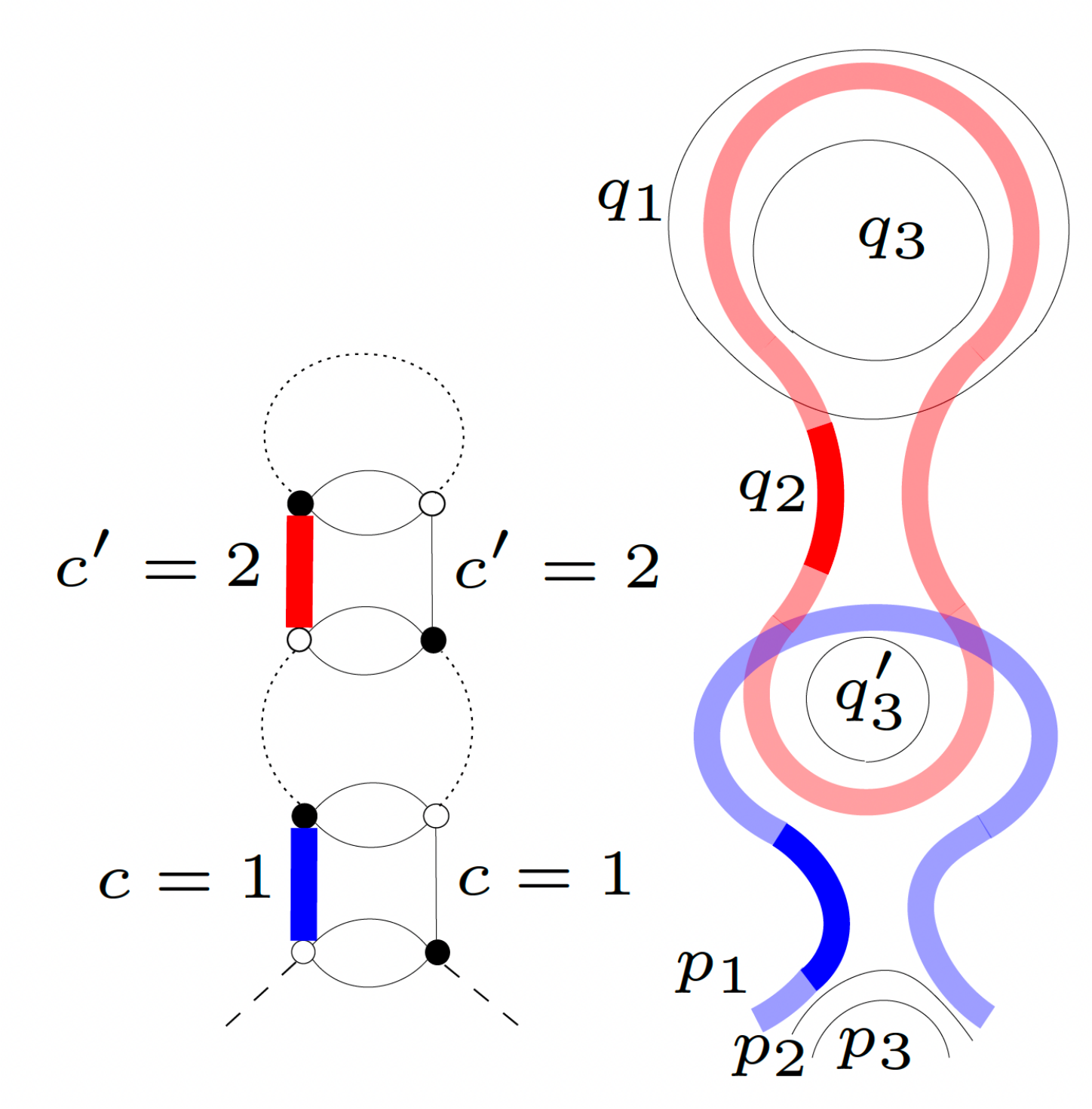}
\caption{  {\small  For $d=3$, $meme_2^{(c=1 \, c'=2)}$ is illustrated. 
$q'_{({\check c}{\check {c'}})}={q'}_{3}$,
$\{q\}=\{ q_{1}, q_{2}, q_{3}\}$,
${|{\bf q}|}^{2 b}={\vert q_{1}\vert}^{2 b} + {\vert q_{2}\vert}^{2 b} + {\vert q_{3}\vert}^{2 b} $,
${|{\bf q}_{({\check c}{\check {c'}})}|}^{2 b}= {\vert q_{3}\vert}^{2 b}$.}}
\label{fig:meme2}
\end{minipage}
\end{figure}
\bea
&&
\sum_{ c' \ne c} 
K_{meme_2^{(c, c')}} S_{meme_2^{(c, c')}} (\{ p\}) 
\crcr
&=&
\sum_{c' \ne c}^{(d-1) \; {\rm terms}}
K_{meme_2^{(c, c')}} 
\Big( - \frac{\lambda_{+}^{(c)}}{2}\Big)
\Big( - \frac{\lambda_{+}^{(c')}}{2}\Big) 
{\vert p_{c}\vert}^{2 a} 
\nonumber
\\
 && \qquad \quad
\sum_{\{q'_{({\check c}{\check {c'}})}\}, \{q\}}
\frac{{\vert q_{c'}\vert}^{2 a}}{({|{\bf q}|}^{2 b} +  \mu) ({|{\bf q'}_{({\check c}{\check {c'}})}|}^{2 b}  +{\vert q_{c'} \vert}^{2 b} + {\vert p_{c}\vert}^{2 b} + \mu)^2}
\eea
where $K_{meme_2^{(c, c')}}=4$, and where we introduced the notation $q'_{({\check c}{\check {c'}})}$ which means that the color of this momentum cannot be neither $c$ nor $c'$.

\end{itemize}

\subsection{Self energy and mass corrections}
\label{app:Sigma+}

We gather here the contributions to the self-energy and wave function renormalization $Z_b$ at second order of perturbation. 

\begin{itemize}
\item
For the graph $m_2^{(c, c')}$, the degree of divergence is given by
$\omega_{d;+}(m_2^{(c, c')})= 0$.
 The class VI in Table \ref{tab:listprim1} includes this graph.
\begin{figure}[H]
\centering
     \begin{minipage}[t]{0.6\textwidth}
      \centering
\includegraphics[angle=0, width=5cm, height=2.5cm]{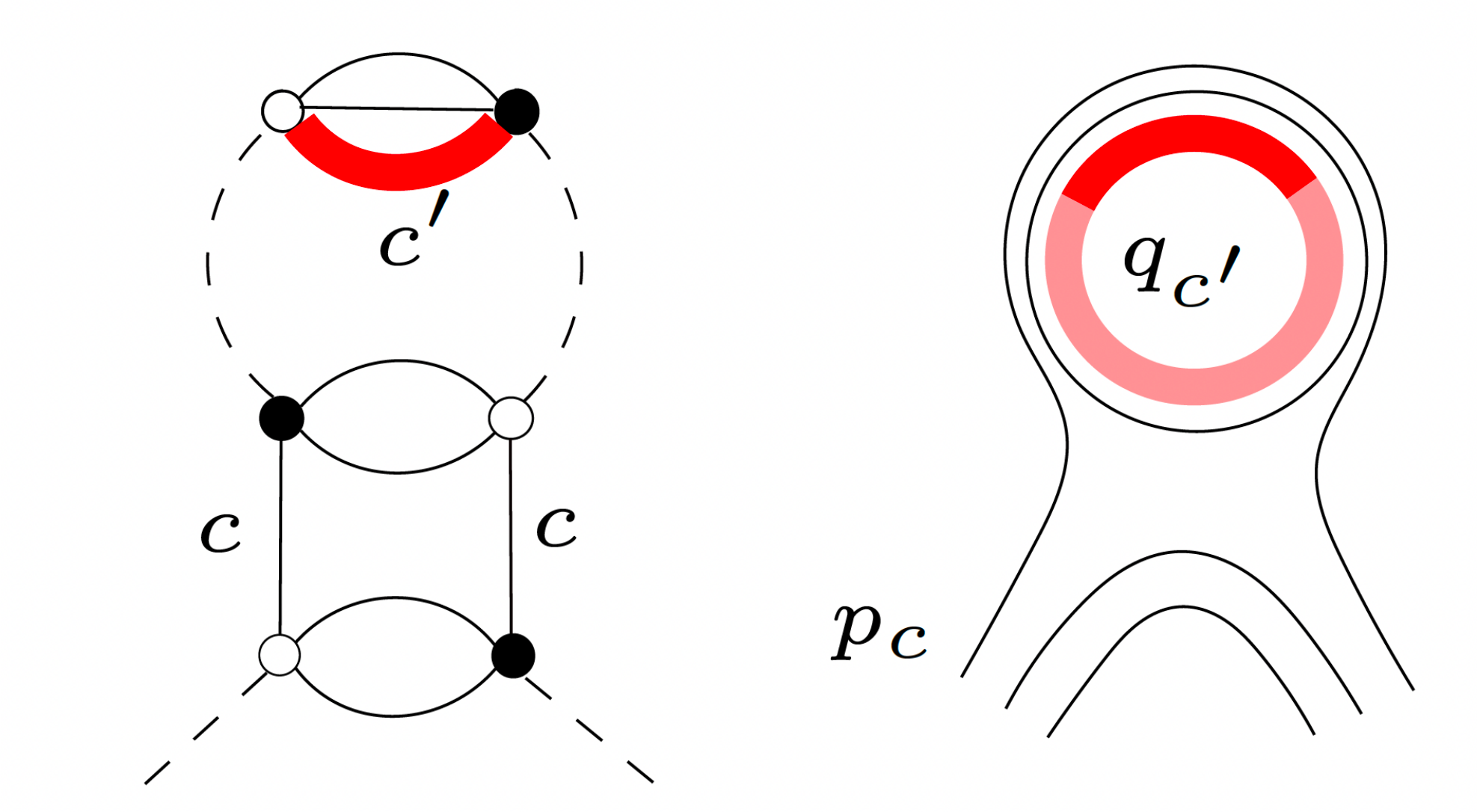}
\caption{{\small $m_2^{(c, c')}$ in colored and stranded representations at $d=3$. ${|{\bf q}_{\check {c}}|}^{2 b} = {|q_{c'}|}^{2 b} + {| q_{1}|}^{2b}$ and $\{q_{\check c}\} =\{ q_{c'}, \; q_{1}\}$.}}
\label{fig:m2}
\end{minipage}
\end{figure}
The contribution of the graph $m_2^{(c, c')}$ shown in Fig. {\ref{fig:m2}} in the amplitude is
\bea
\sum_{c, c' \ne c} K_{m_2^{(c, c')}}S_{m_2^{(c, c')}} (\{ p\}) 
&=&
\sum_{c, c' \ne c}^{d \times (d-1) \; {\rm terms}}
2
\Big[ 
\Big( - \frac{\lambda^{(c)}}{2}\Big) 
\Big( - {Z_a}^{(c')} \Big) 
\Big]
\sum_{\{q_{\check c}\}}
\frac{\vert q_{c'}\vert^{2 a}}{({|{\bf q}_{\check c}|}^{2 b} + \vert p_{c}\vert^{2 b} + \mu)^2} \nonumber \\
&=&
\sum_{c=1}^{d}
\lambda^{(c)} 
\sum_{c' \ne c}
\sum_{\{q_{\check c}\}}
\frac{ {Z_a}^{(c')} \vert q_{c'}\vert^{2 a}}{({|{\bf q}_{\check c}|}^{2 b} + \vert p_{c}\vert^{2 b}+ \mu)^2}
\,,
\eea
where $K_{m_2^{(c, c' \ne c)}} = 2$.

\item
Concerning the graph $mme_2^{(c, c')}$, the superficial degree of divergence is
$\omega_{d;+}(mme_2^{(c, c')}) = \frac{D}{2}$.
It can be found in the class III in Table \ref{tab:listprim1}. 
\begin{figure}[H]
\centering
     \begin{minipage}[t]{0.7\textwidth}
      \centering
\includegraphics[angle=0, width=4.5cm, height=4.5cm]{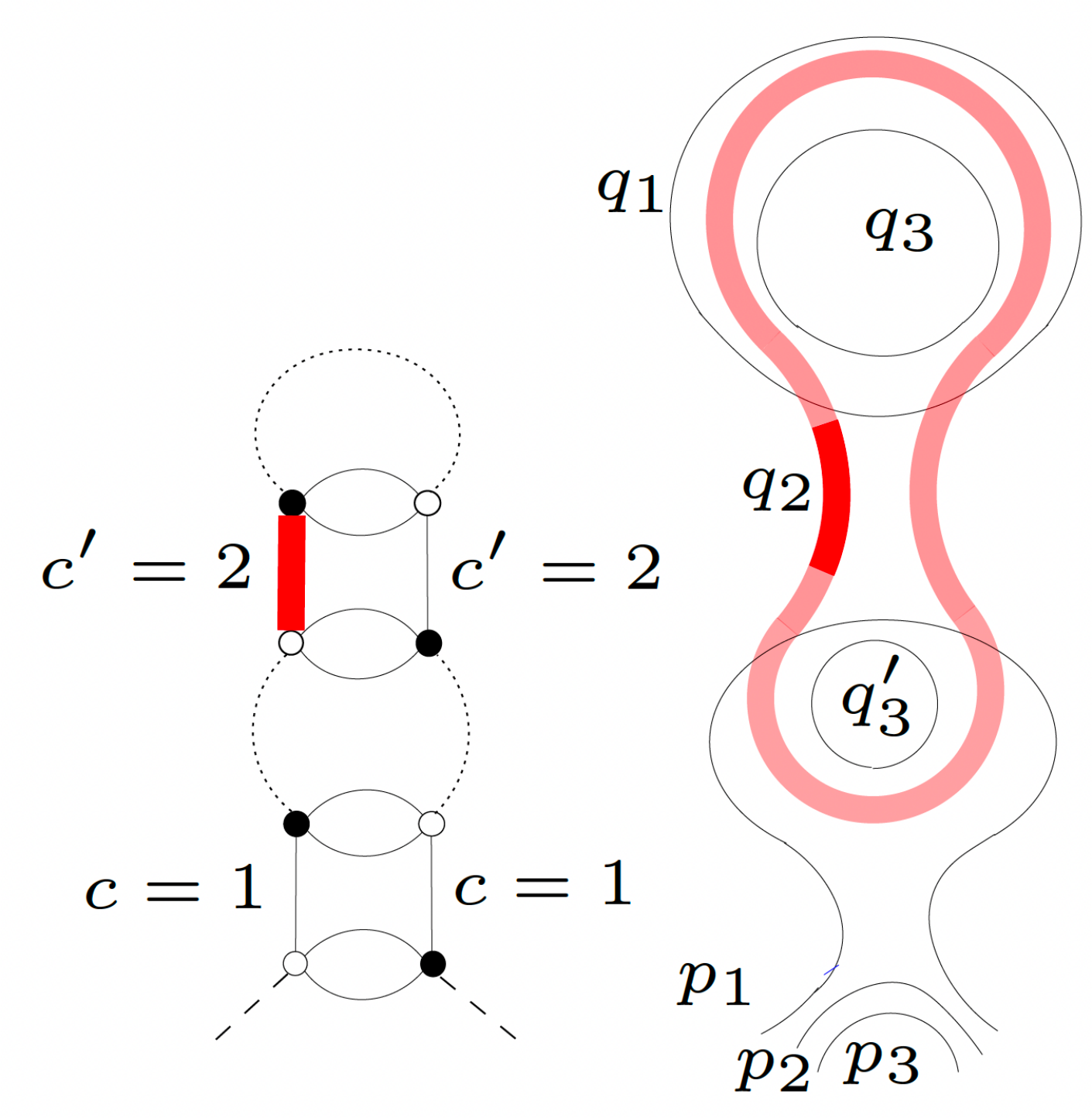}
\caption{{\small  $mme_2^{(c=1, c'=2)}$,  for $d=3$ in colored and stranded representations, 
with 
$q'_{({\check c}{\check {c'}})}={q'}_{3}$, 
${|{\bf q}_{({\check c}{\check {c'}})}|}^{2 b}= {\vert q_{3}\vert}^{2 b}$.}}
\label{fig:mme2}
\end{minipage}
\end{figure}
The amplitude associated with the graph $mme_2^{(c, c')}$ shown in Fig. {\ref{fig:mme2}} is given by
\bea
&&\sum_{c, c' \ne c} K_{mme_2^{(c, c')}} S_{mme_2^{(c, c')}} (\{ p\}) 
\crcr
&=&
\sum_{c, c' \ne c}^{d \times (d-1) \; {\rm terms}}
4
\Big[ 
\Big( - \frac{\lambda^{(c)}}{2}\Big) 
\Big( - \frac{\lambda_{+}^{(c')}}{2}\Big)  
\Big]
\crcr
&&
\qquad \qquad
\sum_{\{q'_{({\check c}{\check {c'}})}\}, \{q\}}
\frac{{\vert q_{c'}\vert}^{2 a}}{({|{\bf q}|}^{2 b} +  \mu) ({|{\bf q'}_{({\check c}{\check {c'}})}|}^{2 b} +{\vert q_{c'} \vert}^{2 b}+ {\vert p_{c}\vert}^{2 b} + \mu)^2}
\crcr
&=&
\sum_{c=1}^{d}
\lambda^{(c)} 
\sum_{c' \ne c}
\sum_{\{q'_{({\check c}{\check {c'}})}\}, \{q\}}
\frac{ \lambda_{+}^{(c')}  {\vert q_{c'}\vert}^{2 a}}{({|{\bf q}|}^{2 b} +  \mu) ({|{\bf q'}_{({\check c}{\check {c'}})}|}^{2 b} +{\vert q_{c'} \vert}^{2 b} + {\vert p_{c}\vert}^{2 b} + \mu)^2}
\,,
\quad
\eea
where $K_{mme_2^{(c, c')}} = 4$.

\item
For the graph $n_2^{(c)}$, 
$\omega_{d;+}(n_2^{(c)}) = 0$, and 
it belongs to the class IV in Table \ref{tab:listprim1}. 
\begin{figure}[H]
\centering
     \begin{minipage}[t]{0.7\textwidth}
      \centering
\includegraphics[angle=0, width=5cm, height=2.5cm]{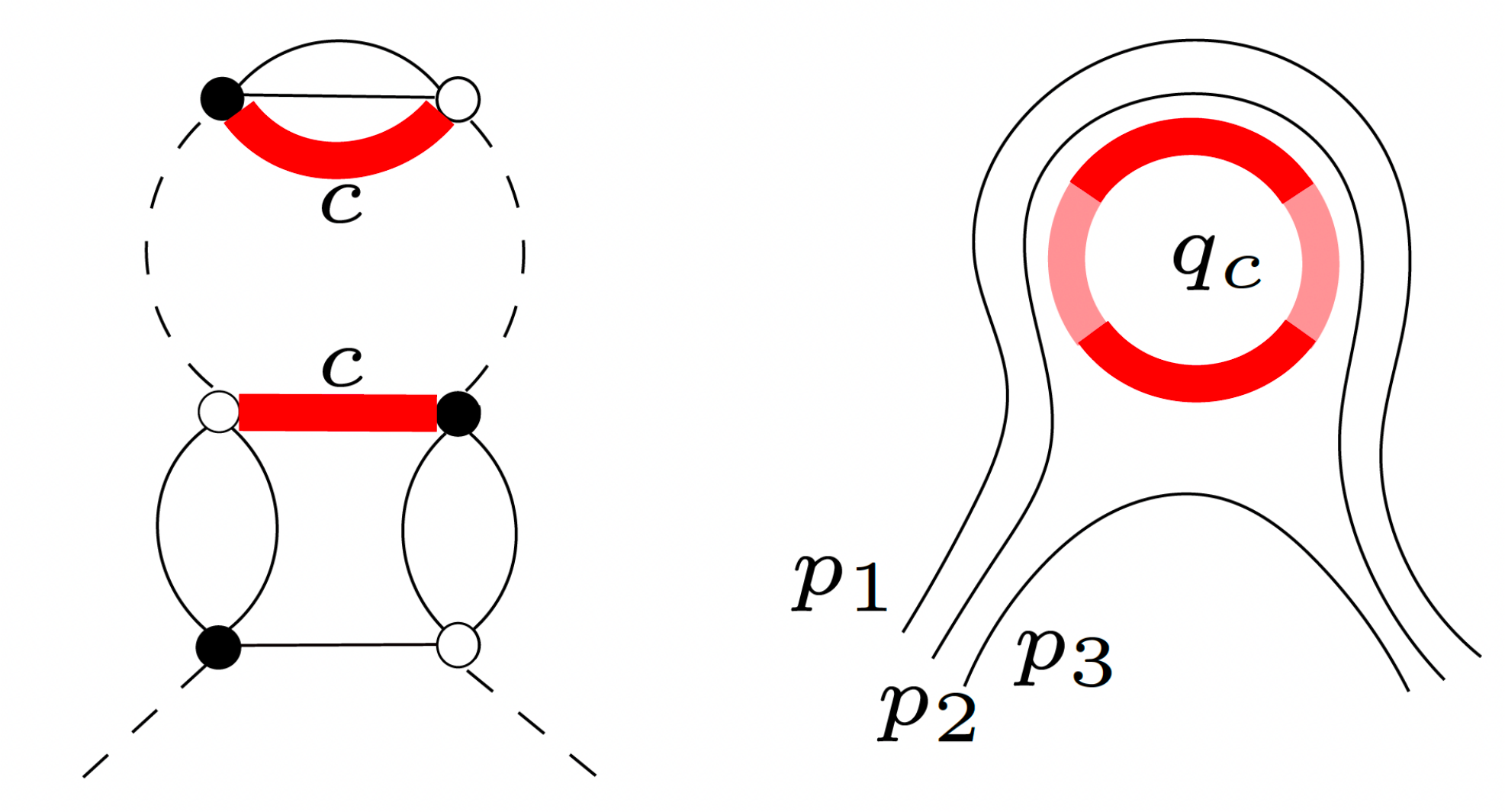}
\caption{ {\small  The illustration of $n_2^{(c)}$ in $d=3$. 
}} 
\label{fig:n2}
\end{minipage}
\end{figure}
Now the amplitude of $n_2^{(c)}$ is
\bea
\sum_c K_{n_2^{(c)}} S_{n_2^{(c)}}( \{p\})
&=&\frac{1}{2}  
\sum_{c=1}^{d} {\lambda_{+}}^{(c)} Z_{a}^{(c)}
\sum_{q_{c}} \frac{(\vert q_{c}\vert ^{2a})^2}{(\vert q_{c}\vert^{2 b} + {|{\bf p}_{\check c}|}^{2 b} + \mu)^2}
\,,
\eea
where  
$K_{n_2^{(c)}} = 1$.

\item
The $nme_2^{(c)}$ 
 belongs to the class I in Table \ref{tab:listprim1}
 and it fulfills 
$\omega_{d;+}(nme_2^{(c)}) = \frac{D}{2}$.
\begin{figure}[H]
\centering
     \begin{minipage}[t]{0.7\textwidth}
      \centering
\includegraphics[angle=0, width=5cm, height=5cm]{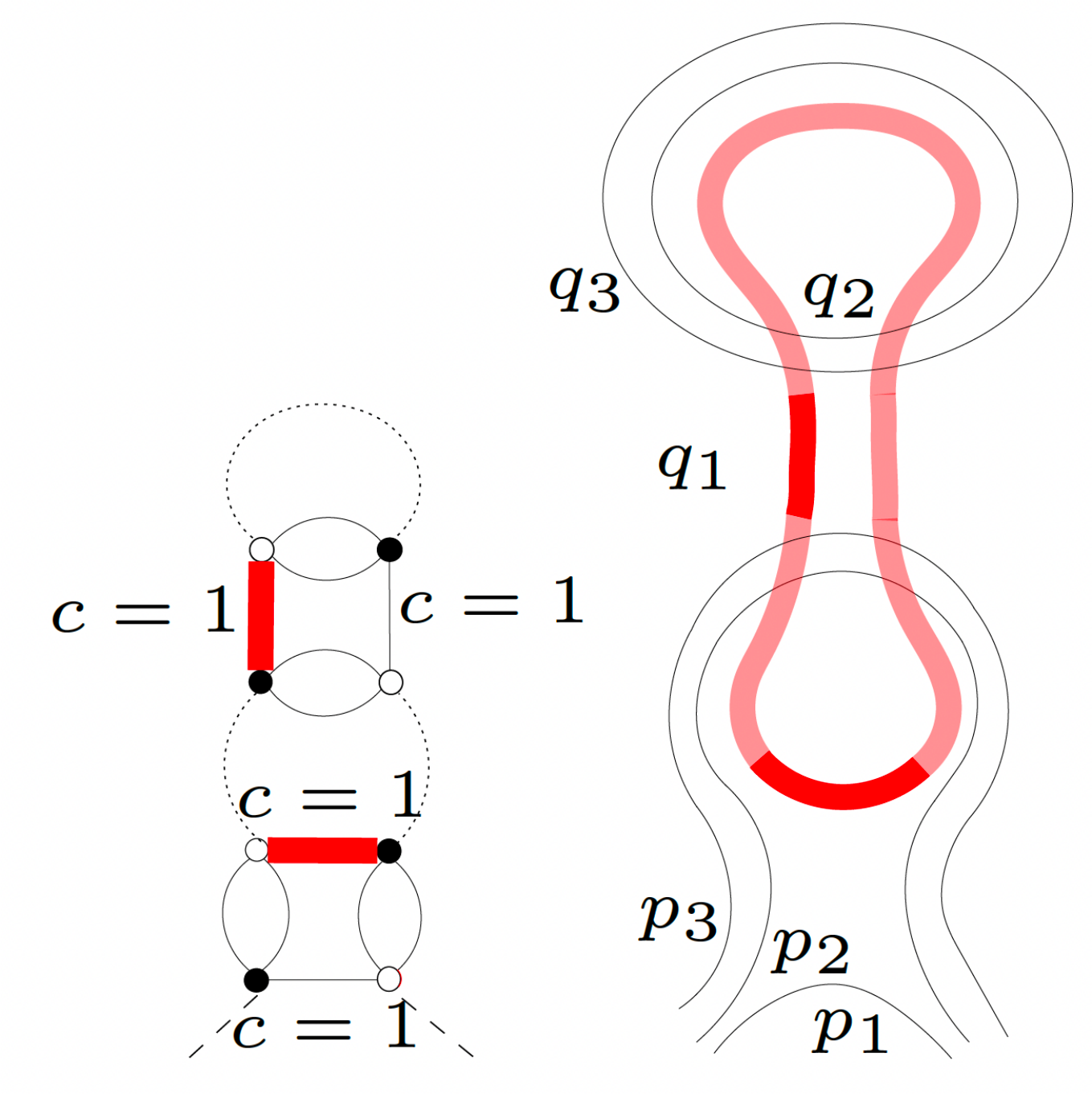}
\caption{ {\small  For $d=3$, the $nme_2^{(c=1)}$ 
with ${|{\bf p}_{\check c}|}^{2 b}={|p_{2}|}^{2 b}+{|p_{3}|}^{2 b}$.
}} 
\label{fig:nme2}
\end{minipage}
\end{figure}
\bea
&&
\sum_c K_{nme_2^{(c)}} S_{nme_2^{(c)}}( \{p\})
\crcr
&=&
\sum_{c=1}^{d}
4
\frac{1}{2!}
\Big [ \Big( - \frac{{\lambda_{+}}^{(c)}}{2}\Big)^2 \Big ]
\sum_{\{q\}} \frac{({\vert q_{c}\vert}^{2a})^2}{({\vert {\bf q}\vert}^{2 b} + \mu)({\vert q_{c}\vert}^{2 b} + {|{\bf p}_{\check c}|}^{2 b} + \mu)^2}
\,,
\crcr
&=&
 \frac{1}{2}
\sum_{c=1}^{d}
({\lambda_{+}}^{(c)})^2 
\sum_{\{q\}} \frac{({\vert q_{c}\vert}^{2a})^2}{({\vert {\bf q}\vert}^{2 b} + \mu)({\vert q_{c}\vert}^{2 b} + {|{\bf p}_{\check c}|}^{2 b} + \mu)^2}
\,,
\eea
where $K_{nme_2^{(c)}} = 4$.

\end{itemize}

\section{Second order perturbation of model $\times$}
\label{app:2ndOx}

\subsection{Self energy and mass corrections}
\label{app:Sigmax}

\begin{itemize}

\item
For $mm_{ee}^{(c, c')}$, the superficial degree of divergence is
$\omega_{d;\times}(mm_{ee}^{(c, c')}) = 0$.
This graph belongs to the class III in Table \ref{tab:listprim2}, and appears in the 
second order $\cO (\lambda \, \lambda_\times)$ in Taylor expansion.

\begin{figure}[H]
\centering
     \begin{minipage}[t]{0.7\textwidth}
      \centering
\includegraphics[angle=0, width=5cm, height=5cm]{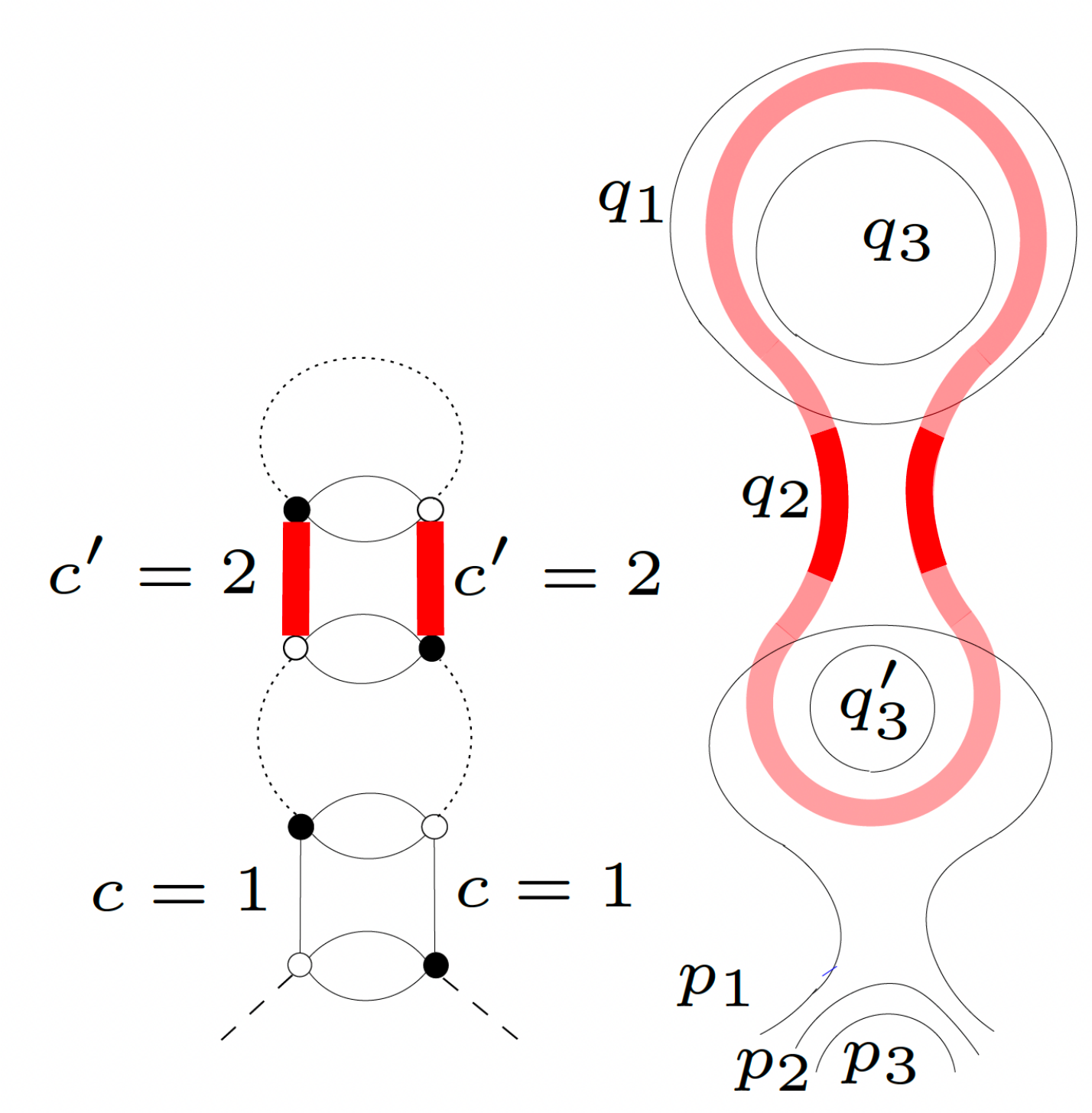}
\caption{{\small  $mm_{ee}^{(c=1, c'=2)}$, for $d=3$  in both colored and stranded representations.   
$\{q'_{({\check c}{\check {c'}})}\}={q'}_{3}$,
${|{\bf q}_{({\check c}{\check {c'}})}|}^{2 b}= {\vert q_{3}\vert}^{2 b}$.
}} 
\label{fig:mmee}
\end{minipage}
\end{figure}
The amplitude associated with the graph $mm_{ee}^{(c, c')}$ shown in Fig. {\ref{fig:mmee}} is given by
\bea
&&\sum_{c, c' \ne c} K_{mm_{ee}^{(c, c')}} S_{mm_{ee}^{(c, c')}} (\{ p\}) 
\crcr
&=&
\sum_{c, c' \ne c}^{d \times (d-1) \; {\rm terms}}
4
\Big[ 
\Big( - \frac{\lambda^{(c)}}{2}\Big) 
\Big( - \frac{\lambda_{\times}^{(c')}}{2}\Big)  
\Big]
\crcr
&&
\qquad \qquad
\sum_{\{q'_{({\check c}{\check {c'}})}\}, \{q\}}
\frac{{\vert q_{c'}\vert}^{4 a}}{({|{\bf q}|}^{2 b} +  \mu) ({|{\bf q'}_{({\check c}{\check {c'}})}|}^{2 b} +{\vert q_{c'} \vert}^{2 b}+ {\vert p_{c}\vert}^{2 b} + \mu)^2}
\crcr
&=&
\sum_{c=1}^{d}
\lambda^{(c)} 
\sum_{c' \ne c}
\sum_{\{q'_{({\check c}{\check {c'}})}\}, \{q\}} 
\frac{  \lambda_{\times}^{(c')}{\vert q_{c'}\vert}^{4 a}}{({|{\bf q}|}^{2 b} +  \mu) ({|{\bf q'}_{({\check c}{\check {c'}})}|}^{2 b} +{\vert q_{c'} \vert}^{2 b} + {\vert p_{c}\vert}^{2 b} + \mu)^2}
\,,
\eea
where  $K_{mm_{ee}^{(c, c')}} = 4$.
%\sum_{c' \ne c}

\end{itemize}

\subsection{$\Gamma_{2;a}$ at second order}
\label{app:Gamma2ax}

We work out the contribution to the flow of $Z_{a}^{(c)}$ at fixed color $c$. 

\begin{itemize}
\item
The graph $n_{ee}m_{ee}^{(c)}$
obeys  
$\omega_{d; \times}(n_{ee}m_{ee}^{(c)}) = 0$
and  belongs to the class I in Table \ref{tab:listprim2}. 
\begin{figure}[H]
\centering
     \begin{minipage}[t]{0.7\textwidth}
      \centering
\includegraphics[angle=0, width=5cm, height=5cm]{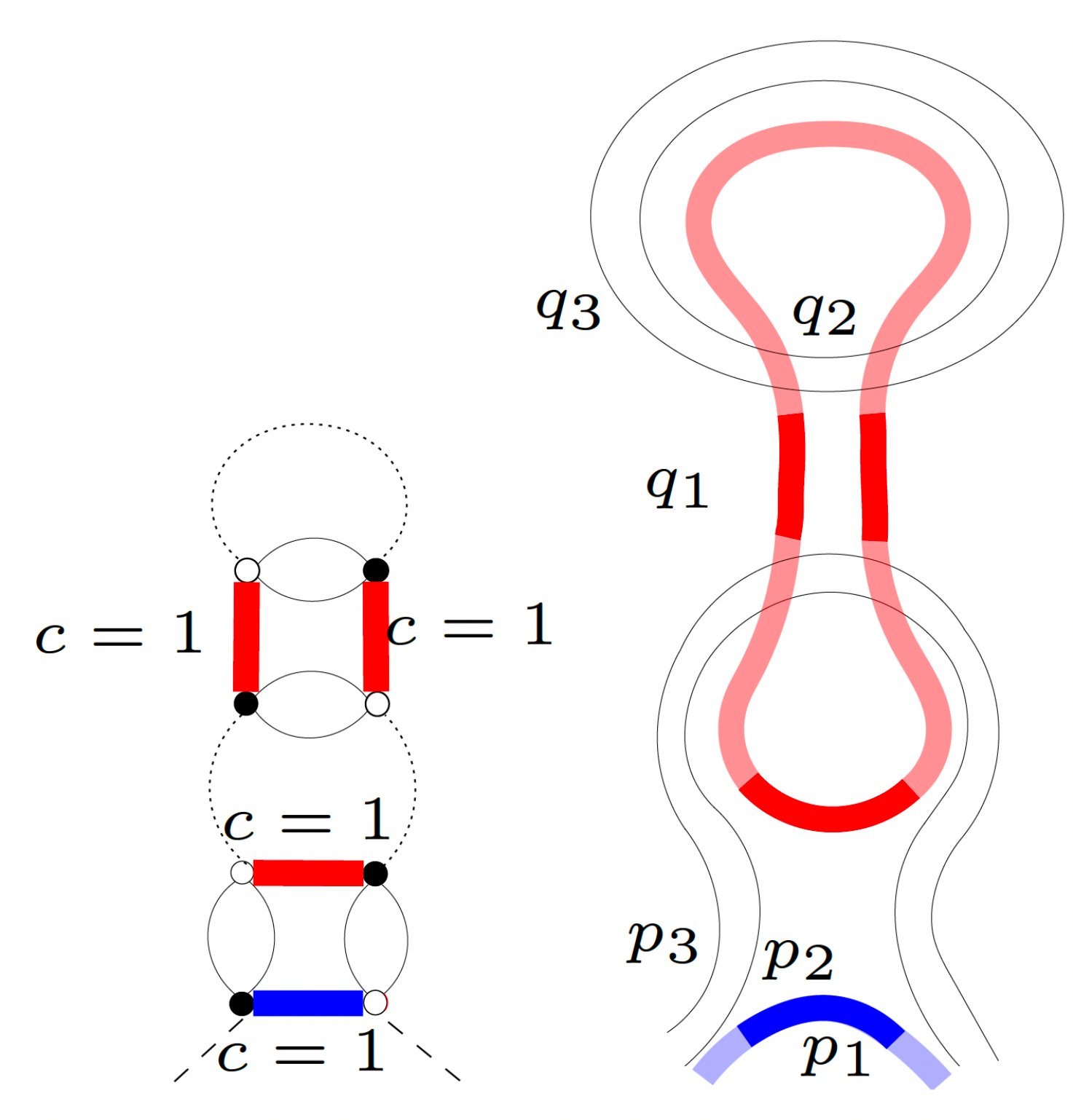}
\caption{ {\small  
The graph  $n_{ee}m_{ee}^{(c=1)}$ 
at $d=3$. ${|{\bf p}_{\check c = \check{1}}|}^{2 b}={|p_{2}|}^{2 b}+{|p_{3}|}^{2 b}$. 
}} 
\label{fig:neemee}
\end{minipage}
\end{figure}
\bea
&&
K_{n_{ee}m_{ee}^{(c)}} S_{n_{ee}m_{ee}^{(c)}}( \{p\})
\nonumber
\\
&=&
8 
\,
\frac{1}{2!}
 \Big( - \frac{{\lambda_{\times}}^{(c)}}{2}\Big)^2 
\vert p_{c}\vert^{2a}
\sum_{\{q\}} \frac{({\vert q_{c}\vert}^{2a})^3}{({\vert {\bf q}\vert}^{2 b} + \mu)({\vert q_{c}\vert}^{2 b} + {|{\bf p}_{(\check c)}|}^{2 b} + \mu)^2}
\, \crcr
&=& 
{\lambda_{\times}}^{(c)}
\vert p_{c}\vert^{2a}
\sum_{\{q\}} \frac{({\vert q_{c}\vert}^{2a})^3}{({\vert {\bf q}\vert}^{2 b} + \mu)({\vert q_{c}\vert}^{2 b} + {|{\bf p}_{(\check c)}|}^{2 b} + \mu)^2}
\, 
\eea
where we set $K_{n_{ee}m_{ee}^{(c)}} = 8$.

\end{itemize}

\subsection{$\Gamma_{2; 2a}$ at second order}
\label{app:Gamma22ax}

We list below the second order contributions to 
the flow of the last coupling $Z_{2a}^{(c)}$. 
We investigate this at given color $c$. 

\begin{itemize}

\item 
The graph $m_{ee}m_{ee}^{(c, c')}$
is part of the class II in Table \ref{tab:listprim2}. 
It has vanishing divergence degree:  
$\omega_{d; \times}(m_{ee}m_{ee}^{(c, c')}) = 0$. 
\begin{figure}[H]
\centering
     \begin{minipage}[t]{0.7\textwidth}
      \centering
\includegraphics[angle=0, width=5cm, height=5cm]{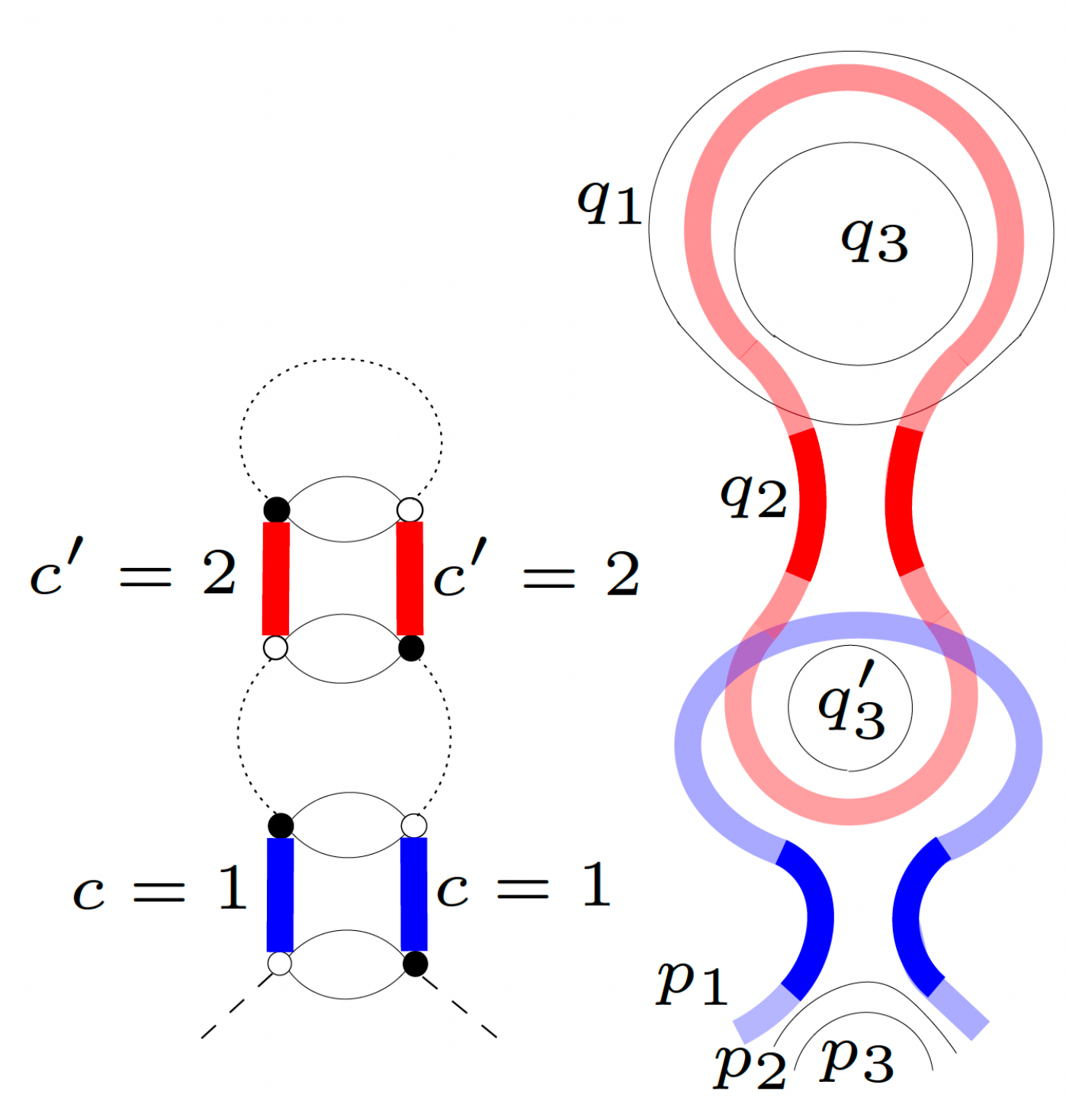}
\caption{ {\small 
The graph $m_{ee}m_{ee}^{(c=1, c'=2)}$ illustrated at $d=3$.  
$q'_{({\check c}{\check {c'}})}={q'}_{3}$,
${|{\bf q}_{({\check c}{\check {c'}})}|}^{2 b}= {\vert q_{3}\vert}^{2 b}$.}}
\label{fig:meemee}
\end{minipage}
\end{figure}
\bea
&&
\sum_{ c' \ne c} 
K_{m_{ee}m_{ee}^{(c, c')}} S_{m_{ee}m_{ee}^{(c, c')}} (\{ p\}) 
\crcr
&=&
\sum_{c' \ne c}^{(d-1) \; {\rm terms}}
4
\Big[ 
\Big( - \frac{\lambda_{\times}^{(c)}}{2}\Big) 
\Big( - \frac{\lambda_{\times}^{(c')}}{2}\Big)  
\Big]
{\vert p_{c}\vert}^{4 a} 
\nonumber
\\
 && \qquad \quad
\sum_{\{q'_{({\check c}{\check {c'}})}\}, \{q\}}
\frac{{\vert q_{c'}\vert}^{4 a}}{({|{\bf q}|}^{2 b} +  \mu) ({|{\bf q'}_{({\check c}{\check {c'}})}|}^{2 b}  +{\vert q_{c'} \vert}^{2 b} + {\vert p_{c}\vert}^{2 b} + \mu)^2}
\crcr
&=&
\lambda_{\times}^{(c)} 
{\vert p_{c}\vert}^{4 a} 
\sum_{c' \ne c} 
\sum_{\{q'_{({\check c})}\}, \{q\}}
\frac{\lambda_{\times}^{(c')}  {\vert q_{c'}\vert}^{4 a}}{({|{\bf q}|}^{2 b} +  \mu) ({|{\bf q'}_{({\check c}{\check {c'}})}|}^{2 b} +{\vert q_{c'} \vert}^{2 b} + {\vert p_{c}\vert}^{2 b} + \mu)^2}
\,,
\quad
\eea
where  $K_{m_{ee}m_{ee}^{(c, c')}}=4$.

\end{itemize}

\end{document}